\documentclass[notitlepage,a4paper,prd,noshowpacs,noshowkeys,nofootinbib,preprintnumbers]{revtex4}
\usepackage[latin1]{inputenc}
\usepackage{ae,aecompl}
\usepackage{amsmath,amssymb,mathrsfs}
\usepackage{graphicx}
\usepackage{enumerate} %\begin{enumerate}[oprtion], option=1,a,A,i,I
\usepackage[usenames,dvipsnames]{color} % p usar nomes de cores
\usepackage{hyperref}
\hypersetup{
   colorlinks=true,       % false: boxed links; true: colored links
    linkcolor=Blue,          % color of internal links
    citecolor=Plum,        % color of links to bibliography
    filecolor=magenta,      % color of file links
    urlcolor=YellowOrange           % color of external links
}
\usepackage{ifthen}

\providecommand{\upq}[1][]{
\ifthenelse{\equal{#1}{1}}{U(1)_{\rm PQ_{1}}}{}
\ifthenelse{\equal{#1}{2}}{U(1)_{\rm PQ_{2}}}{}
}

\usepackage{xspace}
\providecommand{\PQ}{Peccei-Quinn\xspace}

\providecommand{\lag}{\mathscr{L}}

\providecommand{\tG}{\tilde{G}}
\providecommand{\tF}{\tilde{F}}

\providecommand{\om}[1][]{\omega_{#1}}
\providecommand{\eff}{\mathrm{eff}}

\providecommand{\guniv}{g^{(0)}_{A\gamma}}

\providecommand{\by}{\mathbf{y}}
\providecommand{\bz}{\mathbf{z}}
\providecommand{\hy}{\hat{y}}
\providecommand{\hz}{\hat{z}}
\providecommand{\hby}{\hat{\mathbf{y}}}
\providecommand{\hbz}{\hat{\mathbf{z}}}
\providecommand{\bu}{\mathbf{u}}

\providecommand{\eqali}[1]{\begin{equation}\begin{aligned} #1
    \end{aligned}\end{equation}}
\providecommand{\subeqali}[2][]{\begin{subequations}\label{#1}\begin{align} #2
    \end{align}\end{subequations}}
\providecommand{\eq}[1]{\begin{equation} #1 \end{equation}}
\providecommand{\aver}[1]{\langle #1 \rangle}
\providecommand{\ums}[2][1]{\ml{\tfrac{#1}{#2}}} %small fraction in equations
\providecommand{\ml}[1]{\mbox{\large $#1$}}
\providecommand{\mfn}[1]{\mbox{\footnotesize $#1$}}
\providecommand{\ZZ}{\mathbb{Z}}
\providecommand{\ponto}{{\cdot}}

\begin{document}
%%%%%%%%%%%%%%%%%%%%%%%%%%%%%%%%%%%%%%%%%%%%%%%%%
\preprint{DESY 14-020}
\title{
The Quest for an Intermediate-Scale Accidental Axion and Further ALPs
}
\author{A.~G.~Dias$^1$}%
\email{alex.dias@ufabc.edu.br}
\author{A.~C.~B.~Machado$^2$}%
\email{a.c.b.machado1@gmail.com}
\author{C.~C.~Nishi$^1$}%
\email{celso.nishi@ufabc.edu.br}
\author{A. Ringwald$^3$}%
\email{andreas.ringwald@desy.de}
\author{P. Vaudrevange$^4$}%
\email{patrick.vaudrevange@tum.de}
\affiliation{
$^1$Universidade Federal do ABC - UFABC,
Santo Andr\'e, S\~ao Paulo, Brazil\\
$^2$Instituto  de F\'\i sica Te\'orica--Universidade Estadual Paulista,
R. Dr. Bento Teobaldo Ferraz 271, Barra Funda\\ S\~ao Paulo - SP, 01140-070, Brazil\\
$^3$Deutsches Elektronen-Synchrotron, DESY, Notkestr. 85, 22607 Hamburg, Germany\\
$^4$Excellence Cluster Universe, Technische Universit\"at M\"unchen,
Boltzmannstr. 2, D-85748, Garching, Germany\\
}
%\date{March 30, 2013}
%\date{\today}
%%%%%%%%%%%%%%%%%%%%%%%%%%%%%%%%%%%%%%%%%%%%%%%%%
\begin{abstract}
The recent detection of the cosmic microwave background polarimeter
experiment BICEP2 of tensor fluctuations in the B-mode power spectrum basically excludes
all plausible axion models where its decay constant is above $10^{13}$\,GeV.
Moreover, there are strong theoretical, astrophysical, and cosmological motivations for models involving,
in addition to the axion, also
axion-like particles (ALPs), with  decay constants in the
intermediate scale range, between $10^9\,{\rm GeV}$ and $10^{13}\,{\rm GeV}$.
Here, we present a general  analysis of models with an axion and further ALPs and derive bounds on the relative size
of the axion and ALP photon (and electron) coupling.
We discuss what we can learn from measurements of the axion and
ALP photon couplings
about the fundamental parameters of the underlying ultraviolet completion of the
theory.  For the latter we consider extensions of the Standard Model
in which the axion and the ALP(s) appear as pseudo Nambu-Goldstone bosons from the breaking of
global chiral $U(1)$ (Peccei-Quinn (PQ)) symmetries, occurring accidentally as low
energy remnants from
exact discrete symmetries. In such models, the axion and
the further ALP are protected from disastrous explicit symmetry breaking
effects due to Planck-scale suppressed operators. The scenarios considered exploit heavy right
handed neutrinos getting their mass via PQ symmetry breaking and thus explain the small mass of the
active neutrinos via a seesaw relation between the electroweak and
an intermediate PQ symmetry breaking scale. For a number of explicit models, we determine the
parameters of the low-energy effective field theory describing the axion, the ALPs, and their interactions
with photons and electrons, in terms of the input parameters, in particular the  PQ symmetry breaking
scales. We show that these models can accommodate simultaneously an axion dark matter candidate, an
ALP explaining the anomalous transparency of the universe for $\gamma$-rays, and an ALP explaining
the recently reported 3.55 keV gamma line from galaxies and clusters of galaxies, if
the respective decay constants are of intermediate scale.
Moreover, they do not suffer severely from the domain wall problem.
\end{abstract}
%%%%%%%%%%%%%%%%%%%%%%%%%%%%%%%%%%%%%%%%%%%%%%%%%
% \pacs{12.60.Fr, 11.30.Er, 11.30.Fs, 11.30.Qc}
% 14.80.Ec 	Other neutral Higgs bosons
% 14.80.Fd 	Other charged Higgs bosons
% 11.30.Fs 	Global symmetries (e.g., baryon number, lepton number)
% 11.30.Qc 	Spontaneous and radiative symmetry breaking
% \keywords{Higgs, multi-Higgs models, 2HDM, global symmetry, custodial symmetry,
% symmetry breaking}
%\twocolumn
\maketitle
\tableofcontents
%%%%%%%%%%%%%%%%%%%%%%%%%%%%%%%%%%%%%%%%%%%%%%%%%
\section{Introduction}
\label{sec:intro}

\begin{figure}
\begin{center}
\includegraphics[width=0.75\textwidth]{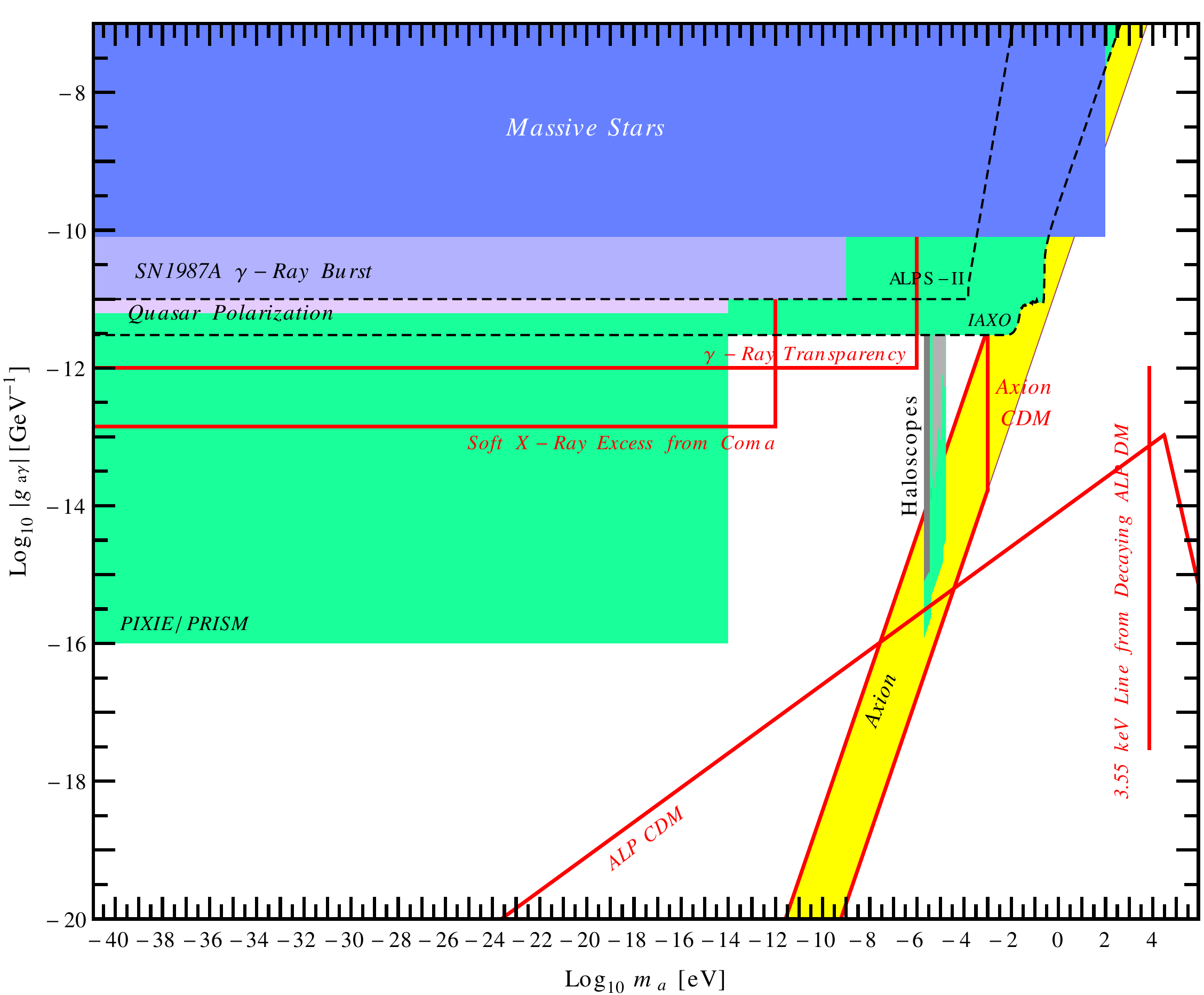}
\caption{Prediction of the axion-photon coupling versus its mass (yellow band). Also shown are excluded regions arising from the non-observation of an anomalous energy loss of massive stars due to axion or ALP emission \cite{Friedland:2012hj}, of a $\gamma$-ray burst (in coincidence with the observed neutrino burst) from SN 1987A
due to conversion of an initial ALP burst in the galactic magnetic field,  of  changes in quasar polarizations due to photon-ALP oscillations, and of dark matter axions or ALPs converted into photons in microwave cavities placed in magnetic fields~\cite{DePanfilis:1987dk,Wuensch:1989sa,Hagmann:1990tj,Asztalos:2001tf,Asztalos:2009yp}.
Axions and ALPs with parameters in the regions surrounded by the red lines may
constitute all or a part of cold dark matter (CDM), explain the cosmic $\gamma$-ray transparency, and the soft X-ray excess from Coma.
The green regions are the projected sensitivities of the light-shining-through-wall experiment ALPS-II, of the helioscope IAXO, of the
haloscopes ADMX and ADMX-HF, and of the PIXIE or PRISM cosmic microwave background observatories.}
\label{ALP_coupling_limits}
\end{center}
\end{figure}

The axion $A$ is a strongly motivated very weakly interacting ultralight particle beyond
the Standard Model (SM). Its existence is predicted in the course of  an elegant
solution to the strong CP problem \cite{Peccei:1977hh,Weinberg:1977ma,Wilczek:1977pj}, that is the non-observation of flavor conserving CP violation originating from the
theta-angle term in the Lagrangian of QCD,
\begin{equation}
\lag_{\rm QCD}\supset
  -\frac{\alpha_s}{8\pi} \,\bar\theta\, G_{\mu\nu}^a {\tilde G}^{a,\mu\nu}
,
\label{qcd}
\end{equation}
involving the gluon field strength $G_{\mu\nu}^a$ and its dual,
${\tilde G}^{a,\mu\nu} \equiv \epsilon^{\mu\nu\lambda\rho} G_{\lambda\rho}^a/2$, with
$\varepsilon^{0123}=1$.
This solution consists in adding to the Standard Model a scalar field theory describing
a (pseudo) Nambu-Goldstone boson arising from the breaking of a
global chiral (Peccei-Quinn (PQ)) $U(1)_{\rm PQ}$ symmetry: that is, the corresponding scalar field
$A(x)$ satisfies a shift symmetry $A(x)\to A(x) + {\rm const.}$ which is
only violated by the anomalous coupling to gluons,
\begin{equation}
\lag \supset -
  \frac{\alpha_s}{8\pi} \,\frac{A}{f_A}\, G_{\mu\nu}^a {\tilde G}^{a,\mu\nu}
.
\label{qcd}
\end{equation}
In fact, the $\bar\theta$-term can then be eliminated by absorbing it into the axion field, $A + \bar{\theta} f_A\to A$. Moreover, non-perturbative QCD effects\footnote{Semiclassical instanton methods
are not reliable to calculate the potential
accurately. One has to use matching to the low-energy chiral Lagrangian instead~\cite{Weinberg:1977ma,Georgi:1986df}.} provide for a non-trivial potential for
the shifted axion field $A$ -- minimized at zero expectation value, $\langle A\rangle =0$, thus wiping out strong CP violation
-- and predict a mass for the particle excitation around this minimum, the axion $A$,
\begin{equation}
m_A = \frac{m_\pi f_\pi}{f_A} \frac{\sqrt{z}}{1+z}
\simeq {6\,  {\rm meV}}
         \times
         \left(
         \frac{10^{9}\, {\rm GeV}}{f_A}\right).
\label{axion_mass}
\end{equation}
in terms of the pion mass, $m_\pi = 135$ MeV, its decay constant, $f_\pi \approx 92$ MeV, the
ratio $z=m_u/m_d \approx 0.56$ of up and down quark masses, and the decay constant $f_A$. For large $f_A$,
the axion is an ultralight particle with very weak interactions with the Standard Model \cite{Kim:1979if,Shifman:1979if,Dine:1981rt,Zhitnitsky:1980tq}. Its
universal and phenomenologically most important interaction with photons \cite{Georgi:1986df,Bardeen:1977bd,Kaplan:1985dv,Srednicki:1985xd},
\begin{equation}
\label{photon_coupling_universal}
\lag \supset
- \frac{g_{A\gamma}}{4}\,A\, F_{\mu\nu} \tilde{F}^{\mu\nu}
; \hspace{6ex}
        | g_{A\gamma} |
\sim \frac{\alpha}{2\pi f_A}
\sim \frac{\alpha}{2\pi} \frac{m_A}{m_\pi f_\pi}
\sim 10^{-12}\ {\rm GeV}^{-1}          \left(
         \frac{10^{9}\, {\rm GeV}}{f_A}\right)
\sim 10^{-12}\ {\rm GeV}^{-1}          \left(
         \frac{m_A}{6\, {\rm meV}}\right)
,
\end{equation}
is tiny (see the yellow band in Fig. \ref{ALP_coupling_limits}), since observations in astrophysics -- in particular the observed duration of the neutrino signal from
supernova SN 1987A -- require a large decay constant $f_A$ (small mass $m_A$)  \cite{Raffelt:2006cw},
\begin{equation}
f_A\gtrsim 4\times 10^8\,{\rm GeV} \Rightarrow
m_A\lesssim 16\,{\rm meV}
.
\end{equation}

A further strong motivation for the axion comes from the fact that, for sufficiently large decay constant $f_A$,
\begin{equation}
 10^9\,{\rm GeV}  \lesssim  f_A\lesssim 10^{13}\,{\rm GeV}  \Rightarrow
 10^{-7}\,{\rm eV}\lesssim  m_A\lesssim 1\,{\rm meV}
,
\label{cosmic_axion_window}
\end{equation}
it is a cold dark matter candidate, being non-thermally produced in the early universe
by the vacuum-realignment mechanism and, in some models and under certain circumstances,
also via the decay of topological defects such as axion strings and domain walls~\cite{Preskill:1982cy,Abbott:1982af,Dine:1982ah,Sikivie:2006ni,Bae:2008ue,Visinelli:2009zm,Wantz:2009it,Hiramatsu:2010yu,Hiramatsu:2012gg}.

The upper bound on the decay constant in Eq. \eqref{cosmic_axion_window} follows from the recent discovery of the cosmic microwave background polarimeter experiment BICEP2 of tensor fluctuations  in the $B$-mode power spectrum \cite{Ade:2014xna}. This implies a high value
for the Hubble scale during inflation,
\begin{equation}
H_I\simeq 1.1\times 10^{14}\,{\rm GeV},
\end{equation}
which, together with constraints on isocurvature fluctuations \cite{Fox:2004kb,Beltran:2006sq,Hertzberg:2008wr,Hamann:2009yf,Ade:2013uln}, rule out plausible scenarios where inflation occurs after PQ symmetry breaking \cite{Higaki:2014ooa,Marsh:2014qoa,Visinelli:2014twa}, that is where
\begin{equation}
f_A>{\rm max}\{T_{\rm GH},T_{\rm max}\},
\end{equation}
with
\begin{equation}
T_{\rm GH}=\frac{H_I}{2\pi }
\end{equation}
 the Gibbons-Hawking temperature during inflation and
\begin{equation}
T_{\rm max}=\epsilon_{\rm eff}\sqrt{\sqrt{\frac{3}{8\pi}}M_{\rm Pl}\, H_I}
\end{equation}
the maximum thermalization temperature after reheating. Here $M_{\rm Pl}=1.22\times
10^{19}$\,GeV is the Planck mass and $0 < \epsilon_{\rm eff} < 1$ is  an efficiency
parameter. Ways out of this conclusion have been put forward in Refs.
\cite{Higaki:2014ooa,Marsh:2014qoa}. The pre-inflationary PQ symmetry breaking
scenario would have allowed, in principle, much higher values of the decay constant
without overshooting the dark matter abundance, by invoking small values of the
initial misalignment angle. After BICEP2, the plausible axion possibilities have
narrowed down to scenarios with post-inflationary PQ symmetry breaking.
A conservative upper bound on the axion decay constant is then
\begin{equation}
f_A < \frac{H_I}{2\pi }\simeq 1.8\times 10^{13}\,{\rm GeV}, \hspace{6ex} m_A > 0.3\ \mu{\rm eV}  .
\end{equation}
A more stringent upper bound can be obtained by requiring that the axion dark matter abundance generated via the vacuum-realignment
mechanism should not exceed the observed dark matter abundance, leading to \cite{Bae:2008ue,Hertzberg:2008wr,Visinelli:2009zm,Wantz:2009it,Visinelli:2014twa}
\begin{equation}
f_A\lesssim     ( 1   \div 10 ) \times 10^{11}\,{\rm GeV}    , \hspace{6ex} m_A\gtrsim (6 \div 60 )\ \mu{\rm eV} .
\end{equation}

To be on the very conservative side, the red region  in Fig. \ref{ALP_coupling_limits} labelled ``Axion DM" comprises still both cases, the disfavored pre- and the favored post-inflationary PQ symmetry breaking.
Two haloscope experiments (ADMX \cite{Asztalos:2011bm} and ADMX-HF) are currently aiming at the direct detection of axion dark matter, based on microwave cavities in the mass region $2\times 10^{-6}\,{\rm eV}\lesssim m_A\lesssim 2\times 10^{-5}\,{\rm eV}$ (see the green regions
labelled ``Haloscopes" in Fig. \ref{ALP_coupling_limits}). Further experiments have
been proposed  to extend this region on both ends of the spectrum \cite{Graham:2011qk,Baker:2011na,Irastorza:2012jq,Horns:2012jf,Graham:2013gfa,Budker:2013hfa,Jaeckel:2013sqa,Jaeckel:2013eha,Horns:2013ira,Sikivie:2013laa,Redondo:2013hca,Rybka:2014cya}.

There is also a theoretical motivation for the existence of additional axion-like particles (ALPs) (for reviews, see \cite{Jaeckel:2010ni,Ringwald:2012hr,Ringwald:2012cu}), emerging as pseudo Nambu-Goldstone bosons from the breaking of other global symmetries,
such as lepton number \cite{Chikashige:1980ui,Gelmini:1980re} or family symmetries \cite{Wilczek:1982rv,Berezhiani:1990wn,Jaeckel:2013uva}.
Most notably, the low-energy effective field theories emerging from string theory generically contain
candidates for the axion  plus possibly a multitude of additional ALPs.
Indeed, when compactifying the six extra spatial dimensions of string theory, Kaluza-Klein
zero modes of antisymmetric form fields -- the latter belonging to the massless spectrum of the
bosonic string propagating in ten dimensions -- appear as axion and further ALP candidates in the low-energy effective action
\cite{Witten:1984dg,Conlon:2006tq,Svrcek:2006yi,Arvanitaki:2009fg,Acharya:2010zx,Cicoli:2012sz}, their
number being determined by the topology of the internal compactified manifold.
Moreover, the axion and a multitude of  additional ALP candidates may also arise
as pseudo Nambu-Goldstone bosons from the breaking of accidental global $U(1)$ symmetries that appear as
low energy remnants of exact discrete symmetries occurring in string
compactifications~\cite{Lazarides:1985bj,Choi:2006qj,Choi:2009jt}.
Intriguingly, very light and weakly coupled generic ALPs $a_i$, with decay constants in the same range as Eq. \eqref{cosmic_axion_window},
share with the axion the property of being cold dark matter candidates  since they are also produced via the vacuum-realignment mechanism   \cite{Arias:2012az,Marsh:2013ywa,Kadota:2013iya}.
In fact, their relic abundance, in the now strongly favored post-inflationary symmetry breaking scenario, is roughly given by\footnote{The red line labelled ``ALP CDM" in Fig. \ref{ALP_coupling_limits} corresponds to the ALP photon coupling, where
an ALP, according to \eqref{eq:omegaalp}, can account for all of the dark matter in the universe, assuming that it couples with strength
$g_{a_i \gamma} = \alpha/(2\pi f_{a_i})$. The region below this line is strongly disfavored after BICEP2 by overdensity constraints \cite{Higaki:2014ooa,Marsh:2014qoa}.}
\begin{equation}
\Omega_{a_i} h^2 \approx 0.053 \times
\left( \frac{m_{a_i}}{\rm eV} \right)^{1/2} \left( \frac{f_{a_i}}{10^{11}\ \rm GeV} \right)^{2} .
\label{eq:omegaalp}
\end{equation}

%%%%%%%%%
%%%%%%%%%%%%

Therefore, there is a strong motivation to search not only for the axion $A$, but also for additional light ALPs
$a_i$, for which the low-energy effective Lagrangian describing their interactions with (the lightest Standard Model particles)
photons can generically be written as
 \begin{equation}
\lag \supset\frac{1}{2}\, \partial_\mu A\, \partial^\mu A
-\frac{1}{2} m_A^2 A^2
- \frac{g_{A\gamma}}{4}\,A\, F_{\mu\nu} \tilde{F}^{\mu\nu}
+\sum_{i=2}^{n_{\rm ax}}
\left( \frac{1}{2}\, \partial_\mu a_i\, \partial^\mu a_i
-\frac{1}{2}  m_{a_i}^2 a_i^2
- \frac{g_{a_i\gamma}}{4}\,a_i\, F_{\mu\nu} \tilde{F}^{\mu\nu}
\right)
.
\label{AALPgen_leff}
\end{equation}
However, unlike the axion, which inherits many of its properties ($m_A$, $g_{A\gamma}$), from
non-perturbative QCD effects associated with chiral symmetry breaking, the  masses $m_{a_i}$ and the
photon couplings $g_{a_i\gamma}$ of the additional ALPs are model dependent.
Thus, there exists the possibility that ALPs are hierarchically more strongly coupled to photons than
axions with the same mass and therefore easier to detect.

Interestingly, there are indications from gamma-ray astronomy, which may point to the existence of at
least one ALP beyond the axion. Gamma-ray spectra from distant active galactic nuclei (AGN) should show
an energy and redshift-dependent exponential attenuation, $\exp (-\tau (E,z))$,
due to $e^+ e^-$ pair production off the extragalactic background light (EBL) -- the stellar and
dust-reprocessed light accumulated during the cosmological evolution following
the era of re-ionization. The recent detection of this effect by Fermi \cite{Ackermann:2012sza} and
H.E.S.S. \cite{Abramowski:2012ry} has allowed to constrain the EBL density.
At large optical depth, $\tau\gtrsim 2$, however, there are hints that
the Universe is anomalously transparent to gamma-rays \cite{Aharonian:2007wc,Aliu:2008ay,Essey:2011wv,Horns:2012fx,Horns:2013pha}.
This may be explained by photon $\leftrightarrow$ ALP oscillations:
the conversion of gamma rays into ALPs in the
magnetic fields around AGNs or in the intergalactic medium, followed by their unimpeded
travel towards our galaxy and the consequent reconversion into photons in the galactic/intergalactic magnetic
fields~\cite{DeAngelis:2007dy,Simet:2007sa,SanchezConde:2009wu,Horns:2012kw,Mena:2013baa}. This explanation requires a very light ALP, which couples to two photons with strength \cite{Meyer:2013pny},
\begin{equation}
 |g_{a\gamma}|\gtrsim  10^{-12} \ {\rm GeV}^{-1}
; \hspace{6ex}
 m_{a}\lesssim 10^{-7}\  {\rm eV}.
\label{dec_const_transp}
\end{equation}

Note that the entire parameter region \eqref{dec_const_transp} has no overlap with the
universal prediction of the axion, see Fig. \ref{ALP_coupling_limits}. In fact, an axion with $m_A\lesssim 10^{-7}$\,eV would have a photon coupling
$|g_{A\gamma}|\lesssim 10^{-16}$\,GeV$^{-1}$. Therefore, if this hint is taken seriously, it points to the existence
of an ALP beyond the axion.

Intriguingly, an observed soft X-ray excess in the Coma cluster may also be explained by the conversion of
a cosmic ALP background radiation (CABR) -- produced by heavy moduli decay and corresponding to an effective number  $\triangle N_{\rm eff}\sim 0.5$ of extra neutrinos species \cite{Cicoli:2012aq,Higaki:2012ar}  -- into photons in the cluster magnetic field
\cite{Conlon:2013txa,Angus:2013sua}.
This explanation requires
that the spectrum of the CABR is peaked in the soft-keV region and that the ALP coupling and mass satisfy
\begin{equation}
|g_{a\gamma}|\gtrsim  10^{-13} \ {\rm GeV}^{-1}\, \sqrt{0.5/\triangle N_{\rm eff}}\,; \hspace{6ex} m_a\lesssim 10^{-12}\ {\rm eV},
\end{equation}
respectively,
overlapping with the
parameter range \eqref{dec_const_transp} preferred by the ALP solution of the gamma-ray transparency puzzle, as is apparent in
Fig. \ref{ALP_coupling_limits}.

Astrophysical bounds on ALPs arising from magnetised white dwarfs \cite{Gill:2011yp} and
from the non-observation of a $\gamma$-ray burst in coincidence with neutrinos from the supernova SN 1987A~\cite{Brockway:1996yr,Grifols:1996id} provide limits close to
$|g_{a\gamma}|\lesssim 10^{-11}~{\rm GeV}^{-1}$, for   masses $m_a\lesssim 10^{-7}~{\rm eV}$
and $m_a\lesssim 10^{-9}~{\rm eV}$, respectively, and thus cut into the parameter range \eqref{dec_const_transp} preferred by the
cosmic $\gamma$-ray transparency anomaly.
Even stronger limits,
$|g_{A\gamma}|\lesssim  6.3 \times 10^{-12}\, {\rm  GeV}^{-1} (n_e / 10^{-5} {\rm cm}^{-3})^{1.3} (2 \mu G / B_{\rm cell})$
for $m_A\lesssim 10^{-14}~{\rm eV}$,
have been obtained by exploiting high-precision measurements of quasar polarizations
\cite{Payez:2012vf,Payez:2013yxa}, where $n_e$ is the electron density and $B_{\rm cell}$ is the
magnetic field in the neighborhood of the quasars.
But there remains still a sizeable region in ALP parameter space motivated by the above anomalies and
at the same time consistent with all astrophysical constraints, as can be seen in Fig. \ref{ALP_coupling_limits}.

At small masses below $10^{-14}$ eV, a part of the region of interest in ALP parameter space will be probed indirectly
by the next generation of cosmic microwave background (CMB) observatories such as PIXIE \cite{Kogut:2011xw} and
PRISM \cite{Andre:2013nfa}
(see the region labelled ``PIXIE/PRISM" in  Fig. \ref{ALP_coupling_limits}, based on the assumption of an extragalactic magnetic field $B$ of nG size; the projected sensitivity scales with the magnetic field as $B^{-1}$),
because resonant photon-ALP oscillations in primordial magnetic fields may lead to observable spectral distortions
of the CMB \cite{Mirizzi:2009nq,Tashiro:2013yea,Ejlli:2013uda}.

A complementary part of the region of interest in parameter space will soon be
probed by a pure laboratory experiment: the light-shining-through-a-wall (LSW)  \cite{Redondo:2010dp}  experiment
Any-Light-Particle-Search II (ALPS-II)
is designed to detect photon--ALP--photon oscillations in the range \cite{Bahre:2013ywa}
(see the green region labelled ``ALPS-II" in Fig. \ref{ALP_coupling_limits})
\begin{equation}
|g_{a\gamma}|\gtrsim  2\times 10^{-11} \ {\rm GeV}^{-1}
; \hspace{6ex}
 m_{a}\lesssim 10^{-4}\  {\rm eV}.
\label{ALPSII}
\end{equation}
Further experimental opportunities covering this region in ALP parameter space will
open if the International Axion Observatory (IAXO), a helioscope searching for solar axions
and ALPs, is realized \cite{Armengaud:2014gea}. Its projected sensitivity is
\begin{equation}
|g_{a\gamma}|\gtrsim  5\times 10^{-12} \ {\rm GeV}^{-1}
; \hspace{6ex}
 m_{a}\lesssim 10\  {\rm meV}.
\label{IAXO}
\end{equation}

The latter instrument has also the possibility to probe the possible coupling of the axion or further ALPs to electrons,
\begin{equation}
\mathcal{L} \supset
\frac{\left( g_{Ae}\partial_\mu A
+ g_{ae}\partial_\mu a \,\right)}
{2 m_e} \,\bar{e} \gamma^\mu\gamma_5 e
,
\end{equation}
via their solar production by atomic axio-recombination, axio-deexcitation,
axio-Bremsstrahlung in electron-ion or electron-electron collisions,
and Compton scattering \cite{Redondo:2013wwa}. This is of considerable interest because of
hints of an extra stellar cooling mechanism not accounted by the Standard Model.
In fact,  the white dwarf (WD) luminosity function seems to require a new energy-loss
channel that can be interpreted in terms of losses due to sub-keV mass axions or ALPs, with Yukawa couplings~\cite{Isern:2008nt},
\begin{equation}
|g_{Ae}|\equiv |C_{Ae}| \,m_e/f_A \sim 10^{-13}
\ {\rm \  and/or\ }\ | g_{ae}|\equiv |\, C_{ae}| m_e/f_a \sim 10^{-13},
\label{wd_energy_loss}
\end{equation}
which are well in the range expected for an intermediate scale axion or further ALPs
(if the model-dependent couplings $C_{Ae}$ and $C_{ae}$ are of order unity).
The same parameter range is preferred to explain the anomalous size of the observed period decrease of the pulsating WDs
G117-B15A and R548 by additional axion/ALP losses \cite{Isern:2010wz,Corsico:2012sh}.
A third independent hint of anomalous stellar losses has recently been found in the
red-giant branch of the globular cluster M5, which seems to be extended to larger brightness than expected within the
Standard Model. A possible explanation of this observation is that the helium cores of red giants lose energy in axions or further ALPs
with electron couplings of the same order as in Eq.  \eqref{wd_energy_loss} \cite{Viaux:2013lha}.

Very recently, two groups have found an unidentified X-ray line signal at 3.55 keV in the stacked spectrum of a number of
galaxy clusters  \cite{Bulbul:2014sua}  and the Andromeda galaxy \cite{Boyarsky:2014jta}. It is tempting to identify this line with the expected signal from two photon decay
of 7.1 keV mass ALP dark matter  \cite{Higaki:2014zua,Jaeckel:2014qea,Lee:2014xua,Cicoli:2014bfa}. To match the observed X-ray flux, but allowing for the likely possibility, that the ALP dark matter makes only a fraction $x_a \equiv \rho_a/\rho_{\rm DM}$ of the total density of dark matter, the required lifetime and thus coupling of the ALP is \cite{Jaeckel:2014qea}
\begin{equation}
\tau_a = \frac{64 \pi}{g^2_{a\gamma} m_a^3} = x_a\times \left( 4\times 10^{27} - 4\times 10^{28}\right)\,{\rm s} \hspace{2ex}
\Rightarrow\hspace{2ex}      3\times 10^{-18}\,{\rm GeV}^{-1} \left( \frac{1}{x_a}\right)^{1/2} \lesssim | g_{a\gamma} | \lesssim
10^{-12}\,{\rm GeV}^{-1} \left( \frac{10^{-10}}{x_a}\right)^{1/2}\,.
\end{equation}

Therefore, it is timely to have a close look onto ultraviolet extensions of the Standard Model featuring,  apart from an  intermediate scale axion, also further  ALPs  and to investigate possible correlations between
the low-energy axion and ALP couplings. In fact, as we will show in  Sec. \ref{sec:correlations}, such correlations inevitably occur if  there
are originally two (or more) Nambu-Goldstone fields coupled to $G\tilde G$.
%We will investigate there also the important phenomenological question whether an
%ALP in the parameter range \eqref{dec_const_transp} -- explaining the cosmic $\gamma$-ray transparency and detectable by ALPS-II and other %LSW experiments -- can co-exist with an axion in the cosmic axion window
%\eqref{cosmic_axion_window}, explaining cold dark matter and detectable by ADMX, ADMX-HF and other haloscopes.
The crucial determination of the decay constants and couplings of axion-like
fields from original high-scale theories to gluons, photons, and electrons will be done in Secs. \ref{axions_ad_hoc} and  \ref{accions_bottom_up}. In the latter section, we construct particularly
well-motivated ultraviolet completions of the Standard model
featuring accidental Peccei-Quinn symmetries arising from exact discrete symmetries and deduce their low-energy parameters, as summarized in Table \ref{summary_models}. As a first
example (model A.1), we present a multi-Higgs model with  discrete $\ZZ_{13}\otimes \ZZ_5\otimes\ZZ_{5}^\prime$ symmetry (Sec. \ref{sec:unificaccion}), which predicts gauge coupling unification near the scale of Peccei-Quinn symmetry breakdown, as well as neutrino mass generation through the seesaw mechanism.  Another example (model B1.1) features a  $\ZZ_{11}\otimes\ZZ_9$ symmetry (Sec. \ref{accion_laccion_model}), has two Higgs doublets and a  photophilic ALP which can fit the astrophysical hints such as  the  anomalous cosmic transparency to gamma-rays, and has also a seesaw mechanism. A third example (model B3) with  $\ZZ_{11}\otimes\ZZ_7$ symmetry  (Sec. \ref{another_singlets_model}) has similar properties as model  B1.1, but with the ALP being a dark matter candidate whose decay into photons can explain the 3.55 keV line, since it has a mass of 7.1 keV generated through effective interactions suppressed by the Planck scale. Last, but not least,
we present in \ref{yet_another_singlets_model} a model with $\ZZ_{11}{\otimes} \ZZ_{9}{\otimes} \ZZ_{7}$ symmetry, which has two photophilic ALPs, besides the axion. The field content and the discrete symmetry of this model is such that one of the ALPs is very light and able to match the astrophysical hints previously mentioned, while the other one has  a coupling to photons and a mass as required to explain the 3.55 keV line.
Finally, we summarize, conclude and give an outlook for further
investigations in Sec. \ref{sec:conclusions}.

\begin{table}
\begin{center}
\begin{tabular}{|c||c|c|c|c|c|c|c|}\hline
&  \multicolumn{7}{c|}{Resulting low-energy parameters} \\ \cline{2-8}
\raisebox{1.5ex}{Model} &     $f_A$ [GeV] & $m_A$ [eV] & $m_a$ [eV]&
$|g_{A\gamma}|$ [GeV]$^{-1}$ & $|g_{a\gamma}|$ [GeV]$^{-1}$ & $|g_{A e}|$ & $|g_{a e}|$ \\ \hline
%\raisebox{1.5ex}
A.1 &  $8.3\times 10^{11}$ & $7.2\times 10^{-6}$  & $1.3\times 10^{-22}$ &  $8.2\times 10^{-16}$ & $5.4\times 10^{-16}$  & $2\times 10^{-16}$ & $1.7\times 10^{-17}$   \\ \hline
%\raisebox{1.5ex}
B1.1 &   $3\times 10^{11}$ & $2\times 10^{-5}$ &$1\times 10^{-7}$  & $1.6\times 10^{-14}$ & $1.4\times 10^{-11}$    & 0 & $1\times 10^{-12}$  \\ \hline
%\raisebox{1.5ex}
B3 &   $5.5\times 10^{11}$ & $1.1\times 10^{-5}$ &$7.1\times 10^{3}$  & $3.4\times 10^{-15}$ & $1.3\times 10^{-12}$    & 0 & $0$  \\ \hline
\end{tabular}
\caption{\label{summary_models} Low-energy parameters for: model A.1  with  $\ZZ_{13}\otimes \ZZ_5\otimes\ZZ_{5}^\prime$ symmetry (Sec. \ref{sec:unificaccion}); model B1.1  with $\ZZ_{11}\otimes\ZZ_9$ symmetry (Sec. \ref{accion_laccion_model}); and model B3 with $\ZZ_{11}\otimes\ZZ_7$ symmetry (Sec. \ref{another_singlets_model}).}
\end{center}
\end{table}

%%%%%%%%%%%%%%%%%%%%%%%%%%%%%%%%%%%%%%%%%%%
\section{Correlations between Axion and ALPs Couplings in Multi-Axion Models}
\label{sec:correlations}

Up to now, most phenomenological studies have considered just one
particle type at a time: either the axion or an ALP different from the axion, without taking into account possible relations
between their low-energy parameters (see, however, Refs. \cite{Kim:1998kx,Kim:1999dc,Kim:2006aq,Choi:2006qj,Choi:2009jt}).  In this section, we will study, in multi-axion models,  the phenomenologically most important axion and ALP couplings to photons and electrons in depth.
We will find that there are possibly strong correlations in these couplings.

The most general low-energy effective Lagrangian of a model with $n_{\rm ax}$ axion-like fields  enjoying shift symmetries $a_i^\prime\to a_i^\prime + {\rm const.}$, i.e. realizing a non-linear representation of
$n_{\rm ax}$ $U(1)_{{\rm PQ}_i}$ symmetries, coupling to
gluons and photons, with field strengths $G$ and $F$, respectively, and  to electrons,
reads \cite{Cicoli:2012sz},
\begin{equation}
\lag =\frac{1}{2}\, \partial_\mu a_i^\prime\, \partial^\mu a_i^\prime
- \frac{\alpha_s}{8\pi} \left(\sum_{i=1}^{n_{\rm ax}} C_{ig} \frac{a_i^\prime}{f_{a_i^\prime}}\right)  G_{\mu\nu}^b \tilde{G}^{b,\mu\nu}
 - \frac{\alpha}{8\pi}
\left(\sum_{i=1}^{n_{\rm ax}}C_{i\gamma} \frac{a_i^\prime}{f_{a_i^\prime}}\right) \, F_{\mu\nu} \tilde{F}^{\mu\nu}
+ \frac{1}{2} \,
\left( \sum_{i=1}^{n_{\rm ax}}C_{ie}\frac{\partial_\mu a_i^\prime}{f_{a_i^\prime}} \,\right)\,\bar{e} \gamma^\mu\gamma_5 e
,
\label{ALP_leff}
\end{equation}
where $f_{a_i^\prime}$ are the decay constants of the axion-like fields $a_i^\prime$.
The anomaly coefficients, $C_{ig}$ and $C_{i\gamma}$, and the electron coupling, $C_{ie}$, are typically of order unity, see Secs. \ref{axions_ad_hoc} and  \ref{accions_bottom_up}  \footnote{There can be large hierarchies in these coefficients in multi-axion models from IIB string flux compactifications \cite{Cicoli:2012sz}.}.

The proper axion field $A$, that is the field which solves the strong CP problem,  is the linear combination of
the axion-like fields in front of the $G\tilde G$ term  in Eq. (\ref{ALP_leff})  \cite{Witten:1984dg},
\begin{equation}
\frac{A}{f_A}\equiv\sum_{i=1}^{n_{\rm ax}} C_{ig} \frac{a_i^\prime}{f_{a_i^\prime}}.
\end{equation}
Its particle excitation, the axion $A$, mixes with the pion, rendering it a pseudo Nambu-Goldstone boson,
whose mass has been given in Eq. \eqref{axion_mass}. To this end, $f_A$ has to be chosen such that the kinetic term of $A$ is normalized canonically. The remaining $n_{\rm ax}-1$
axion-like fields $a_i$, orthogonal to the axion $A$,
are still massless Nambu-Goldstone bosons.

To be more explicit, let us
 specialize first to the phenomenologically well motivated two-axion model, $n_{\rm ax}=2$ (see the appendix of Ref.
 \cite{Cicoli:2012sz} for a further exposition of the multi-axion case) and defer the discussion of the more general case to the end of
this section.  The properly normalized
axion $A$ and the additional ALP $a$ are related to the original axion-like fields $a_i^\prime$ by
\begin{equation}
\frac{A}{f_A} = C_{1g} \frac{a_1^\prime}{f_{a_1^\prime}}+C_{2g} \frac{a_2^\prime}{f_{a_2^\prime}},
\hspace{6ex}
\frac{a}{f_a} = -C_{2g} \frac{a_1^\prime}{f_{a_2^\prime}}+C_{1g} \frac{a_2^\prime}{f_{a_1^\prime}},
\label{mixing1}
\end{equation}
with normalization \cite{Choi:2009jt}
\begin{equation}
\frac{1}{f_A^2} = \frac{1}{f_a^2} = \left( \frac{C_{1g}}{f_{a_1^\prime}}\right)^2+
\left(\frac{C_{2g}}{f_{a_2^\prime}}\right)^2
\label{fA}
.
\end{equation}
The low-energy Lagrangian of this model, below the chiral symmetry breaking scale, is then
\begin{equation}
\lag =\sum_{\phi = A,a} \left( \frac{1}{2}\, \partial_\mu \phi\, \partial^\mu \phi
-\frac{1}{2} m_\phi^2 \phi^2
- \frac{g_{\phi\gamma}}{4}\,\phi\, F_{\mu\nu} \tilde{F}^{\mu\nu}
 +\frac{g_{\phi e}}{2 m_e} \,\partial_\mu \phi\,\bar{e} \gamma^\mu\gamma_5 e \right)
,
\label{AALP_leff}
\end{equation}
where
\begin{eqnarray}
        g_{A\gamma}
&= &\frac{\alpha}{2\pi f_A}
\left( \left(\frac{f_A}{f_{a_1^\prime}}\right)^2 C_{1g}C_{1\gamma}
+ \left(\frac{f_A}{f_{a_2^\prime}}\right)^2 C_{2g}C_{2\gamma}
- {\frac{2}{3}\,\frac{4 +z}{1+z}     }\right),
\label{gAgamma}
 \\
\label{gAe}
g_{Ae} &=&
\frac{m_e}{f_A}\left(
\left(\frac{f_A}{f_{a_1^\prime}}\right)^2 C_{1g}C_{1e}
+ \left(\frac{f_A}{f_{a_2^\prime}}\right)^2 C_{2g}C_{2e}
\right)
,
\\
        g_{a\gamma}
&= &\frac{\alpha}{2\pi}
 \frac{f_A}{f_{a_1^\prime}f_{a_2^\prime}}
\left( C_{1g}C_{2\gamma} - C_{1\gamma}C_{2g} \right),
\label{gagamma}
\\
\label{gae}
g_{ae} &=&
 \frac{m_e f_A}{f_{a_1^\prime}f_{a_2^\prime}}
\left( C_{1g}C_{2e} - C_{1e}C_{2g} \right)
.
\end{eqnarray}
As in a single axion model \cite{Weinberg:1977ma,Georgi:1986df,Bardeen:1977bd,Kaplan:1985dv,Srednicki:1985xd}, the axion $A$ has a universal, model-independent contribution to its  coupling to the photon (the last
term in Eq. \eqref{gAgamma}) which arises from its mixing with the pion (see, e.g., appendix of Ref. \cite{Cicoli:2012sz}).

The parameters in the above expressions for the couplings are however redundant, because $f_A$ is constrained
by Eq. \eqref{fA}. The physics is more obvious if one expresses the transformations in Eq. \eqref{mixing1} in terms
of mixing angles \cite{Choi:2006qj},
\begin{equation}
\label{mixing:delta}
A = a_1^\prime\,\cos\delta + a_2^\prime \,\sin\delta , \hspace{6ex}
a = -\, a_1^\prime\, \sin\delta +  a_2^\prime\, \cos\delta ,
\end{equation}
with
\begin{equation}
\cos\delta =
C_{1g} \frac{f_A}{f_{a_1^\prime}}, \hspace{3ex}
\sin\delta = C_{2g} \frac{f_A}{f_{a_2^\prime}}, \hspace{3ex}
\tan\delta = \frac{C_{2g}}{C_{1g}} \frac{f_{a_1^\prime}}{f_{a_2^\prime}}.
\end{equation}
In terms of this parametrization, the couplings can then be written as
\begin{eqnarray}
        g_{A\gamma}
&= &\frac{\alpha}{2\pi f_A}
\left( \frac{C_{1\gamma}}{C_{1g}} \cos^2\delta
+  \frac{C_{2\gamma}}{C_{2g}} \sin^2\delta
- {\frac{2}{3}\,\frac{4 +z}{1+z}    }\right),
\label{gAgamma_mixing}
 \\
g_{Ae} &=&
\frac{m_e}{f_A}\left(
\frac{C_{1e}}{C_{1g}} \cos^2 \delta
+ \frac{C_{2e}}{C_{2g}} \sin^2\delta
\right)
\label{gAe_mixing}
,
\\
        g_{a\gamma}
&= &\frac{\alpha}{2\pi f_A}
\left( \frac{C_{2\gamma}}{C_{2g}} - \frac{C_{1\gamma}}{C_{1g}} \right) \sin\delta \cos\delta
= \frac{\alpha}{2\pi f_A}
\left( \frac{C_{2\gamma}}{C_{2g}} - \frac{C_{1\gamma}}{C_{1g}} \right) \frac{\tan\delta}{1+ \tan^2\delta}
\,,
\label{gagamma_mixing}
\\
g_{ae} &=&
 \frac{m_e}{f_A}
\left( \frac{C_{2e}}{C_{2g}} - \frac{C_{1e}}{C_{1g}} \right)  \sin\delta \cos\delta
= \frac{m_e}{f_A}
\left( \frac{C_{2e}}{C_{2g}} - \frac{C_{1e}}{C_{1g}} \right)  \frac{\tan\delta}{1+ \tan^2\delta}
\label{gae_mixing}
\,.
\end{eqnarray}
Effectively, the couplings depend on the dimensionful parameter $f_A$,  on the dimensionless  ratios $C_{1\gamma}/C_{1g}$,  $C_{2\gamma}/C_{2g}$, $C_{1e}/C_{1g}$, $C_{2e}/C_{2g}$,
and on the mixing angle $\delta$.

Apart from the
model-independent contribution due to the mixing with the pion, the photon coupling of the axion has
a slight, order one,  model dependence due to the ratio of the anomaly coefficients $C_{1\gamma}/C_{1g}$, $C_{2\gamma}/C_{2g}$, and the mixing angle, cf. Eq. \eqref{gAgamma_mixing}. On the other hand,
the electron coupling of the axion $A$ is very much model dependent: a non-vanishing $g_{Ae}$ requires that at least one of the original
axion-like fields has a non-zero coupling to the electron, $C_{ie}\neq 0$ for $i=1$ and/or $2$,  cf. Eq. \eqref{gAe_mixing}.

This strong model dependence is shared by the photon and electron couplings of the ALP $a$: here one needs that at least one
of the original axion-like fields has a non-zero coupling
to the photon (electron), $C_{i\gamma}\neq 0$ for $i=1$ and/or $2$ ($C_{ie}\neq 0$ for $i=1$ and/or $2$),  otherwise $g_{a\gamma}$ ($g_{ae}$) vanishes, cf. Eq. \eqref{gagamma_mixing} (Eq. \eqref{gae_mixing}).
In this context it is also important to mention that eventual explicit mass terms, $m_{a_i^\prime}^2 a_i^{\prime\,2}/2$, for axion-like fields arising from
breaking of the Peccei-Quinn symmetries, e.g.  by higher dimensional operators suppressed by some high scale (see next section), will have negligible effects on the photon and electron couplings of the axion $A$ and the ALP $a$, as long as the masses are much smaller than the dynamically generated axion mass, $m_{a_i^\prime}\ll m_A$
(cf., e.g., appendix of Ref.  \cite{Cicoli:2012sz}).

We conclude that in the case that the axion $A$ and the ALP $a$ consist of an appreciable mixture of the original axion-like fields, i.e. as long as $|\tan\delta|$ is of order unity,
the couplings are all determined by $f_A$ and are therefore expected
to be approximately of the same size for for the axion $A$ and the ALP $a$, up to order one factors.
A hierarchical difference between the axion and the ALP coupling can possibly only
arrive in the situation where
the axion (and correspondingly also the ALP) originates essentially only from one axion-like field, which occurs for
$\delta \approx 0$ ($\pi/2$), meaning that $\cos\delta \approx 1$ ($\sin\delta = 1$), i.e.  that $a_1^\prime$ ($a_2^\prime$)
constitutes the axion. In fact, concentrating without loss of generality on the former case,
in which
\begin{equation}
\frac{1}{f_A} \approx
 \frac{|C_{1g}|}{f_{a_1^\prime}} , \hspace{6ex}
|\tan\delta | =  \left| \frac{C_{2g}}{C_{1g}} \frac{f_{a_1^\prime}}{f_{a_2^\prime}}\right|
\approx
 \left| C_{2g} \frac{f_A}{f_{a_2^\prime}}\right| \ll 1
,
\end{equation}
the sizes of the axion and ALP couplings appear to decouple from each other,
\begin{eqnarray}
        g_{A\gamma}
&\approx & \frac{\alpha}{2\pi f_A}
\left( \frac{C_{1\gamma}}{C_{1g}}
- {\frac{2}{3}\,\frac{4 +z}{1+z}     }\right)
\approx \frac{\alpha\,C_{1g}}{2\pi f_{a_1^\prime}}
\left( \frac{C_{1\gamma}}{C_{1g}}
- {\frac{2}{3}\,\frac{4 +z}{1+z}    }\right)
,
\label{gAgamma_nomixing}
 \\
g_{Ae} &\approx &
\frac{m_e}{f_A}
\frac{C_{1e}}{C_{1g}} \approx
\frac{m_e}{f_{a_1^\prime}}
\, C_{1e}
\label{gAe_nomixing}
,
\\
      g_{a\gamma}
&\approx &\frac{\alpha}{2\pi f_A}
\left( \frac{C_{2\gamma}}{C_{2g}} - \frac{C_{1\gamma}}{C_{1g}} \right)
\frac{C_{2g}}{C_{1g}} \frac{f_{a_1^\prime}}{f_{a_2^\prime}}
\approx \frac{\alpha}{2\pi f_{a_2^\prime}}
\left( C_{2\gamma} - C_{1\gamma}\,\frac{C_{2g}}{C_{1g}}\right)
,
\label{gagamma_nomixing}
\\
g_{ae} &\approx &
 \frac{m_e}{f_A}
\left( \frac{C_{2e}}{C_{2g}} - \frac{C_{1e}}{C_{1g}} \right)  \frac{C_{2g}}{C_{1g}} \frac{f_{a_1^\prime}}{f_{a_2^\prime}}
\approx \frac{m_e}{f_{a_2^\prime}}
\left( C_{2e} - C_{1e}\,\frac{C_{2g}}{C_{1g}}\right) ,
\label{gae_nomixing}
\end{eqnarray}
the former (latter) being determined by $f_{a_1^\prime}$ ($f_{a_2^\prime}$).
However, as long as $C_{2g}\neq 0$, the ALP couplings still can not be hierarchically larger than the axion couplings,
since the above relations imply an upper bound on the ratio
\begin{equation}
\left| \frac{g_{a\gamma}}{g_{A\gamma}}\right|\sim
\left|
\frac{C_{2\gamma} - C_{1\gamma}\,\frac{C_{2g}}{C_{1g}}}{\frac{C_{1\gamma}}{C_{1g}}
- {\frac{2}{3}\,\frac{4 +z}{1+z}    }}
\right|
\frac{f_A}{f_{a_2^\prime}} \ll
\left|
\frac{\frac{C_{2\gamma}}{C_{2g}} - \frac{C_{1\gamma}}{C_{1g}}}{\frac{C_{1\gamma}}{C_{1g}}
- 1.95 }
\right|.
\end{equation}

A true hierarchy in the couplings, in particular an ALP-photon coupling much larger than an axion-photon coupling,
is only possible if there is no mixing at all, $C_{2g}\equiv 0$, but  still $C_{2\gamma}\neq 0$, i.e.
if the ALP $a_2^\prime$ is photophilic\footnote{
For this conclusion, we exclude the possibility of accidental
cancellation between model dependent contributions and the universal contribution
to the axion-photon coupling in Eq.\,\eqref{gAgamma}.}.
In this case we get
\begin{eqnarray}
        g_{A\gamma}
&= &\frac{\alpha}{2\pi f_A}
\left( \frac{C_{1\gamma}}{C_{1g}}
- {\frac{2}{3}\,\frac{4 +z}{1+z}     }\right),
\label{gAgamma_nomixing_photophilic}
 \\
g_{Ae} &=&
\frac{m_e}{f_A}
\frac{C_{1e}}{C_{1g}}
\label{gAe_nomixing_photophilic}
,
\\
        g_{a\gamma}
&= &\frac{\alpha}{2\pi f_{a_2^\prime}}\, C_{2\gamma},
\label{gagamma_nomixing_photophilic}
\\
g_{ae} &=&
 \frac{m_e}{f_{a_2^\prime}}\, C_{2e},
\label{gae_nomixing_photophilic}
\end{eqnarray}
such that e.g. $|g_{a\gamma}|$
can very well be in the range suggested by the ALP explanation of the cosmic $\gamma$-ray transparency puzzle
and be accessible to the next generation of laboratory experiments,
\begin{equation}
10^{-10}\ {\rm GeV}^{-1}\gtrsim  |g_{a\gamma}|\gtrsim  10^{-12} \ {\rm GeV}^{-1}
; \hspace{6ex}
 m_{a}\lesssim 10^{-7}\  {\rm eV},
\label{dec_const_transp_2}
\end{equation}
provided that $f_{a_2^\prime}/|C_{2g}$ is in the range,
\begin{equation}
10^7\,{\rm GeV} \lesssim f_{a_2^\prime}/|C_{2\gamma}|\lesssim 10^9\,{\rm GeV}\,,
\label{phen_int_range}
\end{equation}
while at the same time
$|g_{A\gamma}|$ can be right in the cosmic axion window,
provided that $f_A = f_{a_1^\prime}/|C_{1g}|$ is
in the range
\begin{equation}
 10^{13}\,{\rm GeV}\gtrsim f_A\gtrsim 10^9\,{\rm GeV} \Rightarrow
10^{-7}\,{\rm eV}\lesssim m_A\lesssim 1\,{\rm meV}
\Rightarrow
10^{-16}\ {\rm GeV}^{-1} \lesssim |g_{A\gamma}|  \lesssim 10^{-12}\ {\rm GeV}^{-1}
.
\label{cosmic_axion_window_2}
\end{equation}

We therefore conclude that in models with multiple axion-like fields it is of very high phenomenological relevance to
know whether several of them couple to $G\tilde G$ simultaneously. This question will
be considered in the following sections.

\bigskip

Let us end this section by noting that even further insight into the general relation between the axion and ALP couplings can be obtained by purely geometrical means.
To this end, we rewrite Eqs.\,\eqref{gAgamma} and \eqref{gagamma} as
\eqali{
g_{A\gamma}&=\frac{\alpha}{2\pi}y\left(
\frac{\by\cdot\bz}{y^2}-C_0\right)
=\frac{\alpha}{2\pi}z(\hby\cdot\hbz) -\guniv\,.
% =\frac{\alpha}{2\pi}z\,\frac{y_iz_i}{yz} -\guniv\,,
\cr
g_{a\gamma}&=\frac{\alpha}{2\pi}\left(
\frac{y_1z_2-y_2z_1}{y}\right)
=\frac{\alpha}{2\pi}z(\epsilon_{ij}\hy_i\hz_j)\,,
}
where
\eqali{
\label{def:vectors}
\by=(y_1,y_2)&\equiv\left(\frac{C_{1g}}{f_{a_1'}},\frac{C_{2g}}{f_{a_2'}}\right)\,,  \hspace{3ex}
\hby=(\hy_1,\hy_2)=\by/y\,, \hspace{3ex} y=|\by| = \sqrt{ \left( \frac{C_{1g}}{f_{a_1^\prime}}\right)^2+
\left(\frac{C_{2g}}{f_{a_2^\prime}}\right)^2} \equiv\frac{1}{f_A},
\cr
\bz=(z_1,z_2)&\equiv\left(\frac{C_{1\gamma}}{f_{a_1'}},\frac{C_{2\gamma}}{f_{a_2'}} \,,
\right)\,,  \hspace{3ex}
\hbz=(\hz_1,\hz_2)=\bz/z \,,  \hspace{3ex}
z=|\bz| = \sqrt{
  \left(\frac{C_{1\gamma}}{f_{a_1'}}\right)^2
  +\left(\frac{C_{2\gamma}}{f_{a_2'}}\right)^2
  }\equiv \frac{1}{f_\gamma}\,.
}
and
\eq{
\guniv\equiv \frac{\alpha}{2\pi f_A}C_0=
\frac{\alpha}{2\pi f_A}\frac{2}{3}\frac{4+z}{1+z}
\approx \frac{\alpha}{2\pi
f_A}\times 1.95\,.
}
Thus $\hby$ points into the direction of the axion, $A(x)=\hy_ia'_i(x)$,
whereas $\hbz$ points into the direction of an ALP
that couples to photons (the orthogonal direction decouples from photons).

\begin{figure}
\begin{center}
\input{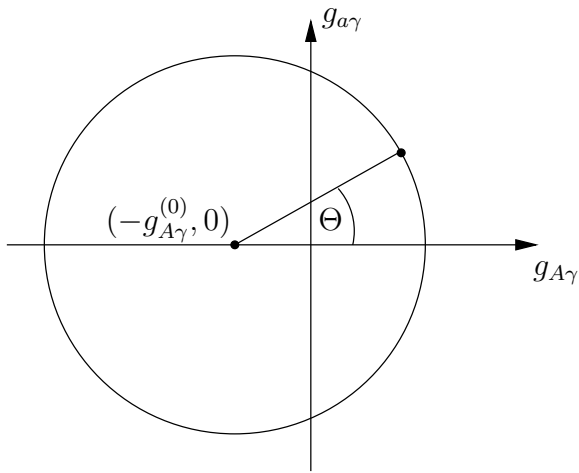}
\caption{The axion-photon coupling $g_{A\gamma}$ and the ALP-photon coupling $g_{a\gamma}$ are constrained to lie on a circle with origin $(-\guniv,0)$ and radius $g_{\gamma\rm max}$, see Eq. \eqref{def:circle}. The case $\guniv \le g_{\gamma\rm max}$ of Eq. \eqref{eq:BoundsgAGamma} is depicted.}
\label{PhotonCouplingPlane}
\end{center}
\end{figure}

Plane geometry ensures us that we can write then
\eq{
g_{A\gamma}= g_{\gamma\rm max}\cos\Theta -\guniv\,,~~
g_{a\gamma}= g_{\gamma\rm max}\sin\Theta\,,
}
where
\eq{
\label{def:gmax}
g_{\gamma\rm max}\equiv
  \frac{\alpha}{2\pi f_\gamma}\,,
}
and
 $\Theta$ is the angle between $\hby,\hbz$, from $\hby$ to $\hbz$,
\eq{
(\hby\cdot\hbz)=\cos\Theta, ~~(\epsilon_{ij}\hy_i\hz_j)=\sin\Theta\,.
}
This implies
\eq{
\label{def:circle}
\left(g_{A\gamma}+\guniv\right)^2+
\left(g_{a\gamma}\right)^2=
  g^2_{\gamma\rm max}\,,
}
see Fig.\,\ref{PhotonCouplingPlane}. In particular, the ALP-photon coupling is constrained by
\eq{
|g_{a\gamma}|\le g_{\gamma\rm max}\,,
}
while the axion-photon coupling satisfies
\eqali{
\label{eq:BoundsgAGamma}
\guniv-g_{\gamma\rm max} \le|g_{A\gamma}|\le \guniv+g_{\gamma\rm max}, &&
  \text{ for $\guniv>g_{\gamma\rm max}$}\,;
\cr
0\le|g_{A\gamma}|\le \guniv+g_{\gamma\rm max}, &&
  \text{ for $\guniv \le g_{\gamma\rm max}$}\,.
}

%%%%%%%%%
%%%%%%%%%%%%

If we take the hierarchy $f_{a_1'}\gg f_{a_2'}$, with order one coefficients
$C_{ig},C_{i\gamma}\neq 0$, we realize that the vectors in
Eqs.\,\eqref{def:vectors} are both approximately aligned along $(0,\pm 1)$.
This situation leads to $\Theta = 0$ or $\pi$ and hence to
\eq{
|g_{A\gamma}|\approx |\guniv\mp g_{\gamma\rm max}|\,,~~
|g_{a\gamma}|\approx 0\,.
}
Then $f_A\sim f_\gamma$ and $|g_{a\gamma}|\ll |g_{A\gamma}|$, barring accidental
cancellations.

The opposite hierarchy, $|g_{a\gamma}|\gg |g_{A\gamma}|$, is more difficult to be
obtained without fine tuning and requires $\guniv \le g_{\gamma\rm max}$, or, equivalently,
$f_\gamma\leq f_A/C_0 \simeq f_A/1.95$.
Analyzing the various possibilities, the only way we can achieve such a hierarchy
from hierarchical scales $f_{a_i'}$ is to require one of the axion-like fields to
be photophilic.

Notice that Eq.\,\eqref{def:circle} can be directly generalized for the case of
$n_{\rm ALP}>1$ ALPs, besides the axion.
The constraint generalizes to
\eq{
\label{def:circle:n>2}
\left(g_{A\gamma}+\guniv\right)^2+
\sum_{i=2}^{n_{\rm ax}}\left(g_{a_i\gamma}\right)^2=
  g^2_{\gamma\rm max}\,,
}
where Eq.\,\eqref{def:gmax} still defines $g_{\gamma\rm max}$ with the
generalization
\eq{
\frac{1}{f_\gamma}\equiv\sqrt{\sum_{i=0}^{n_{\rm
ax}}\left(\frac{C_{i\gamma}}{f_{a'_i}}\right)^2}\,,
}
where $n_{\rm ax}=n_{\rm ALP}+1$ is the number of axion-like fields.
The couplings in this case are given by
\eqali{
g_{A\gamma}&=\frac{\alpha}{2\pi}z\left(\hbz\cdot\hby-\frac{y}{z} C_0\right)\,.
\cr
g_{a_i\gamma}&=\frac{\alpha}{2\pi}z(\hbz\cdot\bu_i)\,,
}
where $\bu_i$ are orthonormal basis vectors spanning the space orthogonal to $\hby$.

An analogous constraint for the electron couplings can be obtained by introducing the angle
$\Theta_e$ between the vector $\by$ in \eqref{def:vectors}, pointing in the axion direction, and
the vector
\eq{
\bz_e\equiv\left(\frac{C_{1e}}{f_{a_1'}},\frac{C_{2e}}{f_{a_2'}}\right)\,,
}
pointing in the direction of
an ALP that couples to electrons. With the help of this, we can rewrite Eqs.\,\eqref{gAe} and \eqref{gae}
in the form
\eq{
g_{Ae}=g_{e\rm max}\cos\Theta_e\,,~~
g_{ae}=g_{e\rm max}\sin\Theta_e\,,
}
where
\eq{
g_{e\rm max}=\frac{m_e}{f_e}\,,
}
with
\eq{
\frac{1}{f_e}\equiv \sqrt{
  \left(\frac{C_{1e}}{f_{a_1'}}\right)^2
  +\left(\frac{C_{2e}}{f_{a_2'}}\right)^2
  }\,.
}
Therefore, the ALP-electron couplings are constrained by
\eq{
\label{def:circle:ge}
\left(g_{Ae}\right)^2+ \left(g_{ae}\right)^2 =  g^2_{e\rm max}\,.
}
%The difference here is that the circle \eqref{def:circle:ge} is centered at the
%origin and then the couplings $g_{Ae},g_{ae}$ can not vanish simultaneously unless
%both axion-like fields are decoupled from electrons, i.e., $C_{1e}=C_{2e}=0$.
This constraint can be generalized also straightforwardly
to the case of more than two axion-like fields.

\section{Axion and Further ALPs in Ad-Hoc Peccei-Quinn SM Extensions}
\label{axions_ad_hoc}

In this section we consider a number of field theoretic extensions of the SM where
an axion or further ALPs occur as Nambu-Goldstone bosons from the breaking of  ad-hoc $U(1)$ Peccei-Quinn (PQ) symmetries.
We will derive the low-energy couplings in terms
of the fundamental parameters of the underlying high-scale theories, with the aim of elucidating, in the following section, to which extent
current and upcoming axion and ALP experiments can probe them.

\subsection{KSVZ-type models}

We start with the so called KSVZ model \cite{Kim:1979if,Shifman:1979if}.  In this construction,  a color triplet, but $SU(2)_L$ singlet heavy vector-like fermion, $Q=(Q_L,Q_R)$, is added to the SM. In addition, a hidden (i.e. SM singlet)
complex scalar field  $\sigma$ is introduced.
The SM particles are assumed to be uncharged under the PQ symmetry, while the hidden scalar field and the exotic color triplet are supposed to transform as
\eq{
\label{U1:PQ}
\sigma\to e^{i\alpha}\sigma\,,\hspace{6ex}
Q_L\to e^{i\alpha/2}Q_L\,,\hspace{6ex}
Q_R\to e^{-i\alpha/2}Q_R\,,
}
respectively.
Correspondingly, the most general Yukawa interactions involving the PQ charge fields and the most general renormalizable scalar potential involving also the SM Higgs field $H$ read
\eq{
\label{yukawa}
\lag_Y \supset
y\overline{Q}_L\sigma Q_R   + h.c.
\,,
}
\eq{
\label{scalar_potential}
V(H,\sigma )= \lambda_H \left( H^\dagger H - \frac{v^2}{2}\right)^2
+\lambda_\sigma \left( |\sigma |^2 - \frac{v_{\rm PQ}^2}{2}\right)^2+
2\lambda_{H\sigma} \left( H^\dagger H - \frac{v^2}{2}\right) \left( |\sigma |^2 - \frac{v_{\rm PQ}^2}{2}\right)
\,,
}
respectively, both being invariant under $U(1)_{\rm PQ}$.
The self couplings in the latter are assumed to satisfy $\lambda_H, \lambda_\sigma >0$ and
$\lambda_{H\sigma}^2 < \lambda_H \lambda_\sigma$, to ensure that the minimum of the scalar potential
is attained at  the vacuum expectation values (vevs)
\eq{
\aver{H^\dagger H} = v^2/2, \hspace{6ex}
\aver{|\sigma |^2}=v_{\rm PQ}^2/2\,,
}
where $v=246$\,GeV.
The real scalar field $a'$ parameterizing the phase of the hidden scalar field $\sigma$ in the expansion around the vev,
\eq{
\label{sigma:}
\sigma (x) =\frac{1}{\sqrt{2}}\big[v_{\rm PQ}+\rho (x)\big]e^{ia'(x)/f_{a'}}
\,,
}
then corresponds to the Nambu-Goldstone boson arising from the breaking of the global PQ symmetry.
Assuming that the vev $v_{\rm PQ}$ is much larger than the electroweak scale, one may integrate out
the heavy fields $\rho$ and $Q$. The low-energy effective Lagrangian of
$a'$ matches Eq. \eqref{ALP_leff}, with $n_{\rm ax}=1$, $a'_1\equiv a'$, and
\eq{
\label{F1}
f_{a'}=v_{\rm PQ}\,.
}
The fermionic current associated with the PQ symmetry is indeed anomalous, leading,
as follows from the general formulae \eqref{Ca'g:gen} in Appendix \ref{sec:lecouplings}, to
axionic couplings to gauge bosons in the low-energy effective Lagrangian \eqref{ALP_leff},
\eq{
\label{C:KSVZ:single}
C_{a'g}
=1
\,, \hspace{6ex}
C_{a'\gamma}
=6\,\left( C^{(Q)}_{\rm em}\right)^2\,
\,,
}
with $C_{{\rm em}_i}^{(Q)}$ denoting a possible electric charge of the exotic color triplet.
The coupling to the electron, on the other hand, vanishes in the KSVZ model,
as follows from the general formula in Appendix \ref{sec:lecouplings},
\eq{
\label{C:KSVZ:single:electron}
C_{a'e}
=0
\,.
}

\subsection{DFSZ-type models}

Along the same lines
one can also construct DFSZ-type ad-hoc models \cite{Dine:1981rt,Zhitnitsky:1980tq}, which
require, apart from the generic hidden complex scalar $\sigma$, at least two Higgs doublets, $H_u$ and $H_d$, giving masses to up-type and down-type
quarks, respectively, and in which the SM quarks, instead of an exotic vector-like color triplet, carry PQ charges.
The fields are supposed to transform as follows under the PQ symmetry,
\eqali{
\sigma &\to e^{i\alpha}\sigma\,,\cr
\label{U1:PQ:DFSZ}
H_d&\to e^{iX_d\alpha}H_d\,,\cr
H_u&\to e^{-iX_u\alpha}H_u\,,\cr
d_{iR}&\to e^{-iX_d\alpha}d_{iR}\,,\cr
u_{iR}&\to e^{-iX_u\alpha}u_{iR}\,,
}
leaving the  Yukawa interactions,
\eq{
\label{yukawa}
\lag_Y=Y_{ij}\bar{q}_{iL}H_d d_{jR}+\Gamma_{ij}\bar{q}_{iL}\tilde H_u u_{jR}
 + h.c.
\,,
}
with $\widetilde{H}_u=\epsilon H_u^\ast$,
as well as the most general renormalizable scalar potential,
\eq{
V(\sigma )= -\mu^2_{\sigma}|\sigma|^2+\lambda_{\sigma}|\sigma|^4
+\lambda_{3}H_d^\dag H_u \sigma^2
\,,
}
invariant under the PQ symmetry, provided that
\eq{
\label{X:du}
X_u + X_d =2
\,.
}

After PQ and EW symmetry breaking at the scales $v_{\rm PQ}\gg v =\sqrt{v_u^2+v_d^2}=246$ GeV, the
Higgs and singlet scalar fields can be parametrized as
\eqali{
\label{phi:12}
H_d^{0}(x)&=\frac{v_d+h_d(x)}{\sqrt{2}}\exp\left[i\left(\frac{\zeta (x)}{v}
  +X_d\frac{a'(x)}{f_{a'}}\right)\right]\,,\cr
H_u^{0}(x)&=\frac{v_u+h_u(x)}{\sqrt{2}}\exp\left[i\left(\frac{\zeta (x)}{v}
  -X_u\frac{a'(x)}{f_{a'}}\right)\right]\,,\cr
\sigma(x)&=\frac{v_{\rm PQ}+\rho(x)}{\sqrt{2}}\exp\left[i\frac{a'(x)}{f_{a'}}\right]\,,
}
The fields in the radial components, $h_d$, $h_u$, and $\rho$, correspond to the two
physical neutral Higgs bosons and the physical scalar singlet, respectively.
The phases of the fields involve the Nambu-Goldstone boson $\zeta$ which is eaten by the $Z^0$ to generate its mass, and
the Nambu-Goldstone boson $a'$ which is an
axion-like field.
Orthogonality of $\zeta$ and $a'$
requires
\eq{
\label{X:du:xi}
X_d=x\xi_v\,,~~X_u=x^{-1}\xi_v\,,
}
where $x\equiv v_u/v_d\equiv \tan\beta$. The condition \eqref{X:du} then determines
that
\eq{
\xi_v=\frac{2}{x+x^{-1}}\,.
}

The low-energy effective Lagrangian of
$a'$ matches Eq. \eqref{ALP_leff}, with $n_{\rm ax}=1$, $a'_1\equiv a'$, and
\eq{
\label{F11}
f_{a'}=\sqrt{f_{\rm PQ}^2+v^2\xi_v^2}\,,
}
the latter arising from the requirement of canonical field normalization of $a'$ using $f_{\rm PQ} = v_{\rm PQ}$.
The fermionic content of the theory is anomalous with respect to $U(1)_{\rm PQ}$,
and the couplings to SM gauge bosons can be obtained from the general formulae in
Appendix \ref{sec:lecouplings}.
The coupling to gluons is given by  (cf. Eq.  \eqref{Ca'g:gen} in Appendix \ref{sec:lecouplings})
\eq{
\label{C:g:DFSZ}
C_{a'g}
=N_g(X_u+X_d)=2\,N_g\,,
}
where $N_g$ is the number of generations.
The couplings to photons (see for example \cite{Cheng:1995fd}) and electrons (see Appendix \ref{sec:lecouplings} for the
general formulae), on the other hand, depend also
on the PQ charge assignment for the
leptons. In fact, one finds
\eq{
\label{Cegamma}
C_{a'\gamma}=
\left\{\begin{array}{ll}
\ums[8]{3}N_g(X_u+X_d)=\ums[16]{3}N_g,&\text{if only $H_d$ couples to $l_{iR}$,}\\[1ex]
\ums[2]{3}N_g(X_u+X_d)=\ums[4]{3}N_g, &\text{if only $H_u$ couples to $l_{iR}$,}\\[1ex]
\ums[2]{3}N_g(4X_u+X_d),&\text{if only a third $H_l$ couples to $l_{iR}$,}
\end{array}
\right.
}
for the coupling to the gluon. Furthermore, for the coupling to the electron, we get
\eq{
C_{a'e}=
\left\{\begin{array}{ll}
X_d,&\text{if only $H_d$ couples to $l_{eR}$,}\\[1ex]
-X_u,&\text{if only $H_u$ couples to $l_{eR}$,}\\[1ex]
0,&\text{if only a third $H_l$ couples to $l_{eR}$.}
\end{array}
\right.
}
On account of Eq. \eqref{X:du:xi}, the physically most relevant ratio of the  electron coupling relative to the gluon coupling,
cf. Eq. \eqref{gae_mixing}, are then finally obtained as
\eq{
\frac{C_{a'e}}{C_{a'g}}=
\left\{\begin{array}{ll}
\frac{1}{N_g}\sin^2\beta,&\text{if only $H_d$ couples to $l_{eR}$,}\\[1ex]
-\frac{1}{N_g}\cos^2\beta,&\text{if only $H_u$ couples to $l_{eR}$,}\\[1ex]
0,&\text{if only a third $H_l$ couples to $l_{eR}$.}
\end{array}
\right.
}

\subsection{Models relating the PQ scale with the seesaw neutrino scale}
\label{axion_majoron}

The Peccei-Quinn breaking scale in these models is however still ad-hoc. In fact,
the strong CP problem is solved for any value of the Peccei-Quinn scale and the
preferred range for it, $10^9\,{\rm GeV}
\lesssim f_A\lesssim 10^{12}\, {\rm GeV}$, arises just from astrophysical
(stellar bounds) and cosmological (dark matter) constraints. Intriguingly, however,
this range overlaps with the preferred one for the breaking scale of lepton number
in seesaw explanations of the
smallness of the active neutrino masses. These scales can indeed be identified
in an extension of the KSVZ-type or DFSZ-type models by the further inclusion of three right handed
Majorana neutrinos $N_{iR}$, $i=1,2,3$ (for similar considerations, see Ref. \cite{Kim:1981jw,Langacker:1986rj}). Considering for example the KSVZ-type model and assigning to the
right handed neutrinos, as well as to the left handed SM lepton doublets $L$
and to the right handed SM lepton singlets $l_R$, the same Peccei-Quinn charges as to the
right-handed exotic color triplet (cf. Eq. \eqref{U1:PQ}),
\eq{
N_R\to e^{-i\alpha/2}N_R\,,\hspace{6ex}
L\to e^{-i\alpha/2}L\,,\hspace{6ex}
l_R\to e^{-i\alpha/2}l_R\,,
}
the most general PQ invariant Yukawa interactions involving the fields charged under the Peccei-Quinn symmetry generalize to
\begin{eqnarray}
\label{lyukseesaw}
{\lag_Y}  =  y \overline{Q}_L \sigma Q_R + G_{ij}\overline{L}_{i } H l_{j R}
 +   F_{ij} \overline{L}_{i
}\widetilde{H} N_{j R}  + y_{ij} \overline{(N_{iR})^c} \sigma N_{jR}
+ h.c. \,.
\end{eqnarray}
Here $H$ denotes the Higgs doublet,
$\widetilde{H}=\epsilon H^\ast$,
and $G_{i j}$, $F_{i j}$, $Y_{ij}$, are arbitrary complex
$3\times 3$ matrices, while $y_{ij}$ is a symmetric  $3\times 3$ matrix.

In this case, the Peccei-Quinn symmetry is an extension of the global lepton number $U(1)_{\rm L}$ in the
SM (including now right-handed neutrinos), which is spontaneously broken when $\sigma$ acquires a vev. Therefore, in this model, the Nambu-Goldstone boson $a'$ is in fact what is usually known as a
majoron \cite{Chikashige:1980ui,Gelmini:1980re,Kim:1981jw,Langacker:1986rj}.  After integrating out the heavy fields,
the low-energy couplings of $a'$ are still given by
\begin{equation}
f_{a'}=v_{\rm PQ},\ C_{a'g}=1,\ C_{a'\gamma}=6 \left( C_{\rm em}^{(Q)}\right)^2,\ {\rm and\ }\
C_{a'e}=0 .
\end{equation}
However, in this case, $v_{\rm PQ}$ is at the same time
the seesaw scale. In fact, below EWSB, the last two terms of Eq. (\ref{lyukseesaw}) give rise
to a neutrino mass matrix of the form
\begin{equation}
M_\nu = \left( \begin{matrix} 0&M_D\\ M_D^T& M_M \end{matrix}\right)
\sim \left( \begin{matrix} 0& F v\\ F^T v & y\, v_{\rm PQ} \end{matrix}\right)\,,
\end{equation}
realizing the seesaw mechanism \cite{seesaw1,seesaw2,Mohapatra:1979ia}, i.e. explaining the smallness of the masses of the left-handed SM active neutrinos by the large mass of the right-handed SM singlet neutrinos,
\begin{equation}
\label{seesaw}
m_{\nu} = - M_D M_M^{-1} M_D^T = -  F\,y^{-1}\,F^T\ \frac{v ^2}{v_{\rm PQ}}
= 0.6\,{\rm eV}  \left( \frac{10^{12}\,{\rm GeV}}{v_{\rm PQ}} \right)
\left( \frac{-  F\,y^{-1}\,F^T}{10^{-2}}\right)
\,.
\end{equation}

Remarkably, for an intermediate scale symmetry breaking scale, $f_{a'}=v_{\rm PQ}\sim 10^{11\div 12}$\,GeV,
such a simple single KSVZ-like extension of the SM can explain simultaneously
\begin{itemize}
\item the non-observation of strong CP violation (due to the coupling of $a'$ to $G\tilde G$),
\item the smallness of the active neutrino masses (seesaw with natural sizes of Yukawa couplings),
\item the nature of cold dark matter (axions),
\item the baryon asymmetry of the universe (due to thermal leptogenesis from $N_{iR}$ decay \cite{Fukugita:1986hr}, requiring $M_{N_{iR}}\gtrsim 10^9$\,GeV \cite{,Buchmuller:2012eb}), and
\item the stabilization of the electroweak vacuum (due to a threshold effect associated with the Higgs portal term proportional to $\lambda_{H\sigma}$ in Eq. \eqref{scalar_potential}
\cite{EliasMiro:2012ay}).
\end{itemize}

Clearly, these single axion models can be easily extended to have two (or more) Peccei-Quinn symmetries,
realizing KSVZ $\times$ KSVZ, DFSZ $\times$ DFSZ, and mixed KSVZ $\times$ DFSZ models. We will present a particularly
well motivated KSVZ $\times$ DFSZ type model in Sec. \ref{sec:unificaccion}.

\subsection{\label{lacion_model} Photophilic models}

We have, however, not yet exhausted our model building possibilities. Both in KSVZ-type as well as in DFSZ-type models, there are always colored fermions carrying PQ charges in a chiral manner -- the exotic color triplets and ordinary SM quarks, respectively. Therefore, in these cases the axion-like fields always have a non-zero axionic coupling to $G\tilde G$, $C_{a'g}\neq 0$. In view of the astrophysical hints for the existence of an ALP different from the axion it is of considerable phenomenological interest to construct also models in which the axion-like field is photophilic, i.e. $C_{a'g}= 0$, but $C_{a'\gamma}\neq 0$, as has been discussed in Sec.  \ref{sec:correlations}.

The simplest model predicting a photophilic ALP is a variant of the KSVZ model
where the exotic fermion is a colorless but electrically charged particle,
which we denote by $E=(E_L,E_R)$.
The hidden complex singlet $\sigma$ is introduced as usual and the PQ symmetry\footnote{Strictly speaking, this symmetry should be dubbed
``PQ-like" rather  than ``PQ" since, unlike in the proper PQ case, the model has no $U(1)_{\rm PQ}\times SU(3)_C\times SU(3)_ C$ chiral anomaly, but
only a  $U(1)_{\rm PQ}\times U(1)_{\rm em}\times U(1)_ {\rm em}$ chiral anomaly.} acts
as
\begin{equation}
\sigma\to e^{i\alpha}\sigma\,,\hspace{6ex}
E_L\to e^{i\alpha/2}E_L\,,\hspace{6ex}
E_R\to e^{-i\alpha/2}E_R\,.
\label{PQ:laxion:ksvz}
\end{equation}
The Yukawa Lagrangian is analogous to the KSVZ model and reads
\begin{equation}
\lag_Y \supset y\bar{E}_L\sigma E_R   + h.c. \,.
\label{yuk:laxion:ksvz}
\end{equation}
The parametrization of the axion-like field $a'$ remains as in Eq. \eqref{sigma:}, with
Eq.\,\eqref{F1} setting the ALP decay constant $f_{a^\prime}$.
The couplings of $a'$ to gluons, photons, and electrons are given by (cf. Appendix
\ref{sec:lecouplings})
\begin{equation}
C_{a'g}=0,\hspace{6ex}
  C_{a'\gamma}=2\, \big(C^{(E)}_{\rm em}\big)^2,  \hspace{6ex}
  C_{a'e}=0\,,
% \label{cgamma12d}
\end{equation}
where $C^{(E)}_{\rm em}$ denotes the electric charge of $E$.
If, however, the interaction in Eq.\,\eqref{yuk:laxion:ksvz} is the only
source of interaction for $E$ at the PQ scale or below, the $E$-number
would be conserved and this exotic lepton would constitute a stable charged
particle which is cosmologically problematic, unless its mass
$m_E = y \, v_{\rm PQ} \lesssim $ TeV (for a review, see Ref. \cite{Perl:2001xi}),
requiring then an unnaturally small Yukawa coupling $y\lesssim 10^{-6}$  if  $v_{\rm PQ}\gtrsim 10^9$ GeV.
One possible remedy for this situation is to set $C^{(E)}_{\rm em}=\pm 1$, so that
either $E_R$ or $E_L^c$ has the SM gauge quantum numbers of $l_{iR}$ and then $E$
is allowed to mix with the SM charged leptons and to decay to SM
particles. A similar mixing mechanism has been proposed to allow for the decay of the
exotic superheavy quark $Q$ in the KSVZ model which shares with the superheavy exotic fermion $E$ the
cosmological problem  \cite{Nardi:1990ku,Berezhiani:1992rk,Dasgupta:2013cwa}.
We present models of this type with both an axion and an additional photophilic axion-like field in
Secs. \ref{singlets_model} and \ref{another_singlets_model}.

Another simple alternative photophilic model exploits, apart from the
usual hidden complex scalar $\sigma$ and similar to the DFSZ model, a second Higgs doublet,
$H_{l}$, which gives rise to the masses of the charged leptons, cf.
 \begin{equation}
{\lag_Y} =Y_{ij}\overline{q}_{iL}\widetilde{H}\, u_{jR}+\Gamma_{ij}\overline{q}_{iL}H\, d_{jR}+G_{ij}\overline{L}_{i}H_{l}l_{jR}+h.c.\label{yuk2d}
\end{equation}
This extra Higgs doublet is necessary for the restriction of  the $U(1)_{\rm PQ}$ symmetry to the
lepton sector.
The fields are supposed to transform as
\begin{align}
l_{iR} & \rightarrow e^{-i\alpha}l_{iR}, \nonumber \\
H_{l} & \rightarrow e^{i\alpha}H_{l}, \label{transf2d}\\
\sigma & \rightarrow e^{i\alpha}\sigma\,,\nonumber
\end{align}
under the $U(1)_{\rm PQ}$ symmetry, such that both the above Yukawa Lagrangian and
the scalar potential,
\begin{equation}
V=V_{\mathcal{H}}+m\, H_{l}^{\dagger}H\,\sigma+h.c.,\label{pot2d}
\end{equation}
where $V_{\mathcal{H}}\equiv V_{\mathcal{H}}\left(H,\, H_{l},\sigma\right)$
stands for the sum of all Hermitian terms in the fields and $m$ is a
mass parameter,  are invariant under PQ transformations.  SM quarks are not charged
under $U(1)_{\rm PQ}$. The axion-like field
$a'$ in this model has a decay constant
\begin{equation}
f_{a}^{\prime}=\sqrt{v_{\rm PQ}^{2}+v_{l}^{2}},
\end{equation}
where $v_l$ is the vev of $H_l$,
and the couplings to gluons, photons, and electrons are (cf. Appendix \ref{sec:lecouplings})
\begin{equation}
C_{a'g}=0,\hspace{6ex} C_{a'\gamma}
  =2\, N_g = 6,\hspace{6ex} C_{a'e}=1\,.
\label{cgamma12d}
\end{equation}
We present a well-motivated model of this type with both an axion and an additional  photophilic axion-like field in Sec. \ref{accion_laccion_model}.
These types of model have the characteristic feature that the presence of a
photophilic axion-like field is always accompanied by possible signals at the
electroweak scale through $H_l$.

\section{Intermediate-Scale accidental axion and further ALPs from Field Theoretic Bottom-Up SM Extensions}
\label{accions_bottom_up}

The ad-hoc Peccei-Quinn extensions of the Standard Model discussed above give some insight into the origin of the decay constants and the sizes of the anomaly coefficients, but still have several drawbacks.
Most importantly, new fields (hidden complex scalar fields, heavy vector-like quarks, extra SM Higgses) and new PQ symmetries have been introduced and imposed by hand.
Moreover, these symmetries are not protected from explicit symmetry breaking effects by Planck-scale suppressed operators --- e.g.
\begin{equation}
\lag \supset \frac{1}{M_{\rm Pl}^{D-4}}{\mathcal O}_D \sim \sigma_1^n \sigma_2^k, \hspace{6ex}
D=n+k>4,
\end{equation}
for a model with two singlets ---  expected to appear generically in  the low-energy effective Lagrangian \cite{Georgi:1981pu,Ghigna:1992iv,Holman:1992us,Kamionkowski:1992mf,Barr:1992qq,Kallosh:1995hi}.
These operators modify the axion potential, eventually shifting its minimum away from zero, thereby destroying the solution of the strong CP problem, cf. Appendix  \ref{sec:effc-op}. Moreover, they induce masses, e.g.
\begin{equation}
m_{12}^{(n,k)}\sim v_1^{(n-1)/2}v_2^{(k-1)/2}/M_{\rm Pl}^{(D-4)/2},
\end{equation}
 lifting also the additional ALPs to pseudo Nambu-Goldstone bosons.

Both drawbacks can be absent in models where the Peccei-Quinn symmetries are not ad-hoc,
but instead automatic or accidental symmetries  \cite{Georgi:1981pu}. In fact, Peccei-Quinn symmetries could be accidental consequences of exact discrete $\ZZ_N$ symmetries, which in addition, if $N=D+1$ is large enough,
i.e.
\begin{equation}
N = D+1 \gtrsim \frac{9}{\left[1-0.1\cdot \log\left(f_A/10^{9}\,{\rm GeV}\right)\right]}+1,
\end{equation}
can protect the axion against semi-classical gravity effects  \cite{Dias:2002hz,Dias:2002gg,Dias:2003zt,Carpenter:2009zs,Harigaya:2013vja,Kim:2013jka}, cf. Appendix  \ref{sec:effc-op}.
In the following, we will construct intermediate scale models with $n_{\rm ax}=2$ accidental axion-like fields.

\subsection{$\ZZ_{13}\otimes \ZZ_5\otimes\ZZ_{5}^\prime$ model with axion-ALP mixing}
\label{sec:unificaccion}

We now  present a SM extension  with two axion-like fields corresponding to (pseudo-) Nambu-Goldstone
bosons from the breaking of two continuous Peccei-Quinn
symmetries $\upq[1]\times \upq[2]$, which themselves are accidental consequences
of exact discrete $\ZZ_N$ symmetries.  The model is a further extension of an $n_{\rm ax}=1$ model proposed in
Ref.~\cite{Dias:2002hz}, where the SM was extended by several fields: not only by a
SM  complex scalar  singlet $\sigma$, but also by
several SM Higgs multiplets and three right-handed SM singlet neutrinos.
These extra fields were primarily included to enable
the formulation of a large enough discrete symmetry ensuring the protection of the QCD axion against de-stabilization by Planck-scale suppressed explicit symmetry-breaking effects.
Among the attractive features of the model are
\begin{itemize}
\item the close connection between the seesaw scale to explain the smallness of the active
neutrino masses and the Peccei-Quinn scale (see also \cite{Kim:1981jw,Kim:1981bb,Mohapatra:1982tc,Shafi:1984ek,Langacker:1986rj,He:1988dm,Berezhiani:1990wn,Berezhiani:1989fp,Arason:1990sg,Dias:2005dn,Gu:2010zv,Altarelli:2013aqa}),
\item the stabilization of the accions and the proton, as well as the non-occurence of flavor changing neutral currents, due to the large discrete symmetries, and
\item the unification of the SM gauge couplings at intermediate scales, $M_{\rm U}\sim 10^{13}$\,GeV
\cite{Dias:2004hy,Dias:2007vx}.
\end{itemize}

The extended $n_{\rm ax}=2$ model has four SM Higgs doublets $H_u$, $H_d$, $H_l$, $H_N$, which give Dirac mass terms for
up-type quarks, down-type quarks, charged leptons and neutrinos, respectively, see Eq. \eqref{lyuk2} below;
an $SU(2)_L$ triplet $T$ with hypercharge $Y=2$; and two SM singlet complex scalar fields
$\sigma_1$ and $\sigma_2$, which carry only
\PQ charges.
The SM fermionic content is augmented by the addition of a vector-like color triplet, ($Q_L,Q_R$),
as in the KSVZ model, and three right-handed neutrinos, $N_{iR}$, $i=1,2,3$.

We impose an exact discrete symmetry $\ZZ_{13}\otimes \ZZ_5\otimes\ZZ_{5}^\prime$, with the
fields transforming according to Table \ref{tableZns}. As mentioned above, the
reason for such a large discrete symmetry is to suppress operators up to some
mass dimension, thereby ensuring the effectiveness of the \PQ mechanism.
The order of each group factor is chosen to be a prime number because it
forbids dangerous operators formed by invariant powers of the singlets
$\sigma_{1,2}^k$, with $k$ less than the group order, as dictated by the Lagrange
theorem.

\begin{table}[h]
\[
\begin{array}{|c|cccccccccccccccc|}
\hline
\psi_i & q_L & u_R & d_R & L & N_{R} & l_R & H_u & H_d & H_l & H_N &
\sigma_2 & T  & Q_L & Q_R & \sigma_1 & \\
\hline
\ZZ_{13} & \omega^5_{13} & \omega^3_{13}  & \omega^{8}_{13} & \omega^{9}_{13} & \omega^3_{13} &
\omega^{7}_{13} & \omega^{11}_{13} &  \omega^{10}_{13} &  \omega^2_{13} &  \omega^{7}_{13} &
\omega^{12}_{13} & \omega^{9}_{13} &  1 & \omega^6_{13} &\omega^{7}_{13}&
\\[.5ex]
\hline
\ZZ_5 & 1 &  \omega_5 &  \omega_5^{4} &  1 &
\omega_5 &  \omega^{4}_5 &  \omega_5  &
\omega_5   & \omega_5  & \omega_5 & 1 &
\omega_5^2 & \omega_5 & \omega^{3}_5 &
\omega^3_5 &
\\[.5ex]
\hline
\ZZ_{5}^{\prime } & 1 & \omega^{4}_5  & 1 & 1 & \omega^{2}_5 &
\omega^{4}_5 & \omega^{4}_5 &  1 &  \omega_5  & \omega^{2}_5 & \omega_5 & \omega^{3}_5 &  1 &
\omega^{4}_5 & \omega_5 & \\[.5ex]
\hline
\end{array}
\]
\caption{\label{tableZns}
$\ZZ_{13}\otimes\ZZ_5\otimes\ZZ^\prime_5$ charges, where $\omega_{13}\equiv e^{i2\pi/13}$
and $\omega_5\equiv e^{i2\pi/5}$.
}
\end{table}

The discrete symmetry in Table \ref{tableZns}  allows for the following
Yukawa interaction terms:
\begin{eqnarray}
\label{lyuk2}
{\lag_Y} & = &  Y_{ij}\overline{q}_{iL}\widetilde{H}_u u_{j R}  +\Gamma_{i
j}\overline{q}_{iL} H_d d_{j R}    + G_{ij}\overline{L}_{i } H_l l_{j R}   \nonumber \\
 &+&     F_{ij} \overline{L}_{i
}\widetilde{H}_N N_{j R}  + y_{ij} \overline{(N_{iR})^c} \sigma_1 N_{jR}
+ y_Q \overline{Q}_L \sigma_1 Q_R \,.
\end{eqnarray}
Here $q_{iL}$, $i=1,2,3$, denote the left-handed $SU(2)_L$ quark doublets,
$\widetilde{H}=\epsilon H^\ast$, and
$Y_{ij}$, $\Gamma_{i j}$, $F_{i j}$ and $G_{i j}$ are complex arbitrary
$3\times 3$ matrices, and $y_{ij}$ is a symmetric $3\times 3$  matrix.
Of phenomenological great importance is the fact that, despite the occurence of multiple Higgs
doublets, no flavor changing neutral currents emerge,
since there is only one Higgs doublet for each type of fermion.

The scalar potential with all renormalizable terms allowed by the discrete symmetry
is $V=V_H+V_{NH}$, where the hermitian terms are
\begin{eqnarray}
\label{VH}
V_H & = &  \mu^2_k \vert H_k\vert^2+\mu^2_T{\rm
Tr}[\vert T\vert^2]+m^2_{a}\vert\sigma_a\vert^2+
\lambda_{kk'}\vert H^\dagger_k H_{k'}\vert^2+\lambda_{T1}({\rm
Tr} \vert T\vert^2)^2+\lambda_{T2} {\rm Tr}[\vert T\vert^4]
\nonumber \\ &+&
\lambda_a\vert\sigma_a\vert^4+\alpha_{k1}\vert H_k\vert^2{\rm Tr}[\vert
T\vert^2]+\alpha_{k2}\vert T\,H_k\vert^2 +\beta_{ka}\vert
H_k\vert^2\vert\sigma_a\vert^2+\gamma_{a}{\rm Tr}[\vert
T\vert^2]\vert\sigma_a\vert^2 ,
\end{eqnarray}
where $k,k'=u,d,l,N$, $a=1,2$, while the nonhermitian terms read
\begin{eqnarray}
\label{VNH}
V_{NH} & = & \tilde{\lambda}_1 (H^\dagger_d H_l)( H^\dagger_N H_l)
+\tilde{\lambda}_2
\widetilde{H}^T_d T\widetilde{H}_u\sigma_2+
\tilde{\lambda}_3(H^{\dagger}_u H_N)(H^\dagger_uH_l)
\nonumber \\ &+& M_1 H^\dagger_d H_u\sigma_2+
M_2\widetilde{H}^T_u T\widetilde{H}_u+
M_3 \widetilde{H}^T_l T\widetilde{H}_N+h.c.
\end{eqnarray}
Note, that if we did not impose the additional  $\ZZ_5\otimes\ZZ_{5}^{\prime }$ symmetry,
dangerous terms of the type
$(H^{\dagger}_l H_N)\sigma_1\sigma_2^*$, $(H^{\dagger}_d
H_l)\sigma_1\sigma_2^*$,  $(H^{\dagger}_u H_d)\sigma_1^2$, and $\sigma_1^2 \sigma_2$,
would still be allowed, breaking the accidental global
symmetries discussed below, in particular the Peccei-Quinn symmetries.

Accidentally,
the model \eqref{lyuk2} - \eqref{VNH} possesses five $U(1)$ (i.e. rephasing) symmetries which are constrained by the non-hermitean terms in \eqref{lyuk2} and \eqref{VNH}:  the imposed $U(1)_Y$ invariance and four
additional accidental global $U(1)$ symmetries. Among the latter, we can identify
two non-chiral ones, namely $U(1)_B$, ensuring baryon number conservation for usual quarks and
$U(1)_Q$, ensuring $Q$-number conservation for the exotic quark.
The two remaining accidental symmetries,
the \PQ symmetries $\upq[1]$ and $\upq[2]$, involve the singlets $\sigma_1$ and $\sigma_2$, and are chiral for quarks.

The latter accidental symmetries can be chosen in a way that $\sigma_1$ is only charged under
$\upq[1]$, while $\sigma_2$ is only charged under $\upq[2]$.
For example, the potential in (\ref{VNH}) is independent of $\sigma_1$ and so a
change of the global phase of this field can be compensated by changing the phases
of $Q_{L,R}$, $N_{iR}$, $L_i$, and $l_{iR}$. We identify this symmetry with
$\upq[1]$.

We denote the charges of a field $\psi$ under
the \PQ symmetries $\upq[1]$ and $\upq[2]$ as $X_\psi$ and $K_\psi$, respectively,
\eqali{
\label{tg}
\upq[1]:& & \psi&\rightarrow e^{i\alpha K_\psi}\psi\,,\cr
\upq[2]:& & \psi&\rightarrow e^{i\alpha X_\psi}\psi\,.
}
If we write $X_{\sigma_1}=(X_u+X_d)$, $X_{u}\equiv - X_{u_R}$, $X_{d}\equiv - X_{d_R}$, and choose the normalization
\eq{
X_u+X_d=1, \hspace{6ex} K_{\sigma_1} = 1
\,,
}
then the transformation of the singlets reads
\eqali{
\label{NM:UPQ:S12}
\upq[1]:& & \sigma_1&\rightarrow e^{i\alpha_1}\sigma_1\,,\cr
\upq[2]:& & \sigma_2&\rightarrow e^{i\alpha_2}\sigma_2\,,
}
and the theory \eqref{lyuk2} - \eqref{VNH} is obviously invariant under the following $\upq[1]$ transformation (displaying only
the non-trivially transforming fields)
\eqali{
\label{U1:PQ2NM}
\upq[1]:&& L_i &\to e^{-i\frac{1}{2}\alpha_1}L_i\,,\cr
        && l_{iR} &\to e^{-i\frac{1}{2}\alpha_1}l_{iR}\,,\cr
        && N_{iR} &\to e^{-i\frac{1}{2}\alpha_1}N_{iR}\,,\cr
        && Q_{L} &\to e^{i\frac{1}{2}\alpha_1}Q_{L}\,,\cr
        && Q_{R} &\to e^{-i\frac{1}{2}\alpha_1}Q_{R}\,.
}
This symmetry is an extension of the SM lepton number, including right-handed
neutrinos, and is spontaneously broken when $\sigma_1$ acquires a vev, cf. Sec. \ref{axion_majoron}.
We show the $\upq[1]$ charges
also in Table \ref{table22}.

The $\upq[2]$ charges are also easily extracted by requiring invariance of \eqref{lyuk2} - \eqref{VNH}
under $\upq[2]$, and we find
\eqali{
\label{U1:PQ1NM}
\upq[2]:
        && H_d &\to e^{iX_d\alpha_2}H_d\,,\cr
        && H_u &\to e^{-iX_u\alpha_2}H_u\,,\cr
        && H_l &\to e^{i(X_L-X_l)\alpha_2}H_l\,,\cr
        && H_N &\to e^{-iX_L\alpha_2}H_N\,,\cr
        && T &\to e^{-i2X_u\alpha_2}T\,,\cr
        && d_{iR} &\to e^{-iX_d\alpha_2}d_{iR}\,,\cr
        && u_{iR} &\to e^{-iX_u\alpha_2}u_{iR}\,,\cr
        && L_i &\to e^{iX_L\alpha_2}L_i\,,\cr
        && l_{iR} &\to e^{iX_l\alpha_2}l_{iR}\,.
}

It should be noted that we have written some charges in terms of $X_u,X_d,X_L,X_l$ but there is only a
freedom to choose $X_u$ or $X_d$, which is subsequently fixed after
electroweak symmetry breaking. The rest of the charges can be found in the second line of Table~\ref{table22} and
they are equivalent to the ones in Ref.~\cite{Dias:2002hz} when the nonhermitian
terms in the potential are the same.
When defining the $\upq[2]$ charges in Eq.\,(\ref{U1:PQ1NM}), we have chosen
$\sigma_1$ and $q_L$ to be uncharged under $\upq[2]$, i.e.,
$X_{\sigma_1}=X_{q_L}=0$, and $X_{Q_L}=-X_{Q_R}$, for convenience.
The first choice is what turns $N_{iR}$ and $Q_{L,R}$ uncharged
under $\upq[2]$. A different choice would lead to a different value for the coefficients
$C_{ig}$ and $C_{i\gamma}$, which are dependent on the choice of $\upq[1]$
and $\upq[2]$. The couplings to the physical axion and ALP, however, are fixed for
a given model and thus independent of the choice of $\upq[1]$ and $\upq[2]$.

The parameters in the scalar potential are assumed to be such that the SM doublet Higgs fields get
non-zero vevs $\langle H_k\rangle=v_k/\sqrt{2}$, $k=u,d,l,N$, at the electroweak scale, $v=\sqrt{\sum_k v_k^2}= 246$\,GeV, while
the SM singlet fields are assumed to attain their vevs, $\langle\sigma_a\rangle =v_a/\sqrt{2}$, $a=1,2$, at much higher scales.
Therefore, the right-handed neutrinos $N_{iR}$ and the extra colored triplet, which get their masses
via Yukawa couplings with $\sigma_{1}$ in Eq. \eqref{lyuk2}, are heavy, decoupling
from the electroweak scale. The first two terms in the second line of Eq. (\ref{lyuk2}) give rise
to a neutrino mass seesaw relation, which now involves the vev $v_N$ of $H_N$,  which might be much smaller than
$v$,
\begin{equation}
\label{seesaw}
m_{\nu} = - M_D M_M^{-1} M_D^T = -  F\,y^{-1}\,F^T\ \frac{v_N ^2}{v_1}
= 0.1\,{\rm eV} \left( \frac{v_N}{100\,{\rm GeV}}\right)^2 \left( \frac{10^{12}\,{\rm GeV}}{v_1} \right)
\left( \frac{-  F\,y^{-1}\,F^T}{10^{-2}}\right)
\,.
\end{equation}

At low energies, two Nambu-Goldstone bosons $a_a^\prime$ with decay constants
\begin{equation}
f_{a_b^\prime}\simeq v_b
\end{equation}
arise from the breaking of the two accidental accidental Peccei-Quinn symmetries,
describing the phases of the singlet fields $\sigma_b$ in the expansion around the vevs.
The Nambu-Goldstone boson $a_1^\prime$ is
thus also a majoron-like particle \cite{Chikashige:1980ui,Gelmini:1980re}.  The
interaction term $y_Q \overline{Q}_L \sigma_1 Q_R$ implies that $\upq[1]$ is in
fact a chiral symmetry for the exotic quark $Q$. As noted previously, the number of
this quark is separately conserved.

\begin{table}[h]
\[
\begin{array}{|c|cccccccccccccccc|}
\hline
\psi & q_L & u_R & d_R & L & N_{R} & l_R & H_u & H_d & H_l & H_N &
\sigma_2 & T  & Q_L & Q_R & \sigma_1 & \\
\hline
K_\psi & 0 & 0 & 0 & -1/2 & -1/2 & -1/2 & 0 & 0 & 0 & 0 & 0 & 0 & 1/2 & -1/2 &
1 &
\\[.5ex]
\hline
X_\psi & 0 & -X_u & -X_d & \frac{1}{3}(4X_u+X_d) & 0 & 2X_u & -X_u & X_d &
  -\frac{1}{3}(2X_u-X_d) & -\frac{1}{3}(4X_u+X_d) & 1 & -2X_u  & 0  & 0 & 0 &
\\[.5ex]
\hline
\end{array}
\]
\caption{\label{table22}
 $U(1)_{\rm PQ_1}$ and $U(1)_{\rm PQ_2}$ charge assignments of the fields in the
SM extension \eqref{lyuk2} - \eqref{VNH}.
}
\end{table}

The ALP-gluon and ALP-photon couplings can be extracted from the charges in
Table~\ref{table22} as (cf. Appendix \ref{sec:lecouplings})
\begin{gather}
\label{values_model_unificaxion}
C_{1g}=1\,,\quad C_{2g}=N_g(X_u+X_d)=N_g =3\,, \\
  C_{ 1\gamma}=6\,\left( C_{\rm em}^{(Q)}\right)^2\,,\quad
C_{2\gamma}=\frac{2}{3}N_g(4X_u+X_d)+2N_g(X_L-X_l)
  =\frac{4}{3}N_g(X_u+X_d)=\frac{4}{3}N_g =4\,,
\end{gather}
leading to
\begin{equation}
\frac{C_{1\gamma}}{C_{1g}}=6\, \left( C_{\rm em}^{(Q)}\right)^2; \hspace{6ex}
\frac{C_{2\gamma}}{C_{2g}}=\frac{4}{3}
\label{ratios_model_unificaxion}
\,.
\end{equation}
The ALP-electron couplings can be extracted in the same way (cf. Appendix \ref{sec:lecouplings}),
\eq{
C_{1e}=0\,,\quad
C_{2e}=X_{H_l}= X_L-X_l=-\frac{2X_u-X_d}{3}\,,
}
so that
\eq{
\frac{C_{2e}}{C_{2g}}=-\frac{1}{3N_g}(2\cos^2\beta-\sin^2\beta) = -\frac{1}{3}\left( 1- \frac{3}{2}\sin^2\beta\right)\,,
}
where $ \tan\beta \equiv v_u/v_d$.

Therefore,  the considered model automatically leads to a low energy theory with two
axion-like fields $a_b^\prime$, both coupling to $G\tilde G$ and $F\tilde F$. Successful implementation of the neutrino seesaw mechanism  with not too much fine-tuning in the Yukawa couplings and thermal leptogenesis prefer the symmetry breaking scale $v_1$ and thus the decay constant $f_{a^\prime_1}$
of the majoron-like particle
$a_1^\prime$ to be in the range
\begin{equation}
10^9\,{\rm GeV}\lesssim v_1\simeq f_{a_1^\prime}\lesssim  10^{13}\,{\rm GeV}.
\end{equation}
The symmetry breaking scale $v_2$ and thus the decay constant $f_{a^\prime_2}$ of the axion-like field $a_2^\prime$, however, is not fixed by bottom up considerations.

%%%%%%%%%
\begin{figure}
\begin{center}
\includegraphics[width=0.7\textwidth]{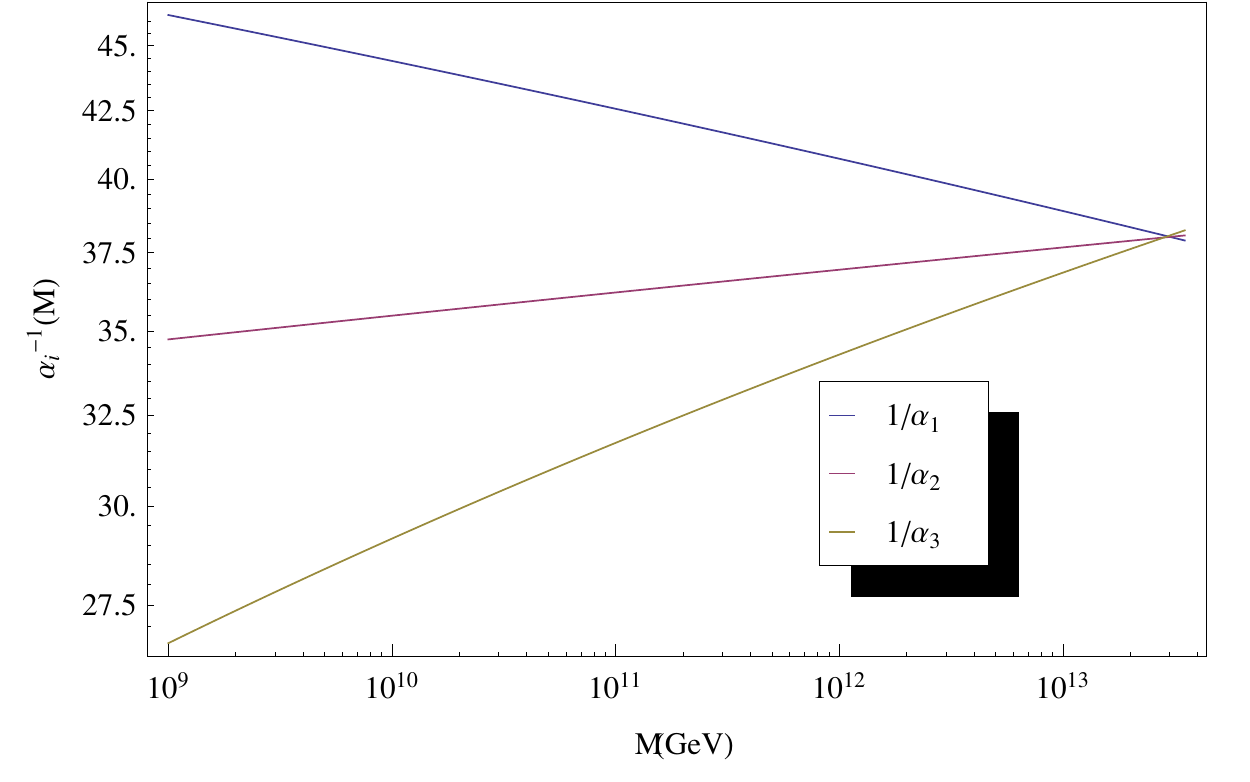}
\caption{Gauge coupling constants unification for the model in sec. \ref{sec:unificaccion} in the one loop approximation.}
\label{unicaxion}
\end{center}
\end{figure}
%%%%%%%%%%%%

Remarkably, the upper end of the above parameter range is distinguished by another observation,
namely gauge coupling constant unification. In fact, if we take both Peccei-Quinn symmetry breaking scales
\begin{equation}
v_b\simeq f_{a_b^\prime}\sim 10^{13}\,{\rm GeV},\
b=1,2,
\end{equation}
then the new fermionic fields, $N_{iR}$, and $(Q_L,\,Q_R)$ are naturally decoupled  and the running of the
gauge couplings from the electroweak scale towards high scales is determined by the SM
fermionic fields plus the electroweak scale doublets, $H_u$, $H_d$, $H_l$,
$H_N$, and the triplet $T$. It was shown in Ref.\,\cite{Dias:2004hy} that, assuming the $SU(5)$ hypercharge normalization, a model with such a content of fields unifies the gauge coupling constants at the energy scale
\begin{equation}
M_{\rm U}\approx 2.8\times 10^{13}\, {\rm GeV},
\end{equation}
see Fig.\,\ref{unicaxion}. This
value is surprisingly close to the PQ scales and indicates that the model can
be embedded in a grand unification group, so pointing to a connection between
such scales \cite{Dias:2007vx}.
In this connection it is important to note that dangerous operators leading to
baryon number violation processes \cite{Weinberg:1979sa} are forbidden by the discrete symmetry $\ZZ_{13}\otimes \ZZ_5\otimes \ZZ_5^\prime$ to very high mass dimensions.
Therefore, in this model proton decay is not observable, despite the low unification scale.

Importantly, as detailed in Appendix  \ref{sec:effc-op}, the discrete symmetry  $\ZZ_{13}\otimes \ZZ_5\otimes\ZZ_{5}^\prime$  is large enough to protect the axion from
destabilization by Planck scale suppressed explicit PQ symmetry breaking operators and that the corresponding induced masses are negligible, as long
as
\begin{equation}
f_{a_b^\prime}\simeq v_b \lesssim 10^{12}\  {\rm GeV},\
b=1,2.
\end{equation}
This is based on the fact, that  the operators of lowest mass-dimension that are
invariant under the discrete symmetries of Table \ref{tableZns}, but not invariant under
the \PQ symmetries of Table \ref{table22} in the low-energy effective Lagrangian
involve high powers of the scalar singlet fields,
\eq{
\label{op:m=1_main}
\frac{1}{M_{\rm Pl}^{10}}\,H_N^\dag H_d\sigma_1^{*5}\sigma_2^7\,,~~
\frac{1}{M_{\rm Pl}^{11}}\,H_l^\dag H_u\sigma_1^{5}\sigma_2^{*8}\,,~~
\frac{1}{M_{\rm Pl}^{11}}\,H_N^\dag H_u\sigma_1^{*5}\sigma_2^{8}
\,.
}

\begin{figure}
\begin{center}
\includegraphics[width=0.75\textwidth]{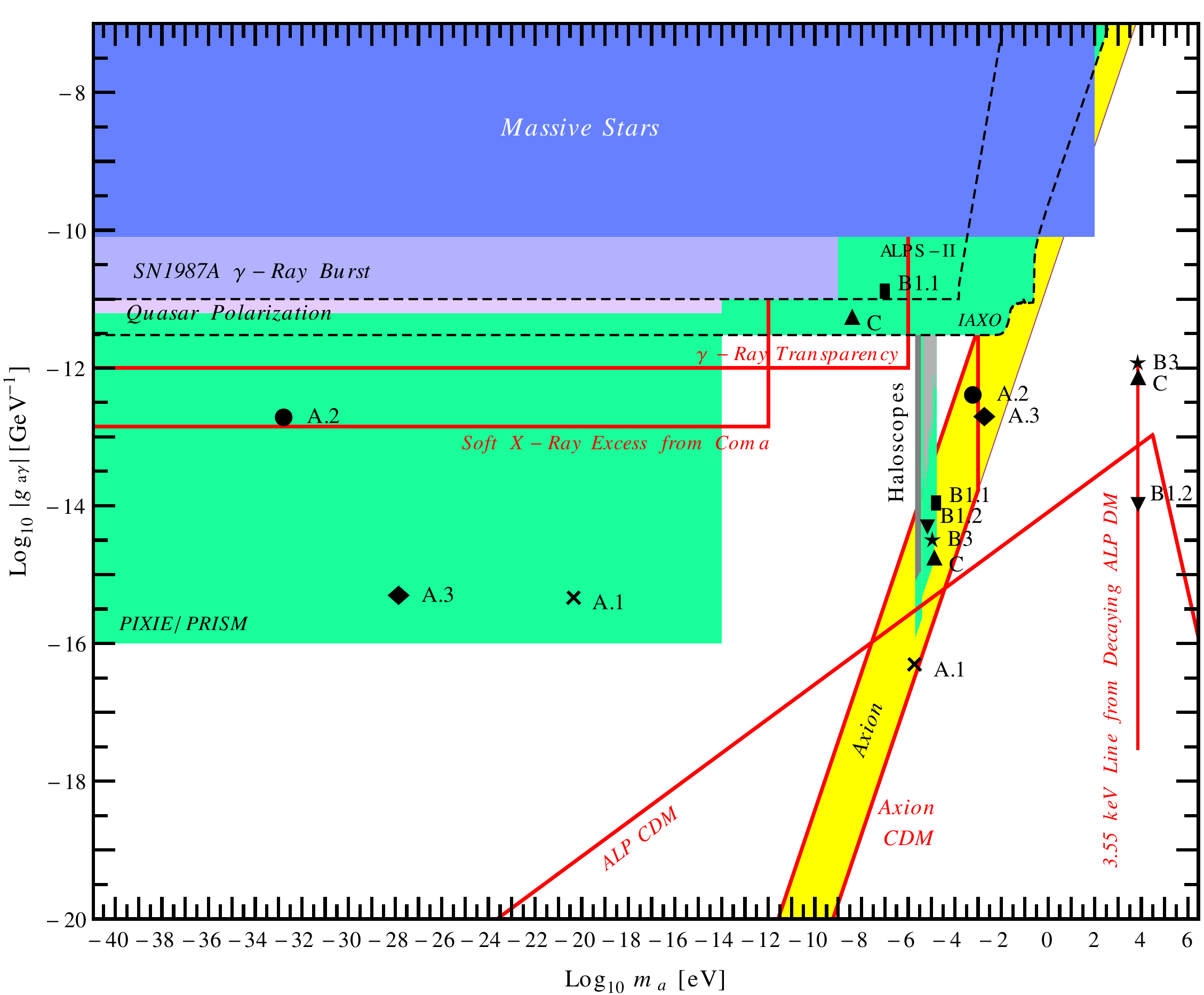}
\caption{Prediction of the axion and ALP photon coupling versus their masses for our models from Sec.  \ref{accions_bottom_up},
compared to the current limits, projected experimental sensitivities and preferred regions for explanations of astrophysical and cosmological
phenomena.}
\label{ALP_coupling_limits_and_models}
\end{center}
\end{figure}

The theoretically most favored value for the decay constant of the axion in this model is then
(cf. Sec. \ref{sec:correlations})
\begin{equation}
f_A = \left( \frac{1}{f_{a_1^\prime}^2} + \frac{9}{f_{a_2^\prime}^2} \right)^{-1/2}
\simeq 8.3\times 10^{11}\ {\rm GeV},
\ {\rm \ for\ }
 f_{a^\prime_1} \simeq v_1  =  10^{13}\ {\rm GeV},\   f_{a^\prime_2}\simeq v_2 =  2.5\times 10^{12}\ {\rm GeV},
\label{acc_model_1_theory_favored}
\end{equation}
leading to the following mass predictions for the axion $A$ and the ALP $a$ (cf. Sec. \ref{sec:correlations} and Appendix \ref{sec:effc-op}),
\begin{equation}
m_A \simeq 7.2\times 10^{-6}\ {\rm eV}, \hspace{3ex}
m_a\simeq \frac{v\  v_1^{5/2} v_2^{7/2}}{2^6 M_{\rm Pl}^{5} f_A} \simeq 1.3\times 10^{-22}\ {\rm eV}
.
\end{equation}
Therefore, this model favors axion dark matter in the mass range probed by ADMX \cite{Asztalos:2011bm}.
The ultralight ALP $a$ in this model, however, will escape detection in current and near future laboratory experiments,
such as  ALPS-II \cite{Bahre:2013ywa} and IAXO \cite{Armengaud:2014gea},
and can not explain the astrophysical puzzles such as the anomalous cosmic $\gamma$-ray transparency, the
cooling excess of stars and the soft X-ray excess from Coma (cf. Sec. \ref{sec:intro}).
In fact, the axion and the ALP consist here of a non-negligible mixture of the original axion-like fields, with mixing
\begin{equation}
\cos\delta \simeq 0.083, \hspace{3ex} \sin\delta\simeq 0.996,  \hspace{3ex} \tan\delta  \simeq  12
.
\end{equation}
Correspondingly, the ALP couplings to photons and electrons are of quite similar size as the respective
axion couplings (for definiteness, we chose here $C^{(Q)}_{\rm em}=1$),
\begin{eqnarray}
\label{acc_model_1_theory_favored_gagamma}
|g_{a\gamma}| &\simeq &\frac{\alpha}{2\pi f_A}
\left| \frac{4}{3} - 6 \right| \cdot \frac{12}{145}
\simeq  5.4\times  10^{-16}\ {\rm GeV}^{-1}, \\
|g_{A\gamma}| &\simeq &\frac{\alpha}{2\pi f_A}
\left| 6 \cdot 0.083^2
+  \frac{4}{3} \cdot 0.996^2
- 1.95     \right| \simeq 8.2\times 10^{-16}\ {\rm GeV}^{-1}
, \\
|g_{ae}| &=&
 \frac{m_e}{f_A}
\left| -\frac{1}{3}\left( 1- \frac{3}{2}\sin^2\beta\right) - 0 \right|  \cdot \frac{12}{145}
\simeq  1.7\times 10^{-17} \left| 1- \frac{3}{2}\sin^2\beta\right|
,
\\
|g_{Ae}| &=&
\frac{m_e}{f_A}\left|
0\cdot  0.083^2
-\frac{1}{3}\left( 1- \frac{3}{2}\sin^2\beta\right) \cdot 0.996^2
\right|
\simeq 2\times 10^{-16} \left| 1- \frac{3}{2}\sin^2\beta\right|
,
\end{eqnarray}
and thus way too small to explain the cosmic $\gamma$-ray transparency (the latter requiring
$|g_{a\gamma}|  \gtrsim 10^{-12}\ {\rm GeV}^{-1}$, cf. Eq. \eqref{dec_const_transp} and Fig. \ref{ALP_coupling_limits}) and the
excess star cooling (the latter requiring $|g_{Ae}|  \gtrsim 10^{-13}$ and/or $|g_{ae}|  \gtrsim 10^{-13}$, cf. Eq. \eqref{wd_energy_loss}).
Moreover, the expected cosmic mass fraction in ALP dark matter is negligible~\cite{Arias:2012az},
\begin{equation}
\Omega_{a} h^2 \sim  10^{-14}
\left( \frac{m_a}{10^{-21}\ {\rm eV}} \right)^{1/2} \left( \frac{f_a}{10^{12}\ \rm GeV} \right)^{2}  ,
\label{eq:omegaalp:2}
\end{equation}
making also the direct laboratory detection of dark matter comprised of the ALP $a$ impossible.
Nevertheless, the indirect detection of an ultra-light ALP which such a small coupling to photons may still be possible
with the next generation of cosmic microwave background (CMB) observatories such as PIXIE and PRISM
(cf. Fig. \ref{ALP_coupling_limits_and_models})
because resonant photon-ALP oscillations in primordial magnetic fields may lead to observable spectral distortions
of the CMB \cite{Mirizzi:2009nq,Tashiro:2013yea,Ejlli:2013uda}.

%%%%%%%%%
%%%%%%%%%%%%

\begin{table}
\begin{center}
\begin{tabular}{|c|c|c||c|c|c|c|c|c|c|}\hline
& \multicolumn{2}{c||}{Input values} & \multicolumn{7}{c|}{Resulting low-energy parameters} \\ \cline{2-10}
\raisebox{1.5ex}{Model} & $v_1=f_{a_{1}^\prime}$ [GeV] & $v_2=f_{a_{2}^\prime}$ [GeV] & $f_A$ [GeV] & $m_A$ [eV] & $m_a$ [eV]&
$|g_{A\gamma}|$ [GeV]$^{-1}$ & $|g_{a\gamma}|$ [GeV]$^{-1}$ & $|g_{A e}|$ & $|g_{a e}|$ \\ \hline
%\raisebox{1.5ex}
A.1 & $1\times 10^{13}$ & $2.5\times 10^{12}$ & $8.3\times 10^{11}$ & $7.2\times 10^{-6}$  & $1.3\times 10^{-22}$ &  $8.2\times 10^{-16}$ & $5.4\times 10^{-16}$  & $2\times 10^{-16}$ & $1.7\times 10^{-17}$   \\ \hline
%\raisebox{1.5ex}
A.2 & $1\times 10^{10}$ & $7.5\times 10^{10}$ & $9.3\times 10^{9}$ & $6\times 10^{-4}$ & $1.8\times 10^{-33}$ & $4\times 10^{-13}$ & $2\times 10^{-13}$ &  $2.5\times 10^{-15}$ & $6.3\times 10^{-15}$    \\ \hline
%\raisebox{1.5ex}
A.3 &  $1\times 10^{13}$ & $1\times 10^{10}$ & $3\times 10^{9}$ & $2\times 10^{-3}$ & $1\times 10^{-28}$  &  $2\times 10^{-13}$ & $5\times 10^{-16}$  & $5\times 10^{-14}$ & $2\times 10^{-17}$  \\ \hline
%\raisebox{1.5ex}
B1.1 & $3\times 10^{11}$& $1\times 10^{9}$ & $3\times 10^{11}$ & $2\times 10^{-5}$ &$1\times 10^{-7}$  & $1.6\times 10^{-14}$ & $1.4\times 10^{-11}$    & 0 & $1\times 10^{-12}$  \\ \hline
%\raisebox{1.5ex}
B1.2 & $9\times 10^{11}$& $1.26\times 10^{12}$ & $9\times 10^{11}$ & $6.7\times 10^{-6}$ &$7.1\times 10^{3}$  & $5.2\times 10^{-15}$ & $1.1\times 10^{-14}$    & 0 & $8\times 10^{-16}$  \\ \hline
%\raisebox{1.5ex}
B3 & $1.1\times 10^{12}$& $1.84\times 10^{9}$ & $5.5\times 10^{11}$ & $1.1\times 10^{-5}$ &$7.1\times 10^{3}$  & $3.4\times 10^{-15}$ & $1.3\times 10^{-12}$    & 0 & $0$  \\ \hline
\end{tabular}
\caption{\label{summary_table}Summary of the low-energy parameters obtained from our models in Secs. \ref{sec:unificaccion},
\ref{accion_laccion_model}, and \ref{another_singlets_model}, for our chosen benchmark values for the PQ vevs $v_1$ and $v_2$. For definiteness, we used
in the electron couplings of the models in Secs. \ref{sec:unificaccion}  and \ref{accion_laccion_model} vev ratios $\beta = v_u/v_d$ and $\beta^\prime = v_l/v$, such that $\left| 1- \frac{3}{2}\sin^2\beta\right| = 1$ and  $\sin^2\beta^\prime = 1$, respectively. Moreover, we took $g=1$ and $g^\prime =1$.}
\end{center}
\end{table}

On the other hand, if we give up unification and allow for Yukawa couplings of order $10^{-4}$ to obtain the
small active neutrino masses even in the case of a lower seesaw scale, we can get an
ALP in the parameter range
favored by the explanation of the soft X-ray excess from the Coma cluster.
Choosing for example
\begin{equation}
v_1 \simeq f_{a_1^\prime} = 1\times 10^{10}\ {\rm GeV}, \hspace{6ex}
v_2 \simeq f_{a_2^\prime} = 7.5\times 10^{10}\ {\rm GeV},
\end{equation}
leads to an axion $A$ with a quite small decay constant and thus rather large mass,
\begin{equation}
f_A \simeq  9.3\times 10^{9}\ {\rm GeV}, \hspace{6ex} m_A\simeq 0.6\ {\rm meV},
\end{equation}
constituted mainly
by the axion-like field $a_1^\prime$ (small mixing $\delta$),
\begin{equation}
\cos\delta \simeq 0.96, \hspace{6ex}
\sin\delta \simeq 0.29, \hspace{6ex} \tan\delta \simeq 0.3 ,
\end{equation}
with a quite sizeable coupling to the photon,
\begin{equation}
|g_{A\gamma}| \simeq \frac{\alpha}{2\pi f_A}
\left( 6 \cdot 0.96^2
+  \frac{4}{3} \cdot 0.29^2
- 1.95     \right) \simeq 4\times 10^{-13}\ {\rm GeV}^{-1}
.
\end{equation}
At the same time, the ALP $a$, with mass $m_a \simeq 1.8\times 10^{-33}\ {\rm eV}$, is constituted mainly by the axion-like field
$a_2^\prime$, and has also a sizeable coupling to the photon,
\begin{equation}
|g_{a\gamma}| \simeq \frac{\alpha}{2\pi f_A}
\left( \frac{4}{3} - 6 \right) \cdot \frac{0.3}{1+0.3^2}
\simeq  2\times  10^{-13}\ {\rm GeV}^{-1}
,
\end{equation}
right in the range of phenomenological interest. The couplings to the electron, however, are in this case still
too small,
\begin{eqnarray}
|g_{Ae}| &=&
\frac{m_e}{f_A}\left|
0\cdot  0.96^2
-\frac{1}{3}\left( 1- \frac{3}{2}\sin^2\beta\right) \cdot 0.29^2
\right|
\simeq 2.5\times 10^{-15} \left| 1- \frac{3}{2}\sin^2\beta\right|
,
\\
|g_{ae}| &=&
 \frac{m_e}{f_A}
\left| -\frac{1}{3}\left( 1- \frac{3}{2}\sin^2\beta\right) - 0 \right|  \cdot \frac{0.3}{1+0.3^2}
\simeq  6.3\times 10^{-15} \left| 1- \frac{3}{2}\sin^2\beta\right|
,
\end{eqnarray}
in order to explain the anomalous cooling of stars.

For illustration, let us take also a third pair of benchmark values for the PQ scales which leads to
a case with very tiny mixing between the axion an the ALP. For definiteness, we chose
\begin{equation}
v_1 \simeq f_{a_1^\prime} = 10^{13}\ {\rm GeV}, \hspace{6ex}
v_2 \simeq f_{a_2^\prime} = 10^{10}\ {\rm GeV},
\end{equation}
corresponding to
\begin{equation}
f_A \simeq  3\times 10^9\ {\rm GeV}, \hspace{3ex} m_A\simeq 2\ {\rm meV}, \hspace{3ex}
m_a \simeq 10^{-28}\ {\rm eV}, \hspace{3ex}
\cos\delta \simeq 3\times 10^{-4}, \hspace{3ex}
\sin\delta \simeq 1, \hspace{3ex} \tan\delta \simeq 3\times 10^3 ,
\end{equation}
with couplings of size
\begin{eqnarray}
|g_{A\gamma}| & \simeq & 2\times 10^{-13}\ {\rm GeV}^{-1} \\
|g_{a\gamma}| &\simeq  & 5\times  10^{-16}\ {\rm GeV}^{-1}, \\
|g_{Ae}| &
\simeq & 5\times 10^{-14} \left| 1- \frac{3}{2}\sin^2\beta\right|
,
\\
|g_{ae}| &\simeq & 2\times 10^{-17} \left| 1- \frac{3}{2}\sin^2\beta\right|
.
\end{eqnarray}

In summary, cf. Table \ref{summary_table} and Fig. \ref{ALP_coupling_limits_and_models}, we have now verified in an explicit example what has been anticipated in the general discussion of Sec. \ref{sec:correlations}: in the case that the axion $A$ and the ALP $a$ consist of an appreciable mixture of the original axion-like fields, i.e. as long as $|\tan\delta|$ is of order unity,
the couplings are all determined by $f_A$ and are therefore expected
to be approximately of the same size for for the axion $A$ and the ALP $a$, up to order one factors. This is clearly the case for models A1
and A2 in Table \ref{summary_table}.
A hierarchical difference between the axion and the ALP couplings can possibly only
arrive in the situation where
the axion (and correspondingly also the ALP) originates essentially only from one axion-like field, which requires
$|\tan\delta|\approx 0$ or $|\tan\delta|\gg 1$. In this case, the (photon and electron) couplings of the ALP are suppressed by the
factor $\tan\delta /(1+\tan^2\delta )$ in comparison to the couplings of the axion and thus hierarchically
smaller than the latter.

\subsection{$\ZZ_n\otimes\ZZ_m$ models with a photophilic ALP}

As has been emphasized in Sec.  \ref{sec:correlations}, in order to arrive at the same time at the prediction of a sizeable amount of
axion dark matter and at the explanation of the astrophysical hints by an additional ALP one needs
\begin{itemize}
\item that only one of the original axion-like fields couples to $G\tilde G$, while the other is photophilic, and
\item that there is a hierarchy between the vevs $\sim$ decay constants of the two PQ symmetries.
\end{itemize}
Therefore, in this subsection we build models featuring one accidental KSVZ-type axion and another accidental
photophilic ALP, which does not couple to $G\tilde G$ through a $U(1)_{\rm PQ_2}\times SU(3)_C\times SU(3)_ C$ chiral anomaly, but, nevertheless, couples to $F\tilde F$ due to a  $U(1)_{\rm PQ_2}\times U(1)_{\rm em}\times U(1)_ {\rm em}$ anomaly. The latter will be
introduced as a  photophilic analog of the DFSZ  or KSVZ axion, cf. Sec. \ref{lacion_model}.

\subsubsection{$\ZZ_{11}\otimes\ZZ_9$ model with extended Higgs sector}
\label{accion_laccion_model}

Our first model has two SM Higgs doublets, $H_q$ and $H_l$, where the first one gives Dirac mass terms for
quarks and neutrinos, while the second one gives them for charged leptons, see Eq. \eqref{model5:yuk} below.
Furthermore, it features two SM singlet complex scalar fields $\sigma_1$ and $\sigma_2$, which carry only
\PQ charges. The phase of $\sigma_1$ will carry a KSVZ-like axion, while the one of $\sigma_2$  will carry a DFSZ-like photophilic ALP.
The SM fermionic content is augmented by the addition of a vector-like color triplet, ($Q_L,Q_R$),
as in the KSVZ model, and three right-handed neutrinos, $N_{iR}$, $i=1,2,3$.

\begin{table}[ht]\centering
\vspace{-\baselineskip}
\eq{\nonumber
\begin{array}{|c|c|c|c|c|c|c|c|c|c|c|c|c|}
\hline\rule[0cm]{0cm}{.9em}
& q_L & u_R & d_R & L & l_R & N_R & H_q & H_l & Q_L & Q_R & \sigma_1 & \sigma_2
\\[-.1ex]
\hline\rule[0cm]{0cm}{1em}
\ZZ_9 & 1 & \om[9]^4 & \om[9]^5 & 1 & \om[9]^3 & \om[9]^4 & \om[9]^4 & \om[9]^6 &
    1 & \om[9]^8 & \om[9] & \om[9]
\\[-.1ex]
\hline\rule[0cm]{0cm}{1em}
\ZZ_{11} & 1 & \om[11]^6 & \om[11]^5 & \om[11]^{10} & \om[11]^4 & \om[11]^5 &
\om[11]^6 & \om[11]^6 & 1 & \om[11]^{10} & \om[11] & 1
\\[-.1ex] \hline
\end{array}
}
\caption{$\ZZ_{11}\otimes\ZZ_9$ charges, where $\omega_9\equiv e^{i2\pi/9}$
and $\omega_{11}\equiv e^{i2\pi/{11}}$.}
 \label{tab:irreps}
\end{table}

Given this field content, we impose an exact discrete $\ZZ_{11}\otimes\ZZ_9$ symmetry, with the
fields transforming according to Table \ref{tab:irreps}. Then the most general Yukawa interactions
are given by
\begin{align}
{\lag_Y} &
=Y_{ij}\overline{q}_{iL}\widetilde{H}_{q}u_{jR}
    +\Gamma_{ij}\overline{q}_{iL}H_{q}d_{jR}
    +G_{ij}\overline{L}_{i}H_{l}l_{jR}+k_{ij}\overline{L}_{i}\widetilde{H}_q N_{jR}
\nonumber\\
    &\quad +\ y_{ij}\overline{N_{iR}^{c}}\sigma_{1}N_{jR}
    +y_{Q}\overline{Q}_{L}\sigma_{1}Q_{R}+h.c.
\label{model5:yuk}
\end{align}
The scalar potential $V=V_H + V_{NH}$ is supposed to be given by the sum of all possible renormalizable hermitian terms involving
the scalar fields and respecting the discrete symmetry in Table \ref{tab:irreps}, $V_H (H_q,H_l,\sigma_1,\sigma_2)$, and the
most general renormalizable non-hermitian term (``DFSZ-type" term),
\eq{
\label{laccion:dine}
V_{NH}=\lambda\sigma_{2}^2H_{l}^{\dagger}H_{q}+h.c.\,.
}

\begin{table}[ht]\centering
\vspace{-\baselineskip}
\eq{\nonumber
\begin{array}{|c|c|c|c|c|c|c|c|c|c|c|c|c|}
\hline\rule[0cm]{0cm}{.9em}
& q_L & u_R & d_R & L & l_R & N_R & H_q & H_l & Q_L & Q_R & \sigma_1 & \sigma_2
\\[-.1ex]
\hline\rule[0cm]{0cm}{1em}
K_\psi & 0 & 0 & 0 & -1/2 & -1/2 & -1/2 & 0 & 0 &
 1/2   & -1/2 & 1 & 0
\\[-.1ex]
\hline\rule[0cm]{0cm}{1em}
X_\psi & 0 & -X_q  & X_q & X_q & X_q-X_l & 0 &
-X_q & X_l & 0 & 0 & 0 & 1
\\[-.1ex] \hline
\end{array}
}\caption{Charge assignments of the fields in the
SM extension of Sec. \ref{accion_laccion_model}. $K_\psi$ and $X_\psi$ are the $U(1)_{\rm PQ_1}$ and $U(1)_{\rm PQ_2}$ charges of the fields, respectively.}
\label{tab:pqmodelaccionlaccion}
\end{table}

The Yukawa interactions \eqref{model5:yuk} feature an accidental $U(1)_{\rm PQ_1} \otimes U(1)_{\rm PQ_2}$ symmetry which is
summarized in Table \ref{tab:pqmodelaccionlaccion}.
In fact, one PQ symmetry, $U(1)_{\rm PQ_1}$, may be defined by the following action on the fields:
\eqali{\label{5:kim}
\sigma_{1} & \longrightarrow e^{i\alpha_{1}}\sigma_{1} &
N_{jR} &\longrightarrow e^{-i\alpha_{1}/2}N_{jR}, \cr
Q_{L} & \longrightarrow e^{i\alpha_{1}/2}Q_{L},  &
Q_{R} & \longrightarrow e^{-i\alpha_{1}/2}Q_{R} , \cr
L_{j} & \longrightarrow e^{-i\alpha_{1}/2}L_j, & l_{jR} & \longrightarrow e^{-i\alpha_{1}/2}l_{jR}
&
,
}
while the rest of the fields transform trivially. This symmetry is anomalous with respect to QCD and gives thus
rise to the axion, whose corresponding field  appears in the phase of $\sigma_1$
in the expansion around its vev $v_1$,
\begin{equation}
\sigma_1(x)=\frac{v_1+\rho_1(x)}{\sqrt{2}}\exp\left[i\frac{a_1'(x)}{f_{a_1'}}\right]\,,
\end{equation}
with
\eq{
\label{F111}
f_{a_1^\prime}=v_1\,.
}

The other accidental PQ symmetry $U(1)_{\rm PQ_2}$ acts on the fields as
\begin{align}
\sigma_{2} & \longrightarrow
e^{i\alpha_{2}}\sigma_{2}\,,
\nonumber
\\
H_l & \longrightarrow e^{iX_l\alpha_{2}} H_l,&
H_q & \longrightarrow e^{-iX_q\alpha_{2}}H_q,
\nonumber
\\
\label{6:kim}
L_{jL} & \to e^{iX_q\alpha_2}L_{jL} &
l_{jR} & \longrightarrow e^{i(X_q-X_l)\alpha_{2}}l_{jR},
\\
u_{jR} & \longrightarrow e^{-iX_q\alpha_{2}}u_{jR}, &
d_{jR} & \longrightarrow e^{iX_q\alpha_{2}}d_{jR}.
\nonumber
\end{align}
The PQ charges $X_l$ and $X_q$ are restricted by \eqref{laccion:dine} to obey
\eq{
\label{X:lN}
X_l+X_q=2\,.
}
This symmetry is non-anomalous with respect to QCD, but anomalous with respect to
QED. Therefore, strictly speaking, $U(1)_{\rm PQ_2}$ is a PQ-like symmetry, rather than a proper PQ symmetry. Its breaking will lead to a photophilic axion-like field $a_2^\prime$ appearing in the phase
$\sigma_2$ in the expansion around its vev $v_2$,
\begin{equation}
\sigma_2(x)=\frac{v_2+\rho_2(x)}{\sqrt{2}}\exp\left[i\frac{a_2'(x)}{f_{a_2'}}\right]\,,
\end{equation}
where
\eq{
\label{F1111}
f_{a_2^\prime}=\sqrt{v_2^2+v^2\xi_v^{\prime 2}}\,,\hspace{3ex}  \xi_v^\prime=\frac{2}{x^\prime+x^{\prime -1}},\hspace{3ex} x^\prime\equiv v_q/v_l.
}

The low-energy couplings to the gluon are found to be (cf. Appendix \ref{sec:lecouplings})
\eq{
\label{C:g:hybrid}
C_{1 g}=1\,, \hspace{6ex}
C_{2 g}=0\,.\quad
}
Therefore, $a_1^\prime$ can be identified with the axion field $A(x)$, while
$a_2^\prime$ is a photophilic axion-like field,.

The couplings to the photon, on the other hand, read
\eq{
C_{1\gamma}=6\left( C_{\rm em}^{(Q)}\right)^2\,,
C_{2\gamma}=2N_g(X_l+X_q)=4N_g =12\,,
}
where  $C_{\rm em}^{(Q)}$ is the
the electric charge of
$Q_L,Q_R$.

The coupling with electrons can be also extracted from Appendix  \ref{sec:lecouplings}      as
\eq{
C_{1 e}=0,\quad
C_{2 e}=X_l,\quad
}
so that
\eq{
\frac{C_{2 e}}{C_{2 \gamma}}=\frac{\sin^2\beta'}{2N_g},
}
where here $\tan\beta'=x^\prime=v_q/v_l$.

As in our GUT example from Sec. \ref{sec:unificaccion},   $U(1)_{\rm PQ_1}$  is in fact a lepton number symmetry
and its breaking will thus give rise to a seesaw neutrino mass relation,
\begin{equation}
\label{seesaw}
m_{\nu} = - M_D M_M^{-1} M_D^T = -  F\,y^{-1}\,F^T\ \frac{v_q^2}{v_1}
= 0.01\,{\rm eV} \left( \frac{v_q}{100\,{\rm GeV}}\right)^2 \left( \frac{10^{13}\,{\rm GeV}}{v_1} \right)
\left( \frac{-  F\,y^{-1}\,F^T}{10^{-2}}\right)
\,,
\end{equation}
where $v_q = {\sqrt{2}}\,\aver{H_k}$ is  naturally of order the electroweak scale, since
$v = \sqrt{v_q^2 + v_l^2}=246$ GeV.  For definiteness and in order to get an axion well in the mass range favored after
BICEP2 (cf. Sec. \ref{sec:intro}), we allow for a modest fine-tuning in the Yukawa's and chose the vev $v_1$ to be
\begin{equation}
v_1 =  3\times 10^{11}\ {\rm GeV},
\end{equation}
leading to an axion decay constant, mass, and photon coupling of order (we take again $C^{(Q)}_{\rm em}=1$)
\begin{equation}
f_A = f_{a_1}/|C_{1g}| = v_1 = 3\times 10^{11}\ {\rm GeV}, \hspace{3ex}
m_A = 20\ {\rm \mu eV}, \hspace{3ex}
g_{A\gamma}
= \frac{\alpha}{2\pi f_A}
\left( 6 - 1.95     \right)\simeq 1.6\times 10^{-14} \, {\rm GeV}^{-1}
,
\end{equation}
right in the ballpark to explain and detect dark matter in terms of axions.

The ALP couplings,  on the other hand, can be simultaneously in the range of highest phenomenological interest,
$|g_{a\gamma}|\sim 10^{-12}$ GeV$^{-1}$,  $|g_{ae}|\sim 10^{-13}$,
if we chose the vev $v_2$ of the second PQ symmetry much smaller,
\begin{equation}
v_2 \sim 10^{9}\ {\rm GeV}.
\end{equation}
In fact, then
\begin{eqnarray}
        g_{a\gamma}
&= &\frac{\alpha}{2\pi f_{a_2^\prime}}\, \cdot 12\simeq 1.4\times 10^{-11} \, {\rm GeV}^{-1},
\label{gagamma_nomixing_photophilic_model}
\\
g_{ae} &=&
 \frac{m_e}{f_{a_2^\prime}}\, \sin^2\beta^\prime \simeq  10^{-12}\, \sin^2\beta^\prime
.
\label{gae_nomixing_photophilic_model}
\end{eqnarray}

In this context it is also important to note that the photophilic ALP has a quite sizeable mass, for the chosen
discrete symmetry and PQ symmetry breaking scale. In fact,
the lowest mass dimension operator violating $U(1)_{\rm PQ_2}$, but respecting the discrete symmetry,
occurs already at mass dimension $D=9$,
\begin{equation}
\label{op:laxion:mass}
\lag \supset
\frac{g}{M_{\rm Pl}^{5}} \, (\sigma_{2})^9,
\end{equation}
and therefore induces a mass which does not suffer from too much Planck-suppression (cf. Appendix \ref{sec:effc-op}),
\begin{equation}
\label{ma:IV.B}
m_{a}\simeq
 |g|^{1/2}
\frac{9}{2^{9/4}}
\left(\frac{v_2}{M_{\rm Pl}}\right)^{7/2} M_{\rm Pl} \simeq 1\times 10^{-7}\,{\rm eV}\
 |g|^{1/2}
\left( \frac{v_2}{10^{9}\ \rm GeV} \right)^{7/2} .
\end{equation}

Finally, the axion in this model is protected from destabilization by Planck suppressed operators. In fact, the lowest mass dimension operator violating $U(1)_{\rm PQ_1}$, but respecting the discrete symmetry,
\begin{equation}
\lag \supset
\frac{1}{M_{\rm Pl}^{8}} \, (\sigma_{1})^{11}(\sigma^*_{2}),
\end{equation}
has no destructive effect on the axion properties, as long as $(v_1^{11} v_2)^{1/12}\lesssim 10^{12}$ GeV  (cf. Appendix \ref{sec:effc-op}), which is the case for the favored choice of parameters.

We have summarized the values predictions of masses and couplings of this model for the chosen input values of the
PQ breaking scales in Table  \ref{summary_table}   and Fig.    \ref{ALP_coupling_limits_and_models}, labelled as ``Model B1.1".
We see that they are well in a region to explain the astrophysical puzzles and to be probed in the next generation of experiments.

Interestingly, for another set of initial input values, which we label as ``Model B1.2"  in Table  \ref{summary_table}   and Fig.    \ref{ALP_coupling_limits_and_models}, namely
\begin{equation}
v_1 = 9\times 10^{11}\, {\rm GeV},  \hspace{6ex}
v_2 = 1.26\times 10^{12}\,{\rm GeV},
\end{equation}
the axion may still be the dominant part of dark matter, while the ALP will have a mass and coupling
\begin{equation}
m_a\simeq 7.1\,{\rm keV}\
 |g|^{1/2}
\left( \frac{v_2}{1.26\times 10^{12}\ \rm GeV} \right)^{7/2} \,, \hspace{6ex}
  g_{a\gamma}
\simeq 1.1\times 10^{-14} \, {\rm GeV}^{-1}  \left( \frac{1.26\times 10^{12}\ \rm GeV}{v_2} \right) ,
\end{equation}
cf. Fig. \ref{ALP_coupling_limits_and_models}.
In this case, it may in fact explain the recently reported unidentified 3.55 keV line from galaxies and galaxy clusters
\cite{Bulbul:2014sua,Boyarsky:2014jta}, mentioned in Sec. \ref{sec:intro},
in terms of two photon decay   \cite{Higaki:2014zua,Jaeckel:2014qea,Lee:2014xua,Cicoli:2014bfa},
 if its fractional matter density
$x_a\equiv \rho_a/\rho_{\rm CDM}$ is of order  $x_a\simeq 10^{-8}-10^{-7}$.

\subsubsection{$\ZZ_{11}\otimes\ZZ_9$ model with minimal SM Higgs sector}
\label{singlets_model}

As a further example, we consider a model which features, like the KSVZ model, a minimal SM Higgs sector and where all fields beyond the
 SM are $SU(2)_L$ singlets. These are:  two scalar singlet fields $\sigma_1$, $\sigma_2$, an exotic quark singlet field $Q$, a colorless and electrically charged fermionic field $E$, and three types of right-handed neutrinos, $N_{iR}$. The imposed discrete $\ZZ_{11}\otimes\ZZ_9$ symmetry transformations are given in Table \ref{tab:znms}.

\begin{table}[ht]\centering
\vspace{-\baselineskip}
\eq{\nonumber
\begin{array}{|c|c|c|c|c|c|c|c|c|c|c|c|c|c|}
\hline\rule[0cm]{0cm}{.9em}
  & q_L & H & u_R & d_R & L  & l_R  & N_R  & \sigma_1 & Q_L &Q_R &\sigma_2 & E_L & E_R
\\[-.1ex]
\hline\rule[0cm]{0cm}{1em}
 \ZZ_9 & 1 & \omega_{9}^5 & \omega_{9}^5 & \omega_{9}^4 & \omega_{9}^6 & \omega_{9} & \omega_{9}^2  & \omega_{9}^5 & 1 & \omega_{9}^8 & \omega_{9} & \omega_{9}^3 & 1  \\[-.1ex]
% \hline
\hline\rule[0cm]{0cm}{1em}
 \ZZ_{11} & 1  & \omega_{11}^3 & \omega_{11}^3  & \omega_{11}^8 & \omega_{11}^2 & \omega_{11}^{10}  & \omega_{11}^5  & \omega_{11} & \omega_{11}^9 & \omega_{11}^7 & 1 & 1 & \omega_{11}^{10}  \\[-.1ex] \hline
\end{array}
}\caption{$\ZZ_{11}\times\ZZ_9$ charges for the fields in the model of Sec. \ref{singlets_model}.}
\label{tab:znms}
\end{table}
The $\ZZ_{11}\otimes\ZZ_9$ symmetry allows the following Yukawa interaction terms

\begin{align}
{\lag_Y} &
=Y_{ij}\overline{q}_{iL}\widetilde{H}u_{jR}
    +\Gamma_{ij}\overline{q}_{iL}H d_{jR}
    +G_{ij}\overline{L}_{i}H l_{jR}
    +k_{ij}\overline{L}_{i}\widetilde{H} N_{jR}
    +\ y_{ij}\sigma_{1}\overline{N_{iR}^{c}}N_{jR}
\nonumber\\
    &\quad +y_{i}\frac{\sigma_{1}}{M_{Pl}}\overline{q}_{iL}H Q_{R}+y_{Q}\frac{\sigma_{1}^2}{M_{Pl}}\overline{Q}_{L}Q_{R}
    +k_{i}\frac{\sigma_{2}}{M_{Pl}}\overline{L}_{i}H E_{R}+k_{E}\frac{\sigma_{1}\sigma_{2}}{M_{Pl}}\overline{E}_{L}E_{R}+h.c.
\label{modelsing:yuk}
\end{align}
In the scalar potential all interaction terms are hermitian. Eq. (\ref{modelsing:yuk}) has an automatic $U(1)_{\rm PQ_1}\otimes U(1)_{\rm PQ_2}$ symmetry in which the  fields  carry charges as shown in Table \ref{tab:pqmodelsing}. Also, there is a seesaw mechanism for the neutrinos working like in the previous models.

\begin{table}[ht]\centering
\vspace{-\baselineskip}
\eq{\nonumber
\begin{array}{|c|c|c|c|c|c|c|c|c|c|c|c|c|c|}
\hline\rule[0cm]{0cm}{.9em}
  & q_L & H & u_R & d_R & L  & l_R  & N_R  & \sigma_1 & Q_L & Q_R &\sigma_2 & E_L & E_R
\\[-.1ex]
\hline\rule[0cm]{0cm}{1em}
 K_\psi & 0 & 0 & 0 & 0 & -1/2 & -1/2 & -1/2 & 1  & 1 & -1 & 0 & 1/2 & -1/2  \\[-.1ex]
% \hline
\hline\rule[0cm]{0cm}{1em}
 X_\psi & 0  & 0 & 0  & 0 & 0 & 0  & 0  & 0  & 0 & 0 & 1 & 0 & -1  \\[-.1ex] \hline
\end{array}
}\caption{Charge assignments of the fields in the
SM extension of Sec. \ref{singlets_model}. $K_\psi$ and $X_\psi$ are the $U(1)_{\rm PQ_1}$ and $U(1)_{\rm PQ_2}$ charges of the fields, respectively.}
\label{tab:pqmodelsing}
\end{table}

The axion-like fields in this case appear only in the singlets fields  parametrization of Eq. (\ref{sigma:}), such that
\begin{equation}
f_{a'_1}=v_1, \hspace{6ex} f_{a'_2}=v_2.
\end{equation}

The low-energy couplings to the gluon they are found to be (cf. Appendix \ref{sec:lecouplings})
\eq{
\label{C:g:hybridc}
C_{1 g}=2\,, \hspace{6ex}
C_{2 g}=0\,,\quad
}
rendering $a_2^\prime$ a photophilic axion-like field, while $a_1^\prime$ can be identified with the axion field $A(x)$. The couplings to the photon, on the other hand, read
\eq{
C_{1\gamma}=12\left( C_{\rm em}^{(Q)}\right)^2+2\left( C_{\rm em}^{(E)}\right)^2=\frac{10}{3}\,,\,\,\,\,\,
C_{2\gamma}=2\left( C_{\rm em}^{(E)}\right)^2=2\,,
}
where  $C_{\rm em}^{(Q)}=-1/3$ and $C_{\rm em}^{(E)}=-1$ are the electric charge of $Q$ and $E$, respectively.

The couplings to the electron can be also extracted from Appendix  \ref{sec:lecouplings}      as
\eq{
C_{1 e}=0,\quad
C_{2 e}=0,\quad
}
such that
\eq{
\frac{C_{1 e}}{C_{1 \gamma}}=0.
}

Again, the active neutrino masses favor $v_1\gtrsim 10^{12\div 14}$ GeV, while the explanation of the astrophysical anomalies favor
$v_2 \sim 10^{8\div 10}$ GeV. As in the previous model, the lowest mass dimension operator violating $U(1)_{\rm PQ_2}$, but respecting the discrete symmetry, is of the form \eqref{op:laxion:mass},
inducing an ALP mass of size
\begin{equation}
m_{a}\simeq
1\times  10^{-7}\,{\rm eV}\
 |g|^{1/2}
\left( \frac{v_2}{10^{9}\ \rm GeV} \right)^{7/2} .
\end{equation}
Moreover, as in the previous model, the axion is protected from destabilization by Planck suppressed operators (cf. Appendix \ref{sec:effc-op}).

The masses of  $E$ and $Q$ are given by
\begin{eqnarray}
\label{Emass}
& & m_{E}=k_E\frac{v_{1}v_{2}}{2M_{\rm Pl}}\approx k_E 10^3\,\rm GeV\,,\\ \nonumber
& & m_{Q}=y_Q\frac{v_{1}^2}{2M_{\rm Pl}}\approx y_Q 10^7\,\rm GeV\,.
\end{eqnarray}
We see in this scenario  $E$ could naturally have a mass around the TeV scale or below. By means of a mixing  proportional to
$\,v_2/M_{\rm Pl}\approx 10^{-10}$ we may have a decay of $E$ into  SM charged leptons, $l_i$, plus the Higgs boson, $h$, according to $E\rightarrow l_i+h$. This could leave a signal in  Drell-Yan $E$ pair production at a hadron collider such as LHC. For $Q$, its mass would be out of the present direct searches unless there is a fine tuning of order $y_Q\approx 10^{-4}$. Assuming the latter, a QCD pair production process of $Q$, followed by decays into SM $d$-type quarks plus a Higgs boson, $Q\rightarrow d_i+h$, is possible at the LHC.

\subsubsection{$\ZZ_{11}\otimes\ZZ_7$ model with minimal SM Higgs sector}
\label{another_singlets_model}

Motivated by the possibility to explain the 3.55 keV line by decaying ALP dark matter, we introduce yet another model with a somewhat smaller discrete symmetry to get a keV scale ALP mass with a PQ symmetry breaking scale of order $10^9$ GeV, instead of the $\sim 10^{12}$ GeV
required in the model of Sec. \ref{accion_laccion_model}. It is based on the fact that, with the same field content as in the model from Sec. \ref{singlets_model}, we can also realize a $\ZZ_{11}\otimes\ZZ_7$ symmetry as given in Table \ref{tab:znms2}.

\begin{table}[ht]\centering
\vspace{-\baselineskip}
\eq{\nonumber
\begin{array}{|c|c|c|c|c|c|c|c|c|c|c|c|c|c|}
\hline\rule[0cm]{0cm}{.9em}
  & q_L & H & u_R & d_R & L  & l_R  & N_R  & \sigma_1 & Q_L &Q_R &\sigma_2 & E_L & E_R
\\[-.1ex]
\hline\rule[0cm]{0cm}{1em}
 \ZZ_7 & 1 & \omega_{7}^4 & \omega_{7}^4 & \omega_{7}^3 & \omega_{7}^6 & \omega_{7}^2 & \omega_{7}^3  & \omega_{7} & \omega_{7}^4 & \omega_{7}^2 & \omega_{7}^2 & \omega_{7}^4 & 1  \\[-.1ex]
% \hline
\hline\rule[0cm]{0cm}{1em}
 \ZZ_{11} & 1  & \omega_{11}^3 & \omega_{11}^3  & \omega_{11}^8 & \omega_{11}^2 & \omega_{11}^{10}  & \omega_{11}^5  & \omega_{11} & \omega_{11}^9 & \omega_{11}^7 & 1 & 1 & \omega_{11}^{10}  \\[-.1ex] \hline
\end{array}
}\caption{$\ZZ_{11}\otimes\ZZ_7$ charges for the fields in the model of Sec. \ref{another_singlets_model}.}
\label{tab:znms2}
\end{table}

In this case the lowest dimensional operator breaking  $U(1)_{\rm PQ_2}$ and involving just $\sigma_{2}$ is
\begin{equation}
\lag \supset\frac{g^\prime}{M_{\rm Pl}^{3}}  (\sigma_{2})^7,
\end{equation}
and this induces the following mass for the ALP (cf. Appendix \ref{sec:effc-op}),
\begin{equation}
\label{ma:IV.C2}
m_{a}\simeq
 |g^\prime|^{1/2}
\frac{7}{2^{7/4}}
\left(\frac{v_2}{M_{\rm Pl}}\right)^{5/2} M_{\rm Pl} \sim 7.1\  {\rm keV}\ |g^\prime|^{1/2}
\left( \frac{v_2}{1.84\times 10^9\ \rm GeV} \right)^{5/2} .
\end{equation}
Therefore, an explanation for the 3.55 keV line from galaxies can be  found with this setup of fields and the $\ZZ_{11}\otimes\ZZ_7$ symmetry,
cf. Table  \ref{summary_table}   and Fig.    \ref{ALP_coupling_limits_and_models}, labelled ``Model B3".

%         {v1   ,     v2}                fA    {mA     ,      gA}
%       {ma      ,      ga}
%B1.1  {5.*10^13, 1.*10^9} 5*10^13 {{1.2*10^-7,9.4*10^-17},
%{1.*10^-7,1.4*10^-11}}
%B1.2  {2.*10^13, 1.26*10^12} 2.*10^13{{3.*10^-7,2.35*10^-16},
%{7100.,1.1*10^-14}}
%B3    {0.8*10^13, 1.8*10^9} 4.*10^12{{1.5*10^-6,-4.7*10^-16}, {6706,1.3*10^-12}}

The lowest dimensional operator involving $\sigma_1$ and breaking $U(1)_{\rm PQ_2}$ is the following one
\begin{equation}
\lag \supset
\frac{1}{M_{\rm Pl}^{9}} (\sigma_{1})^{11}(\sigma^*_{2})^2,
\end{equation}
Therefore, there are no dangerous Planck suppressed operators destabilizing the axion for $v_1=10^{13}$ GeV. Also, the mass generation mechanisms for neutrinos, $Q$ and $E$ keep the same as before.

\subsection{$\ZZ_{11}{\otimes} \ZZ_{9}{\otimes} \ZZ_{7}$ model with two photophilic ALPs}
\label{yet_another_singlets_model}

Finally, we construct a model with two photophilic ALPs: one very light, to explain the astrophysical hints
such as the cosmic gamma ray transparency, and one with a mass of 7.1 keV, to  explain the recently found 3.55 keV line from
Andromeda and galaxy clusters by its decay in two photons.
Clearly, such a  model will involve three complex scalar singlet fields $\sigma_i$, $i=1,2,3$. Also, besides an exotic quark singlet field $Q$, two colorless and electrically charged fermionic fields, $E$, $E^\prime$, and three right-handed neutrinos, $N_{iR}$, are introduced.

\begin{table}[ht]\centering
\vspace{-\baselineskip}
\eq{\nonumber
\begin{array}{|c|c|c|c|c|c|c|c|c|c|c|c|c|c|c|c|c|}
\hline\rule[0cm]{0cm}{.9em}
& q_L & u_R & d_R & L & l_R & N_R & H &  Q_L & Q_R & \sigma_1 & \sigma_2& E_L & E_R & \sigma_3 & E^\prime_L & E^\prime_R
\\[-.1ex]
\hline\rule[0cm]{0cm}{1em}
\ZZ_7 & 1 & \om[7]^3 & \om[7]^4 & 1 & \om[7]^4 & \om[7]^3 & \om[7]^3 & \om[7]^5 &
    \om[7]^3 & \om[7] & 1 & \om[7]^5  & \om[7]^4 & \om[7]^1 & \om[7]^5 & \om[7]^3
\\[-.1ex]
\hline\rule[0cm]{0cm}{1em}
\ZZ_9 & 1 & \om[9]^5 & \om[9]^4 & \om[9]^6 & \om[9] & \om[9]^2 & \om[9]^5 & 1 &
    \om[9]^8 & \om[9]^5 & \om[9] & \om[9]^6  & 1 & 1 & \om[9]^1  & \om[9]^5
\\[-.1ex]
\hline\rule[0cm]{0cm}{1em}
\ZZ_{11} & 1 & \om[11]^3 & \om[11]^8 & \om[11]^{2} & \om[11]^{10} & \om[11]^5 &
\om[11]^3 & \om[11]^9 & \om[11]^7 & \om[11] & 1 &1 & \om[11]^{10} & 1 &  \om[11]^{10}  & \om[11]^{9}
\\[-.1ex] \hline
\end{array}
}
\caption{$\ZZ_7\times\ZZ_9\times\ZZ_{11}$ charges for the fields in the model of Sec. \ref{yet_another_singlets_model}.}
 \label{tab:irreps2}
\end{table}

The field content of our model is specified in Table \ref{tab:irreps2}, which displays also the field transformations which
we impose to ensure an exact discrete $\ZZ_{11}{\otimes} \ZZ_{9}{\otimes} \ZZ_{7}$ symmetry.
Based on this, the most general  interactions involving scalars and fermionic fields
are given by
\begin{align}
{\lag_Y} &
=Y_{ij}\overline{q}_{iL}\widetilde{H} u_{jR}
    +\Gamma_{ij}\overline{q}_{iL}H d_{jR}
    +G_{ij}\overline{L}_{i}H l_{jR}+k_{ij}\overline{L}_{i}\widetilde{H} N_{jR}
 +\ y_{ij}\overline{N_{iR}^{c}}\sigma_{1}N_{jR}
 \nonumber\\
    &\quad +y_{i}\frac{\sigma_{1}}{M_{\rm Pl}}\overline{q}_{iL}H Q_{R}
    +y_{Q}\frac{\sigma_{1}^2}{M_{\rm Pl}} \overline{Q}_{L} Q_{R}+k_{i}\frac{\sigma_{2}}{M_{\rm Pl}}\overline{L}_{i}H E_{R}
      + k_{12}\frac{\sigma_{1}\sigma_{2}}{M_{\rm Pl}}\overline{E}_{L}E_{R}  \nonumber\\
    &\quad +k_{13}\frac{\sigma_{1}\sigma_{3}}{M_{\rm Pl}} \overline{E^\prime}_{L}E^\prime_{R}  + k_{23}\frac{\sigma_{2}\sigma_{3}}{M_{\rm Pl}}\overline{E^\prime}_{L}E_{R} +k_{11}\frac{\sigma_{1}^2}{M_{\rm Pl}}\overline{E}_{L}E^\prime_{R}  + h.c.
\label{model7:yuk}
\end{align}
This interaction Lagrangian features an accidental $U(1)_{\rm PQ_1} \otimes U(1)_{\rm PQ_2} \otimes U(1)_{\rm PQ_3}$ symmetry shown in Table \ref{tab:pq7911}.

\begin{table}[ht]\centering
\vspace{-\baselineskip}
\eq{\nonumber
\begin{array}{|c|c|c|c|c|c|c|c|c|c|c|c|c|c|c|c|c|}
\hline\rule[0cm]{0cm}{.9em}
& q_L & u_R & d_R & L & l_R & N_R & H &  Q_L & Q_R & \sigma_1 & \sigma_2& E_L & E_R & \sigma_3 & E^\prime_L & E^\prime_R
\\[-.1ex]
\hline\rule[0cm]{0cm}{1em}
K_{\psi} & 0 & 0 & 0 & -1/2 & -1/2 & -1/2 & 0 & 1 & -1 & 1 & 0 & 1  & 0 & 0 & 0 & -1
\\[-.1ex]
\hline\rule[0cm]{0cm}{1em}
X_{\psi} & 0 & 0 & 0 & 0 & 0 & 0 & 0 & 0 & 0 & 0 & 1 & 0  & -1 & 0 & 0 & 0
\\[-.1ex]
\hline\rule[0cm]{0cm}{1em}
Z_{\psi} & 0 & 0 & 0 & 0 & 0 & 0 & 0 & 0 & 0 & 0 & 0 & 0 & 0 & 1 & 1 & 0
\\[-.1ex] \hline
\end{array}
}
\caption{Charge assignments of the fields in the SM extention of Sec. \ref{yet_another_singlets_model}. $K_{\psi}$, $X_{\psi}$, and $Z_{\psi}$ are the $U(1)_{\rm PQ_1}$, $U(1)_{\rm PQ_2}$, and $U(1)_{\rm PQ_3}$ charges of the fields, respectively.}
 \label{tab:pq7911}
\end{table}

The lowest mass dimension operator violating $U(1)_{\rm PQ_2}$, but respecting the discrete symmetry, occurs already at mass dimension $D=9$,
\begin{equation}
\lag \supset
\frac{g}{M_{\rm Pl}^{5}} \, (\sigma_{2})^9.
\end{equation}
Thus, the mass of the respective ALP $a_2$ will be the same as in Eq. \eqref{ma:IV.B},
\begin{equation}
m_{a_2}\simeq
 1\times 10^{-7}\,{\rm eV}\
 |g|^{1/2}
\left( \frac{v_2}{10^{9}\ \rm GeV} \right)^{7/2} .
\end{equation}
The lowest mass dimension operator violating $U(1)_{\rm PQ_3}$, but respecting the discrete symmetry, occurs even at lower mass dimension, $D=7$,
\begin{equation}
\lag \supset
\frac{g^\prime}{M_{\rm Pl}^{3}} \, (\sigma_{3})^7,
\end{equation}
leading to a mass (cf. Eq.  \eqref{ma:IV.C2})
\begin{equation}
m_{a_3}\simeq
7.1\  {\rm keV}\ |g^\prime|^{1/2}
\left( \frac{v_3}{1.84\times 10^9\ \rm GeV} \right)^{5/2} .
\end{equation}
The mass of  $Q$  is the same as in Eq. \eqref{Emass}, and the mass matrix involving $E^\prime$ and $E$ on the basis $\{E^\prime_{L},\,E_{L}\}$ is
\begin{equation}
M_{E^\prime E}^2=\frac{v_1}{2M_{\rm Pl}}\left(
\begin{array}{cc}
 k_{13}v_3 & k_{23}\frac{v_2 v_3}{v_1}  \\
  &   \\
 k_{11}v_1 & k_{12}v_2   \\
\end{array}
\right)\,,
\end{equation}
whose eigenvalues are
\begin{eqnarray}
M_{\pm} =
\frac{v_1 (k_{12} v_2 + k_{13} v_3 \pm \Delta M)}{4 M_{\rm Pl}}
\,,
\label{me12}
\end{eqnarray}
where
$$ \Delta M = \sqrt{(k_{12} v_2 -k_{13} v_3)^2  + 4 k_{11} k_{23} v_2 v_3 }\,.$$
These masses can be naturally in the $10^2$ -- $10^3$ GeV scale without severe fine tuning. As an example, taking $k_{11}=0.5$, $k_{12}=0.8$, $k_{13}=0.8$, $k_{23}=0.5$, and $v_1=10^{13}$ GeV, $v_2=10^{9}$ GeV, $v_3=1.8\times 10^{9}$ GeV, we obtain $M_{+}\approx 932$ GeV and $M_{-}\approx 188$ GeV. We see that, as in the model of Sec. \ref{singlets_model}, the exotic fermions in this model also have masses eventually in reach of colliders.

The axion in this model is protected from destabilization by Planck suppressed operators. In fact, the lowest mass dimension operator violating $U(1)_{\rm PQ_1}$, but respecting the discrete symmetry,
\begin{equation}
\lag \supset
\frac{1}{M_{\rm Pl}^{11}} \, (\sigma_{1})^{11}(\sigma^{*}_{2})(\sigma_{3})^{3},
\end{equation}
has no destructive effect on the axion properties, as long as $(v_1^{11} v_2 v_3^3)^{1/15}\simeq 10^{12}$ GeV  (cf. Appendix \ref{sec:effc-op}), which is
marginally the case for the favored choice of parameters.

The low-energy couplings to the gluon they are found to be (cf. Appendix \ref{sec:lecouplings})
\eq{
\label{C:g:hybridc}
C_{1 g}=2\,, \hspace{6ex}
C_{2 g}=0\,,\hspace{6ex}
C_{3 g}=0 \,.
}
The couplings to the photon, on the other hand, read
\eq{
C_{1\gamma}=12\left( C_{\rm em}^{(Q)}\right)^2+2\left( C_{\rm em}^{(E)}\right)^2+2\left( C_{\rm  em}^{(E^\prime)}\right)^2=\frac{16}{3}\,,\,\,\,\,\,
C_{2\gamma}=2\left( C_{\rm em}^{(E)}\right)^2=2\,,\,\,\,\,\,
C_{3\gamma}=2\left( C_{\rm em}^{(E^\prime)}\right)^2=2\,,
}
where  $C_{\rm em}^{(Q)}=-1/3$, $C_{\rm em}^{(E)}=-1$, and $C_{\rm em}^{(E^\prime)}=-1$ are the electric charge of $Q$, $E$ and $E^\prime$, respectively.

The couplings to the electron can be also extracted from Appendix  \ref{sec:lecouplings}      as
\eq{
C_{1 e}=0,\quad
C_{2 e}=0,\quad
C_{3 e}=0.
}

For
\begin{equation}
\label{benchmark_C}
v_1 = 9 \times 10^{11}\ {\rm GeV}\,,\hspace{3ex} v_2 = 4 \times 10^{9}\ {\rm GeV}\,,\hspace{3ex} v_3 = 3 \times 10^{9}\ {\rm GeV}
\,,
\end{equation}
the axion $A=a^\prime_1$ has a mass in the regime where it could be the dominant part of dark matter, and the
two photophilic ALPs, $a_2$ and $a_3$ have masses and couplings in the regime of interest of the astrophysical hints
(gamma ray transparency; X-ray line), see Table  \ref{summary_table_C}.  In fact, as can be seen
in Fig. \ref{ALP_coupling_limits_and_models}, it is easy to accommodate all the hints
in our model.

\begin{table}
\begin{center}
\begin{tabular}{|c|c|c|c|c|c|c|c|c|c|}\hline
 $f_A$ [GeV] & $m_A$ [eV] & $m_{a_2}$ [eV]& $m_{a_3}$ [eV]&  $|g_{A\gamma}|$ [GeV]$^{-1}$ & $|g_{a_2 \gamma}|$ [GeV]$^{-1}$ &  $|g_{a_3 \gamma}|$ [GeV]$^{-1}$ & $|g_{A e}|$ & $|g_{a_2 e}|$ & $|g_{a_3 e}|$ \\ \hline
 4.5 $\times10^{11}$ & 1.3 $\times 10^{-5}$ & 4.0 $\times10^{-9}$&
7.1 $\times 10^3$ &1.9$\times 10^{-15}$ & 5.8$\times10^{-12}$& $7.7\times10^{-13}$  & 0 & $0$ & $0$ \\ \hline
\end{tabular}
\caption{\label{summary_table_C}Summary of the low-energy parameters obtained from the model in Sec. \ref{yet_another_singlets_model}
for our chosen benchmark values \eqref{benchmark_C} for the PQ vevs $v_i$,  $i=1,2,3$. For definiteness, we used
$g=1.0$ and $g^\prime =0.28$.}
\end{center}
\end{table}

%%%%%%%%%%%%%%%%%
\section{Summary  and Outlook}
\label{sec:conclusions}

Motivated by theory, cosmology, astrophysics, and upcoming terrestrial experiments, we have undertaken a general
analysis of models featuring the axion $A$ plus a further ALP $a$. In these models, the former may constitute dark
matter, while the latter may explain simultaneously the anomalous cosmic $\gamma$-ray transparency, the
anomalous cooling of white dwarfs and red giants, and the soft X-ray excess from the Coma cluster
(in case that there is a cosmic ALP background radiation peaked in the soft-keV regime). To this end, the ALP-photon coupling
is required to be of order $|g_{a\gamma}|\gtrsim 10^{-12}$ GeV$^{-1}$, for sufficiently small
ALP mass, cf. Sec. \ref{sec:intro} and Fig. \ref{ALP_coupling_limits}. Fortunately,  the next generation
of light-shining-through-a-wall experiments (ALPS-II) and helioscopes (IAXO) will be sensitive in this
range of ALP parameter space. Intriguingly, a further ALP, decaying into two photons with a  similar coupling strength, but with a much larger mass of 7.1 keV, may explain the recently reported 3.55 keV line from Andromeda and galaxy clusters if it
constitutes a tiny part of cold dark matter, cf. Sec. \ref{sec:intro} and Fig. \ref{ALP_coupling_limits}.

In models with several axion-like fields, $a_i^\prime$, $i=1,..,n_{\rm ax}$, coupling to the topological charge density of QCD,
\begin{equation}
{\lag}  \supset  -
 \left(\sum_{i=1}^{n_{\rm ax}} C_{ig} \frac{a_i^\prime}{f_{a_i^\prime}}\right)
\frac{\alpha_s}{8\pi} \,G \tilde{G},
\end{equation}
%the canonically normalized fields corresponding to the proper QCD axion, $A$, and to the further ALP, $a$, are orthogonal mixtures,
the field $A$ corresponding to the proper QCD axion is a mixture, cf. Sec. \ref{sec:correlations},
\begin{equation}
\frac{A}{f_A} =  \left(\sum_{i=1}^{n_{\rm ax}} C_{ig} \frac{a_i^\prime}{f_{a_i^\prime}}\right)   \,    , \hspace{3ex} {\rm with\ }\
\frac{1}{f_A^2}  \equiv \sum_{i=1}^{n_{\rm ax}}\left( \frac{C_{ig}}{f_{a_i^\prime}}\right)^2
\,.
\end{equation}
In general, if the axion-like field associated to the lowest PQ scale
couples to all gluons, photons and electrons, the ALP couplings to
photons and electrons can not be hierarchically larger than the respective axion couplings (although they can be hierarchically smaller).
In fact, the couplings satisfy the constraints,
\begin{eqnarray}
 \left(g_{A\gamma}+\frac{\alpha}{2\pi f_A}\frac{2}{3}\frac{4+z}{1+z}\right)^2+
 \sum_{i=2}^{n_{\rm ax}}  \left(g_{a_i\gamma}\right)^2 &=&
\left(  \frac{\alpha}{2\pi f_\gamma}\right)^2\,,  \hspace{3ex} {\rm with\ }\
 \frac{1}{f_\gamma^2}  \equiv  \sum_{i=1}^{n_{\rm ax}}   \left( \frac{C_{i\gamma}}{f_{a_i^\prime}}\right)^2\,, \\
\left(g_{Ae}\right)^2+
 \sum_{i=2}^{n_{\rm ax}}  \left(g_{a_ie}\right)^2 &=&  \left( \frac{m_e}{f_e}\right)^2 \,, \hspace{3ex} {\rm with\ }\
\frac{1}{f_e^2}  \equiv  \sum_{i=1}^{n_{\rm ax}}  \left( \frac{C_{ie}}{f_{a_i^\prime}}\right)^2 \,,
\end{eqnarray}
which generalize Eqs. \eqref{def:circle} and \eqref{def:circle:ge}  in Sec.  \ref{sec:correlations} to more
than two axion-like fields.
Barring accidental cancellations and considering order one coefficients, we have typically $f_A\sim f_\gamma\sim f_e$,
$|g_{a_i\gamma}|\lesssim |g_{A\gamma}|\sim \alpha/(2\pi f_A)$, and $|g_{a_ie}|\lesssim |g_{Ae}|\sim m_e/f_A$.

This has important consequences for phenomenology. In particular, in models featuring mixing,
the discovery of an ALP in the next generation of laboratory experiments would imply
$f_A\sim \alpha/(2\pi |g_{a\gamma}|)\sim 10^9\ {\rm GeV}$ and thus an axion mass in the
$m_A\sim {\rm meV}$ range --- unfavorably large for axion dark matter, cf. Fig. \ref{ALP_coupling_limits}. Conversely, the detection of axion dark matter with a mass in
the favored ($0.1$ -- $100$)\,$\mu$eV range would imply, in these models,  a tiny  ALP-photon coupling,
making the latter unaccessible to the next generation laboratory experiments.  Micro-eV mass axion dark matter and a large  ALP-photon coupling, $|g_{a\gamma}|\gtrsim 10^{-12}$ GeV$^{-1}$,
can occur simultaneously only in models where the
axion-like field dominantly coupled to photons (lowest scale)
does not couple to gluons, i.e., it is photophilic (e.g. $C_{1g}\neq 0$, $C_{2g}=0$, $C_{2\gamma}\neq 0$).

We have reviewed in Sec. \ref{axions_ad_hoc} plausible ultraviolet completions of
the Standard Model in which axions and more general  ALPs occur as Nambu-Goldstone
bosons from the breaking of  $U(1)$ Peccei-Quinn (PQ) symmetries.  Parameters
occurring in the low-energy effective Lagrangian (decay constants $f_{a^\prime}$ and
dimensionless couplings $C_{i j}$)  were then determined in terms of the fundamental
parameters of the underlying high-scale theories (PQ symmetry breaking scales
$v_{{\rm PQ}}$ and  PQ charges). We have emphasized
two especially well motivated classes of models:
{\em i)} Models, in which the PQ symmetry coincides with the lepton number symmetry and thus the PQ breaking scale with the seesaw scale, favoring therefore a decay constant of order
$f_{a^\prime}\sim v_{\rm PQ}\sim 10^{11\div 15}\ {\rm GeV}$, to explain the active neutrino masses,
$m_\nu \sim v^2/v_{\rm PQ}\sim 0.01\ {\rm eV}$ (the spread arises from our ignorance of the relevant Yukawa couplings).
Intriguingly,  such a simple, minimalistic extension of the SM can explain at the same time and with the same PQ breaking scale also the
nature of dark matter (axions), the  baryon asymmetry of the universe (through leptogenesis) and the stability of the electroweak vacuum, as has been outlined in Sec.  \ref{axion_majoron}.
{\em ii)} Models, which feature a photophilic axion-like field which then possibly has a much larger coupling to photons than the axion.

The high-scale models introduced in Sec. \ref{axions_ad_hoc} can be combined to construct low-energy theories
featuring simultaneously an axion and one or more further  ALPs. However, such ad-hoc models in which the new PQ symmetries are
introduced and imposed
by hand, have the drawback that these symmetries are not protected from explicit symmetry breaking by Planck-scale suppressed operators, which possibly spoil the solution of the strong CP problem. Therefore, in Sec.   \ref{accions_bottom_up},
we have considered several classes of models
in which the Peccei-Quinn symmetries are not ad-hoc,
but instead accidental symmetries originating from large exact discrete symmetries: $\ZZ_{13}\otimes \ZZ_5\otimes\ZZ_{5}^\prime$,
$\ZZ_{11}\otimes\ZZ_{9}$, $\ZZ_{11}\otimes\ZZ_{7}$, and $\ZZ_{11}\otimes \ZZ_9\otimes\ZZ_{7}$, respectively.
In the first three classes, the SM field content was enlarged by two hidden complex scalars,
by extra Higgs doublets, and an exotic colored fermion; in the fourth class there was an additional third hidden complex scalar introduced.
In the considered models, both the right amount of axion dark matter, viable
neutrino masses, and the baryon asymmetry of the universe can be obtained by
properly choosing the vev $v_{{\rm PQ}_1}\simeq f_{a^\prime_1}\sim 10^{11\div
13}\,{\rm GeV}$
of one of the hidden scalars. Only in the first, $\ZZ_{13}\otimes \ZZ_5\otimes\ZZ_{5}^\prime$,  model the two axion-like fields
turn out to mix (both $C_{1g}$ and $C_{2g}$ are non-vanishing), while in the other model classes, the remaining axion-like fields are photophilic ($C_{i g}= 0$,
$C_{i\gamma}\neq 0$, for $i=2,3$). Remarkably, this model predicts the unification of the SM gauge couplings at
$M_{\rm U}\approx 2.8\times 10^{13}\, {\rm GeV}$, while proton decay is very heavily suppressed due to the large discrete symmetry.
For all classes of models we have given benchmark parameter values for the vevs of the hidden complex scalars and
determined the corresponding low-energy parameters, exploiting the general expressions obtained in Sec. \ref{sec:correlations}.
In particular, we have determined the ALP mass(es) induced by the explicit symmetry
breaking operators. In the $\ZZ_{13}\otimes \ZZ_5\otimes\ZZ_{5}^\prime$ model,
it is very small because it arises from a term in the Lagrangian suffering a suppression
by a rather high power of the Planck mass,
\begin{equation}
m_a\sim \frac{v\  v_1^{5/2} v_2^{7/2}}{2^6 M_{\rm Pl}^{5} f_A} \sim 10^{-21}\ {\rm eV}\,, \hspace{3ex}\ {\rm for\ }\
v_1\sim v_2\sim f_A\sim 10^{13}\ {\rm GeV}\,,
\end{equation}
while, in the $\ZZ_{11}\otimes\ZZ_{9}$ and $\ZZ_{11}\otimes\ZZ_{7}$ models, it is much larger,
\begin{equation}
\label{mass2}
m_{a}\sim
\frac{9}{2^{9/4}}
\left(\frac{v_2}{M_{\rm Pl}}\right)^{7/2} M_{\rm Pl} \sim 10^{-7}\,{\rm eV} \,, \hspace{3ex}\ {\rm for\ }\
v_2\sim 10^{9}\ {\rm GeV}\,,
\end{equation}
 and
\begin{equation}
\label{mass3}
m_{a}\sim
\frac{7}{2^{7/4}}
\left(\frac{v_2}{M_{\rm Pl}}\right)^{5/2} M_{\rm Pl} \sim \  {\rm keV}
\,, \hspace{3ex}\ {\rm for\ }\
v_2\sim 10^{9}\ {\rm GeV}\,,
\end{equation}
respectively, even for smaller PQ vevs. The $\ZZ_{11}\otimes \ZZ_9\otimes\ZZ_{7}$ model has two ALPs: an ultralight ALP whose mass
is given by \eqref{mass2} and which may explain e.g. the cosmic $\gamma$-transparency, and a much heavier ALP with mass
according to Eq. \eqref{mass3} which may explain the 3.55 keV line.
These results have been summarized in Tables \ref{summary_table} and \ref{summary_table_C} and in
Fig. \ref{ALP_coupling_limits_and_models}.
For the chosen benchmark values of the PQ breaking scales, these models indeed predict an axion and ALP(s)
in the astrophysically, cosmologically, and
experimentally
favored parameter regions.

We leave the study of possible tests of this class of models by looking for signatures of the extra --  from the assumed field content beyond the SM -- particles at the LHC for future work.
A detailed study of the early cosmology of these models, in particular for the case where the reheating temperature of inflation is above the PQ phase transition, was beyond the scope of the present investigation, but is certainly
of high interest.  In this context, it is important to note that the models considered in Sec.  \ref{accions_bottom_up} do not suffer from the cosmological domain wall problem, as is shown in Appendix \ref{sec:domain-walls}.
This is crucial in view of the fact that the recent discovery of the tensor modes in the $B$-mode power spectrum by BICEP2
strongly disfavor scenarios where the PQ phase transition occurs before inflation, depriving the universe from the possibility to wipe out
topological defects created at the phase transition, such as  domain walls,  by inflation.

There are still ad-hoc features in these scenarios, however. In particular, the extension of the SM field content and the choice of the
discrete symmetry group are mostly motivated from low-energy phenomenology.
Intriguingly, it appears that large discrete symmetries able to give rise
to multiple accidental Peccei-Quinn symmetries are a generic feature of top-down motivated heterotic string scenarios.
In fact, orbifold compactifications of the heterotic string are known to predict a plenitude of hidden complex scalars and
vector-like exotic particles and a multitude of discrete symmetries ($R$ symmetries from the broken $SO(6)$ symmetry of the
compactified space and stringy symmetries from the joining and splitting of strings) which are exact at the perturbative level.
We plan to scan the mini-landscape of certain heterotic orbifolds for models with accidental axions and further ALPs along the
lines pioneered in Ref. \cite{Choi:2006qj,Choi:2009jt}.
Here, the phenomenologically most pressing question is whether there exist models where the vevs of the accidental
PQ symmetries are naturally in the range of the intermediate scale, $\sim 10^{10\div 13}$ GeV, rather than in the range of the heterotic string scale, $M_s\sim M_{\rm Pl}$.

%%%%%%%%%%%%%%%%%
\acknowledgments

The work of A.G.D and C.C.N are supported  by the grant 2013/22079-8, S\~ao Paulo Research Foundation (FAPESP), and by the  grants 303094/2013-3 (A.G.D) and 311792/2012-0 (C.C.N), National  Council of Scientific and Technological Development (CNPq).
The work of A.C.B.M. is supported by CAPES. P.V. was supported by DFG SFB grant 676 in the initial phase of this project and is now supported by
the DFG cluster of excellence ``Origin and Structure of the Universe'' (www.universe-cluster.de).
We thank Joseph Conlon, Babette D\"obrich, Joerg Jaeckel, Axel Lindner, Alexandre Payez, and Javier Redondo for a careful reading of the manuscript and
various comments and suggestions.

\appendix

\section{Axion and ALP Couplings to Gluons, Photons, and Electrons}
\label{sec:lecouplings}

We outline here the steps necessary to extract the coefficients
$C_{ij}$, $j=g,\gamma,e$, in Eq.\,\eqref{ALP_leff}, which determine the
couplings of the axion-like fields $a'_i$ to gluons, photons and electrons.

The effective Lagrangian in Eq.\,\eqref{ALP_leff} is written in a convenient form
such that the axion and ALPs interact with electrons (and other fermions) only
through the last term, which is of derivative type.
Without a convention, several equivalent Lagrangians can be used to describe the
same physics.
The different equivalent forms are related by chiral rotations on fermionic
fields which result in a change of the functional integration measure induced by the
chiral anomalies of QCD and QED.
The change in measure is summarized by the anomaly formula due to Fujikawa \cite{Fujikawa:1979ay,Fujikawa:1980eg},
\eq{
\label{fujikawa}
\lag(q_{kR})-\theta\,\frac{\alpha_s}{8\pi}G \tG\sim
\lag(e^{-i\alpha_k}q_{kR})
-\Big(\theta+\sum_k\alpha_k\Big)\frac{\alpha_s}{8\pi}G \tG
-\Big[3\sum_k\alpha_k\big(C_{\rm em}^{(k)}\big)^2\Big]\,
\frac{\alpha_{\rm em}}{4\pi} F\tF
\,,
}
when we rewrite the whole functional integral (including the measure) in terms of
$q_{iR}'=e^{i\alpha_k}q_{kR}$ and then drop the relabeling prime. Here we use the convention that
$G\tG=\frac{1}{2}\varepsilon^{\mu\nu\alpha\beta}
G^a_{\mu\nu}G^a_{\alpha\beta}$, with $\varepsilon^{0123}=1$.
Left-handed fields $\psi_{kL}$ should be treated as $\psi_{kR}^c$.
Note that chiral rotations on electrically charged colorless fermions contribute additionally
to the third term.

In all models we treat in this paper, Yukawa terms connect the flavor universal PQ
charges of the chiral fermions to scalar fields (SM singlet fields $\sigma$ or SM Higgs fields $H_k$)
whose phases carry the axion-like fields corresponding to
Nambu-Goldstone bosons. In KSVZ like models involving an exotic colored fermion $Q$, the
axion-like field appears as the phase of the SM singlet field $\sigma$ in
the
Yukawa term
\eq{
\label{yukawa:Q}
\bar{Q}_LQ_R\sigma\,.
}
In models involving SM quarks,
leptons, and right handed singlet neutrinos, the axion-like fields appear as phases of SM Higgs
doublets $H_k$ in the Yukawa terms
\eq{
\label{yukawa:gen}
\bar{q}_{L}H_dd_{R}+
\bar{q}_{L}\tilde{H}_uu_{R}+
\bar{L}H_ll_R+
\bar{L}\tilde{H}_NN_R\,.
}
Here we have suppressed the Yukawa coefficients for simplicity and summation of fermion
fields of the same type is implied.
Depending on the model some of the doublets $H_k$ can be identified such as, e.g.,
$H_l=H_N=H_d$ in one of the versions of the DFSZ axion model.
However, quarks and Higgs doublets can carry anomalous PQ charges only if $H_d\neq
H_u$ and similar considerations apply for leptons.

Let us consider only one PQ symmetry which acts on the scalar fields $H_k$ and
$\sigma$ as
\eq{
\label{ap:PQ}
H_k\to e^{iX_{H_k}\alpha}H_k,~~
\sigma\to e^{i\alpha}\sigma\,.
}
Multiple PQ symmetries can be treated analogously.
The PQ charges $X_{H_k}$ and $X_\sigma =1$ of the scalar fields fix the parametrization
of the axion-like field $a'$ as
\eq{
\label{param:H,sigma}
H_k\to \frac{v_k}{\sqrt{2}}\begin{pmatrix}0\cr1\end{pmatrix}
e^{iX_{H_k}\frac{a'}{f_{a'}}},~~
\sigma\to \frac{v_\sigma}{\sqrt{2}}e^{i\frac{a'}{f_{a'}}}\,.
}
This parametrization induces couplings between axion-like fields and
fermions in Eqs.\,\eqref{yukawa:Q} and \eqref{yukawa:gen}, which can be removed by
the chiral rotations
\eqali{
\label{chiral}
Q_R&\to e^{-i\alpha}Q_R\,,\cr
d_R&\to e^{-iX_{H_d}\alpha}d_R\,,\cr
u_R&\to e^{+iX_{H_u}\alpha}u_R\,,\cr
l_R&\to e^{-iX_{H_l}\alpha}l_R\,,\cr
N_R&\to e^{+iX_{H_N}\alpha}N_R\,,
}
for $\alpha=a'(x)/f_{a'}$.
After these chiral rotations we obtain the second and third
term of Eq.\,\eqref{ALP_leff} through Eq.\,\eqref{fujikawa} with coefficients
\eqali{
\label{gen_formula}
C_{a'g}&= 1+N_g(X_{H_d}-X_{H_u})\,,\\
C_{a'\gamma}&=6 \big(C^{(Q)}_{\rm em}\big)^2 +
2N_g\Big[X_{H_d}\,3\big(\ums[-1]{3}\big)^2-X_{H_u}\,3\big(\ums[2]{3}\big)^2
    +X_{H_l}\Big]
\,.
}
Here, the first terms on the right hand sides are due to the exotic quark $Q$, while the
second terms are due to the SM quarks and leptons.
The coupling of $a'$ with electrons come from the kinetic term
$\bar{l}_Ri\gamma^\mu\partial_\mu l_R$, after the chiral rotation \eqref{chiral},
which induces
\eq{
(\bar{l}_R\gamma^\mu l_R) X_{H_l} \frac{\partial_\mu a'}{f_{a'}}
\sim
\ums{2}(\bar{l}\gamma^\mu\gamma^5 l)X_{H_l}\frac{\partial_\mu a'}{f_{a'}}\,.
}
We have discarded a total derivative in the last equivalence.
Therefore, we obtain the coefficient
\eq{
C_{a'e}=X_{H_l}\,,
}
in Eq.\,\eqref{ALP_leff}.
For example, for the DFSZ axion, we have
$X_{H_d}=X_d$, $X_{H_u}=-X_u$ and
$X_{H_l}=X_d,-X_u,0$ for the three types of model in Eq.\,\eqref{Cegamma},
respectively.

All the supplied formulae are written in terms of the PQ charges of the scalars.
We can rewrite them in terms of the PQ charges of fermions as follows:
\subeqali[Ca':gen]{
\label{Ca'g:gen}
C_{a'g}&=\sum_{i={\rm colored}}X_{iL}-X_{iR}\,, \\
\label{Ca'gamma:gen}
C_{a'\gamma}&=2\sum_{i={\rm charged}}(X_{iL}-X_{iR})(C^{(i)}_{\rm em})^2\,,
\\
C_{a'e}&=X_{e_L}-X_{e_R}\,.
}
Formula \eqref{Ca'g:gen} assumes the summation of all colored fermions $\psi_i$
which are in the fundamental representation of $SU(3)_c$. Analogously,
the summation in Eq.\,\eqref{Ca'gamma:gen} goes over all electrically charged
fermions such that quarks should be counted for each color.
The formulae in \eqref{Ca':gen} have the advantage that they are independent of the
specific implementation of the model (e.g. Yukawa terms) but depend only on the PQ
charges of the fermions. We emphasize that the PQ charges are defined by
relations such as \eqref{ap:PQ} and we are assuming the parametrization
\eqref{param:H,sigma}. The decay constant $f_{a'}$ is fixed by canonical
normalization of the kinetic term $(\partial a')^2$ as in Eq.\,\eqref{ALP_leff} so
that rescaling of PQ charges also leads to rescaling of $f_{a'}$, keeping the
terms in Eq.\,\eqref{ALP_leff} the same.
We have normalized the PQ charges such that the smallest of the PQ charges of the
singlet scalars is set to $X_\sigma=1$, which usually leads to
$f_{a^\prime} =\sqrt{2}\,|\aver{\sigma}|=v_\sigma$.
This normalization is also relevant for the domain wall problem; see
Appendix \ref{sec:domain-walls}.

In more general situations where we have colored fermions in representations other
than the fundamental one, Eq.\,\eqref{Ca'g:gen} generalizes to
\eqali{
\label{Cag}
   C_{a^\prime g} = 2 \left( {\rm Tr}[X_{q_L}\,T_b^2]-{\rm Tr}[X_{q_R}\,T_b	^2]\right)\,,
}
where $b$ is a fixed index without summation.
The SU(3)$_c$ generators should be normalized such that ${\rm Tr}[T_a
T_b]=I(R)\,\delta_{ab}$, with $I(R)=\frac{1}{2}$ for the fundamental representation.

\section{Effects from Gravity through Planck-Scale Suppressed Operators}
\label{sec:effc-op}

At this point we want to estimate the impact of high dimensional operators
suppressed by the Planck scale, $M_{\rm Pl}$, on the Peccei-Quinn solution for the
strong CP problem. We generalize the arguments in
Refs.~\cite{Ghigna:1992iv,Holman:1992us,Kamionkowski:1992mf,Barr:1992qq,Kallosh:1995hi} for the case involving more
Higgs fields, like
the models we worked out here. We consider a generic multi-Higgs model with two Peccei-Quinn
symmetries as in Sec. \ref{accions_bottom_up}.
We assume  that
the triplet $T$ does not have a vev so that it will not contribute to the axion and
ALP low-energy effective potential. A vev for $T$ which is
much lower than those for the doublets, i.e., $\langle T\rangle$ $\ll$  $\langle
H_b\rangle=\langle v_b\rangle/\sqrt2$, could also be considered without impacting
significantly such effective potential, but we will not take it into account.
Thus, we will only consider Planck-scale suppressed
interactions involving Higgs doublets and singlets.
Let us assume then that the operator with the lowest dimension that is suppressed by $M_{\rm Pl}$ and breaks the
Peccei-Quinn symmetries gives a contribution to the high-scale effective Lagrangian according to
\begin{eqnarray}
\lag\supset
 \frac{g}{M_{\rm Pl}^{D-4} }\,\mathcal{ O}_D =\frac{g}{M_{\rm Pl}^{D-4} }\,(H_b^\dagger H_c)^m \sigma_1^n
  \sigma_2^k+h.c.\,,
 \label{vg}
\end{eqnarray}
where $b,c=u,d,l,N$. The integer $D=2m+n+k>4$ is the operator dimension, and
$g=|g|e^{i\Delta}$ is the effective dimensionless coupling
supposedly produced by gravitational interactions, which may violate CP
($\Delta\not=0$).

The operator in Eq.~(\ref{vg}) contributes to the effective potential for the axion
and the ALP as follows
\eqali{
\label{Veff}
 V_{\eff}  \approx m_A^2 f_A^2 \left( 1 -  \cos\left(\frac{A}{f_A}\right) \right)
-\Lambda_{(D)}^4
    \cos\left(C_A^{m,n,k}\frac{A}{f_A}+C_a^{m,n,k}\frac{a}{f_A}
      + \Delta_D\right)\,,
}
where
\eqali{
\label{CACaD}
 && \Lambda_{(D)}^4 = & \,|g|\frac{(v_b v_c)^m f_{a_1^\prime}^n f_{a_2^\prime}^k}{(\sqrt2)^{D-2}
M_{\rm Pl}^{D-4}}\,,\cr
%  && C_A^{m,n,k} = & \, \left( m(X_c-X_b)\pm n\right)\frac{\cos^2 \delta}{C_{1g}}\pm
% k \frac{\sin^2 \delta}{C_{2g}}\,,\cr
 && C_A^{m,n,k} = & \frac{C_1^{m,n}}{C_{1g}}\cos^2 \delta
    +\frac{C_2^{m,k}}{C_{2g}}\sin^2\delta\,,\cr
%  && C_a^{m,n,k} = & \, \left[-\left(m(X_c-X_b)\pm n\right)C_{2g}\pm k
% C_{1g}\right]\frac{\sin\,\delta\,\cos\,\delta}{C_{1g}C_{2g}}\,,\cr
 && C_a^{m,n,k} = &\left[-\frac{C_1^{m,n}}{C_{1g}}
    +\frac{C_2^{m,k}}{C_{2g}}\right]\sin\delta\cos\delta\,,\cr
 && \Delta_D= &
\,\Delta -C_A^{m,n,k}\overline{\theta}\,,
}
and
\eqali{
\label{C1C2}
C_1^{m,n}&\equiv m(K_c-K_b)\pm n\,,\cr
C_2^{m,k}&\equiv m(X_c-X_b)\pm k\,,
}
The minus sign in $\pm n$ ($\pm k$) denotes $\sigma_1^*$ ($\sigma_2^*$) instead of
$\sigma_1$ ($\sigma_2$) in Eq. ($\ref{vg}$),
$K_{b,c}$ ($X_{b,c}$) is the PQ charge of $H_{b,c}$
associated to the PQ symmetry for which $\sigma_1$ ($\sigma_2$) is charged
and $\delta$ is the mixing angle introduced in
Eq.\,\eqref{mixing:delta}.
The first term in Eq. ($\ref{Veff}$)
is the QCD instanton potential whereas the second terms arises from the dominant
non-renormalizable operator, Eq. ($\ref{vg}$), from gravity effects breaking the
Peccei-Quinn symmetries.
We should check if these effects do not spoil the solution to the strong CP problem
by destabilizing the effective potential of Eq.\,($\ref{Veff}$).
Experimental bounds on the strong CP violation parameter require
$\big|\langle A\rangle/f_A\big|<10^{-10}$. As the
minimum of $V_{\eff}$ in the ALP direction requires
\eqali{
\label{dva}
 C_A^{m,n,k}\frac{\langle A\rangle}{f_A}+C_a^{m,n,k}\frac{\langle
a\rangle}{f_A}+\Delta_D =0\,,
}
this implies that the minimum of $V_{\eff}$ still occurs for $\langle
A\rangle=0$.

We should note that if there is more than one operator below a ``critical''
dimensionality, the above observation may not be valid anymore. In this case there
is not a single condition like Eq. ($\ref{dva}$) and some symmetry suppression must
take place. The critical dimensionality is obtained by considering that there is no
significant destabilization of $V_{\eff}$ due to operators like Eq. ($\ref{vg}$)
that break the Peccei-Quinn symmetries. This is ensured, if we require
\eqali{
\label{lcd}
C_A^{m,n,k} \Lambda_{(D)}^4 < 10^{-10} m_A^2 f_A^2\approx 10^{-10} m_\pi^2 f_\pi^2\,,
}
where we take
\eqali{
\label{sind}
\sin\left(C_A^{m,n,k}\frac{\langle A\rangle}{f_A}+C_a^{m,n,k}\frac{\langle
a\rangle}{f_A}+\Delta_D \right)\approx 1
}
to define the lowest dimensional operators that could be allowed.

The vevs of the Higgs doublets can be naturally assumed as
$v_b\approx 10^2$ GeV. So, taking $m_A f_A\simeq (0.1\,\rm GeV)^2$, $f_{a_1^\prime}\approx
f_{a_2^\prime}\approx f_A$ and $M_{\rm Pl}=10^{19}$ GeV, the condition in Eq.\,(\ref{lcd}) tell us
that only operators of the form (\ref{vg}) with dimension
\eqali{
\label{Dmin}
 D \gtrsim \frac{90.3 -\left( 20 +\log\left(\frac{f_A}{10^{12}\,{\rm GeV}}\right)\right) m +\log
(C_A^{m,n,k}|g|)}{7.15-\log\left(\frac{f_A}{10^{12}\,{\rm GeV}}\right)}\,,
}
will not affect the constraint $\big|\langle A\rangle/f_A\big|<10^{-10}$. For instance, if
$C_A^{m,n,k}|g|\approx 1$, $f_A\approx 10^{12}$ GeV, Eq. (\ref{Dmin}) requires $D\gtrsim 13$ for $m=0$, $D\gtrsim 10$ for $m=1$, or $D\gtrsim 7$ for $m=
2$.

We have verified that the model we have presented in Sec. \ref{sec:unificaccion} is safe from gravitational effects
since the dangerous operators defined above are all forbidden by the discrete
symmetries. The operators of the form \eqref{vg} with lowest dimension and $m=1$,
allowed by the symmetries of Table \ref{tableZns}, are
\eq{
\label{op:m=1}
\mathcal{O}_{14}=H_N^\dag H_d\sigma_1^{*5}\sigma_2^7\,,~~
\mathcal{O}_{15}=H_l^\dag H_u\sigma_1^{5}\sigma_2^{*8}\,,~~
\mathcal{O}'_{15}=H_N^\dag H_u\sigma_1^{*5}\sigma_2^{8}
\,,
}
except for the term $H_d^\dag H_u\sigma_2$ in \eqref{VNH} which is invariant under
$\upq[2]$ (and $\upq[1]$). We then seek all operators of dimension equal or lower
than 15 with larger $m$. For $m=2,4,5$ there are no operators besides $(H_d^\dag
H_u\sigma_2)^n$, which we discard.
For $m=3,6$, we can find the operators:
\eqali{
\label{op:m>1}
% \text{$m=2$: }&(H_d^\dag H_u)^2\sigma_2^2\,,\cr
\text{$m=3$: }&(H_l^\dag H_u)^3\sigma_2\,,(H_N^\dag H_u)^3\sigma_2^{*}\,,
  (H_l^\dag H_d)^3\sigma_2^{*2}\,,(H_N^\dag H_l)^3\sigma_2^{*2}\,,
  (H_N^\dag H_d)^3\sigma_2^{*4}\,,
  \cr
% \text{$m=4$: }&(H_d^\dag H_u)^4\sigma_2^4\,,
\text{$m=6$: }&(H_l^\dag H_u)^6\sigma_2^2\,,(H_N^\dag H_u)^6\sigma_2^{*2}\,.
}
We have again dropped the $\upq[1]$ invariant terms $(H_d^\dag H_u\sigma_2)^n$.
We can see that the operators in \eqref{op:m>1} are harmless because they
are also invariant under the \PQ symmetries.
Therefore, the operators in \eqref{op:m=1} are the lowest order operators that are
invariant under the discrete symmetries of Table \ref{tableZns} but not invariant under
the \PQ symmetries.

An estimate for the mass corrections induced by gravity for both the axion and the ALP can
be obtained from $V_{\eff}$ in Eq.\,(\ref{Veff}). The leading operator breaking
the \PQ symmetries is the first one, $\mathcal{O}_{14}$, in Eq. (\ref{op:m=1}). Such
operator leads to the following contribution to the mass matrix of the axion $A$
and ALP $a$:
\[
M_{Aa}^2=\left(
\begin{array}{cc}
 m_A^2+C_A^2\frac{\Lambda_{(D)}^4}{f_A^2} & C_A\,C_a\frac{\Lambda_{(D)}^4}{f_A^2}  \\
  &   \\
 C_A\,C_a\frac{\Lambda_{(D)}^4}{f_A^2} & C_a^2\frac{\Lambda_{(D)}^4}{f_A^2}   \\
\end{array}
\right)\,,
\]
where $C_{A,a}\equiv C_{A,a}^{1,7,5}$, defined in Eq.\,(\ref{CACaD}), are not
expected to be much greater than unity. Thus, the correction
$\delta m_A$ for the axion mass, and the generated ALP mass, $m_a$, will be of order
\eqali{
\label{magc}
\delta m_A\approx m_a\approx \mathcal{O}\left(\frac{\Lambda_{(14)}^2}{f_A}\right)
\approx 10^{-23}\,{\rm eV}\,,
}
for the same set of values for the scales used above. Hence, no significant mass
correction arises for both the axion and the ALP. The mixing is also quite small
and can be neglected. Therefore, although the ALP gain a mass from gravitational
interactions, turning it into a pseudo-Goldstone boson, its feeble mass highly
suppresses any physical effect such as the ALP decay into photons.

For models containing a photophilic ALP, the contribution from operators of the
form \eqref{vg} to the effective potential \eqref{Veff} should be analyzed
differently and needs to be separated into two parts: (a) the axion potential and
(b) the photophilic ALP mass.
The contribution for (a) should be still negligible and we seek here a photophilic
ALP mass satisfying the constraints \eqref{dec_const_transp}.
Note that in this kind of models the two scales associated to the axion and to the
photophilic ALP are not related in general.

Let us assume $\sigma_1$ carries the axion, $A=a_1'$, whereas $\sigma_2$ carries the
photophilic ALP, $a=a_2'$. To avoid a substantial $A$-$a$ mixing, we need to ensure
that the mixing terms are negligible compared to the dominant contributions giving
mass to the axion (QCD potential) and to the photophilic ALP.
The latter should be given by an operator \eqref{vg} with $n=0$.
Taking also $m=0$, for simplicity, the operator has the form
\eq{
\lag \supset
\frac{g}{M_{\rm Pl}^{k-4}}\mathcal{O}_k=
\frac{g}{M_{\rm Pl}^{k-4}}\sigma_2^k\,.
}
This operator induces for the photophilic ALP a mass
\eq{
\label{laxion:ma}
m_a=|g|^{1/2}\frac{k}{2^{k/4}}\left(\frac{f_{a_2'}}{M_{\rm
Pl}}\right)^{\mfn{\frac{k-2}{2}}}M_{\rm Pl}
\sim
  |g|^{1/2}\frac{k}{2^{k/4}}10^{28-5(k-2)}\,{\rm eV}\,
  \left(\frac{f_{a_2'}}{10^9{\rm GeV}}\right)^{\mfn{\frac{k-2}{2}}}
\,.
}

Mixed operators \eqref{vg}, for which $n,k\neq 0$, are responsible for $A$-$a$
mixing.
We assume terms with $k=0$ are negligible.
The induced mass term has the form
\eq{
\frac{1}{2}\mu^2_{12}
 \bigg(a'_2+\frac{C_1^{m,n}}{C_2^{m,k}}\frac{f_{a'_2}}{f_{a_1'}}a'_1+\delta\bigg)^2
}
where $C_1^{m,n},C_{2}^{m.k}$ were introduced in Eq.\,\eqref{C1C2} and
\eqali{
\label{mixed:12}
\mu_{12}^2&=|g|\frac{\big(C_2^{m,k}\big)^2}{2^{D/2}}
  \left(\frac{v_bv_c}{M_{\rm Pl}^2}\right)^{m}
  \left(\frac{f_{a_1'}}{M_{\rm Pl}}\right)^{n}
  \left(\frac{f_{a_2'}}{M_{\rm Pl}}\right)^{k-2}
  M_{\rm Pl}^2\,,\cr
&\sim |g|\frac{\big(C_2^{m,k}\big)^2}{2^{D/2}}
\Big(10^{28-17m-7n/2-5(k-2)}\,{\rm eV}\Big)^2
  \left(\frac{f_{a_1'}}{10^{12}{\rm GeV}}\right)^{n}
  \left(\frac{f_{a_2'}}{10^9{\rm GeV}}\right)^{k-2}\,.
}
The QCD potential and the photophilic ALP mass \eqref{laxion:ma} are not disturbed
if
\eq{
\mu_{12}^2\ll m_A^2,\,m_a^2\,.
}
We have assumed that $f_{a_2'}\le f_{a_1'}$, so that the $A$-$a$ mixing is further
suppressed by $f_{a_2'}/f_{a_1'}$.

For the models presented in Secs.\,\ref{accion_laccion_model} and \ref{singlets_model}, the lowest mass
dimension operator violating $U(1)_{\rm PQ_2}$, but respecting the discrete
symmetry is essentially
\begin{equation}
\mathcal{O}_9=\sigma_{2}^9\,.
\end{equation}
Its contribution to the photophilic ALP mass is given by Eq.\,\eqref{laxion:ma} and
results in Eq.\,\eqref{ma:IV.B}.
On the other hand, the lowest mass dimension operator with $m=0$ violating
$U(1)_{\rm PQ_1}$, but compatible with the discrete symmetry, is
\eq{
\mathcal{O}_{13}=\sigma_1^{11}\sigma_2^{*2}\,.
}
This operator contributes to $A$-$a$ mixing as \eqref{mixed:12}, with
\eq{
\mu_{12}\sim |g|^{1/2}10^{-10.5}\,{\rm eV}\,
  \left(\frac{f_{a_1'}}{10^{12}{\rm GeV}}\right)^{11/2}\,,
}
which is much smaller than both $m_a$ in Eq.\,\eqref{ma:IV.B} or $m_A$ in
\eqref{axion_mass}.
The next operators with $m>0$ that violates either $U(1)_{\rm PQ_1}$ or $U(1)_{\rm
PQ_2}$ are
\eqali{
\label{laxion:op:m>0}
m=1&: H_l^\dag H_q\sigma_2^{*7},~H_l^\dag H_q\sigma_1^{11},~H_l^\dag
H_q\sigma_2^{11}\,;
\\
m=2&: (H_l^\dag H_q)^2\sigma_2^{*5}\,;
\\
m=3&: (H_l^\dag H_q)^3\sigma_2^{*3}\,.
}
We can see their contribution to the QCD potential or to the photophilic ALP mass
are negligible. For example, the first two operators in the first line of
Eq.\,\eqref{laxion:op:m>0} give, respectively,
\eqali{
\delta m_a&\sim |g|^{1/2} 10^{-14}\,{\rm eV}
  \left(\frac{f_{a_2'}}{10^9{\rm GeV}}\right)^{5/2}\,,
\cr
\delta m_A&\sim |g|^{1/2} 10^{-20.5}\,{\rm eV}
  \left(\frac{f_{a_1'}}{10^{12}{\rm GeV}}\right)^{9/2}\,.
}
Other operators give even smaller contributions.

\section{Domain-Wall Problem in the Models}
\label{sec:domain-walls}

Although the PQ symmetry is broken by QCD instantons, the vacuum state may still be symmetric under a discrete group $Z_N\subset U(1)_{\rm PQ}$. As a result a vacuum degeneracy leading to topological defects such as domain wall might occur causing cosmological problems for certain axion models \cite{Sikivie:1982qv}.

It was  observed long ago  that in some theories with spontaneous symmetry breakdown a domain structure for the vacuum is expected \cite{Zeldovich:1974uw}. The reason is that when there are degenerated vacuum states the fields might settle down in different low energy configurations in causally disconnected regions, as the Universe temperature decreases below the energy scale where a symmetry is broken. In the case of axion models this symmetry is the $U(1)_{PQ}$. Thus, domains with distinct vacuum configurations can be form in different regions of the Universe. The boundary between two domains is a domain wall, and it is a solution of minimal energy field configurations interpolating the neighboring vacua (for a review see \cite{Vilenkin:1984ib,Shifman:2012zz}). Stable domain walls bring a cosmological problem because their contribution to the total energy density would be too large for a flat, homogeneous isotropic Universe \cite{Zeldovich:1974uw}, \cite{Vilenkin:1984ib}, \cite{Sikivie:2006ni}. Many
studies and solutions of the domain wall problem in axion models were performed  \cite{Sikivie:1982qv,Lazarides:1982tw,Georgi:1982ph,Barr:1982uj,Choi:1985iv,Barr:1986hs}.  One of the simplest solution is to ensure that the PQ phase transition occurs before inflation. In this case -- which seems however to be strongly disfavored by the recent discovery of the tensor modes in the $B$-mode power spectrum by BICEP2, as has been discussed in Sec. \ref{sec:intro} -- topological defects like domain walls created at the phase transition would be wiped out by inflation.

Before assessing the domain wall problem in the models in Sec. \ref{accions_bottom_up}  we highlight the vacuum degeneracy which occurs generically in axion models. We take into account at first just one $U(1)_{\rm PQ}$ symmetry and assume its only explicit break is due to the QCD instantons. Thus, there is a potential $V(\theta)$ as a result of the effective interaction
\eq{
\label{AGG}
\lag_{\rm eff}= \frac{\alpha_s}{8\pi}C_{a^\prime g}^\prime\theta\, G\ponto\tG
\,,
}
where $\theta(x)={a^\prime(x)}/{f_{a^\prime}}$, and $C_{a^\prime g}^\prime$ an integer number obtained
from Eq. (\ref{Cag}). The potential is not invariant under arbitrary axion field shift transformation
$a^\prime\rightarrow a^\prime+\alpha\,f_{a^\prime}$ realizing the $U(1)_{\rm PQ}$ symmetry.

For the purposes of exposing the domain wall problem we use the parametrization of the
singlet field as  (omitting the heavy fields)
\eqali{
\label{sigdw}
  \sigma(x) = \frac{v_{\rm PQ}}{\sqrt2} e^{iX_\sigma\theta}\,,
}
with the PQ charge $X_\sigma$ of $\sigma$ normalized so that all fields have integer PQ charges.  We assume the periodicity $\theta\equiv \theta + 2\pi$, or,  equivalently, the periodicity of $a^\prime$ field is $2\pi f_{a^\prime}$. It means that changing $\theta\rightarrow \theta+2\pi$ all fields that depend on  $\theta$ return to the same value. Although the shift symmetry
$\theta\rightarrow \theta+ \alpha$ is broken for arbitrary values of
$\alpha$, a shift with any discrete value in the set $\alpha_k=2\pi{k}/{|C_{a'g}|}$,
with $k=0\,,...,\,|C_{ a^\prime g}|-1$, remains as a symmetry. It means that if
$C_{ a^\prime g}\not=1$ there are different vacuum expectation values configurations
which are all degenerated with respect to the minimum of the potential, i. e.,
\eqali{
\label{Vperi}
V\left(\langle\theta\rangle \right)=V\left( \langle\theta\rangle +
\alpha_k C_{ a^\prime g}\right)\,.
}
Therefore, the vacuum is invariant under a discrete symmetry subgroup of $U(1)_{\rm PQ}$
\begin{equation}
Z_N\equiv e^{i2\pi\frac{k}{N}Q}\subset U(1)_{\rm PQ},
\end{equation}
where $Q$ is the $U(1)_{PQ}$ charge operator. The order of the discrete group, $N$, is in fact here the domain wall number $N_{\rm DW}$ and is given by $N_{\rm DW}=|C_{ a^\prime g}|$, if there is no subgroup of $Z_N$ which acts trivially on the vacuum. When there is a subgroup $Z_M$ acting trivially on the vacuum its symmetry group is $Z_{N/M}\equiv Z_N/Z_M$, and the domain wall number equal to $N_{\rm DW}/M$.

This is the essence of the domain wall problem in axion models with a single $U(1)_{\rm PQ}$ symmetry.
The analysis of this problem for the models in Sec. \ref{accions_bottom_up} requires additional observations due the fact they have two global chiral symmetries. For these cases  the domain wall number is better defined as the disconnected degenerated vacuum configurations. A treatment for the domain wall problem in theories with more than one global chiral symmetry is developed  in \cite{Barr:1982uj}, \cite{Choi:1985iv}, \cite{Barr:1986hs}. In what follows we disregard the mixing and mass corrections due nonrenormalizable effective operators generated by gravitational interactions.

We work on the same basis of the global chiral symmetries for the models in Sec. \ref{accions_bottom_up}, with $K_{\sigma_1}$ being the $U(1)_{\rm PQ_1}$ charge of $\sigma_1$, and $X_{\sigma_2}$ being the $U(1)_{\rm PQ_2}$ charge of $\sigma_2$. But differently of what we have taken before we choose here $K_{\sigma_1}$ and $X_{\sigma_2}$ not equal to one.

For the model in Sec. \ref{sec:unificaccion} we take $K_{\sigma_1}=2$ and all $U(1)_{\rm PQ_1}$ charges are integers as shown Table \ref{table22a}. This is the same as multiplying by two all  $U(1)_{\rm PQ_1}$ charges in Table \ref{table22}. Concerning the $U(1)_{\rm PQ_2}$ charges, starting with the ones defined in Table \ref{table22} we can use the hypercharge  gauge transformation to redefine the PQ charges of all fields according to $X_\psi\rightarrow X_\psi+Y_\psi X_u$, with $Y_\psi$ the hypercharge of the field $\psi$. Using the baryon number symmetry we can still redefine the quark PQ  charges by a shift $X_q\rightarrow X_q-\frac{1}{3} X_u$. Finally, using the normalization $X_u+X_d=X_{\sigma_2}=3$, all the $U(1)_{\rm PQ_2}$ are then integers as shown in Table \ref{table22a}.

\begin{table}[h]
\[
\begin{array}{|c|cccccccccccccccc|}
\hline
\psi & q_L & u_R & d_R & L & N_{R} & l_R & H_u & H_d & H_l & H_N &
\sigma_2 & T  & Q_L & Q_R & \sigma_1 & \\
\hline
K_\psi & 0 & 0 & 0 & -1 & -1 & -1 & 0 & 0 & 0 & 0 & 0 & 0 & 1 & -1 &
2 &
\\[.5ex]
\hline
X_\psi & 0 & 0 & -3 & 1 & 0 & 0 & 0 &  3 &
  1 & -1 & 3 & 0  & 0  & 0 & 0 &
\\[.5ex]
\hline
\end{array}
\]
\caption{\label{table22a}
 Equivalent $U(1)_{\rm PQ_1}$ and $U(1)_{\rm PQ_2}$ charge assignments of the fields for the
model in Sec. \ref{sec:unificaccion}.
}
\end{table}

The sets of charges in Table \ref{table22a} are equivalent to the previous ones in Table \ref{table22} in the sense that although the previous anomaly coefficients $C_{ig}$ obtained  in Sec. \ref{sec:unificaccion} change now to $C_{1g}^\prime= K_{\sigma_1}C_{1g}$ and $C_{2g}^\prime=X_{\sigma_2}C_{2g}$  the scales also change according to $f_{a^\prime_1}^\prime= f_{a^\prime_1} K_{\sigma_1}$ an $f_{a^\prime_2}^\prime= f_{a^\prime_2} X_{\sigma_2}$, so that the ratios $C_{ig}/f_{a^\prime_i}$ do not change.

The axion field $A$ is defined as, according Eq. (\ref{mixing1}),
\begin{equation}
\frac{A}{f_A} = C_{1g}^\prime \theta_1 + C_{2g}^\prime \theta_2,
\label{Amadw}
\end{equation}
with  $C_{1g}^\prime=K_{\sigma_1}=2$,  $C_{2g}^\prime=N_g X_{\sigma_2} = 9$, and  $\theta_i(x)={a_{i}^\prime(x)}/{f_{a_{i}^\prime}^\prime}$. It can be seen from the parametrization of the scalar fields that both $\theta_1$ and $\theta_2$ have period $2\pi$, i. e., all scalar fields return to the same value when shifting $\theta_{1,2}\rightarrow \theta_{1,2}+2\pi$.  The $U(1)_{\rm PQ_{1}}\times U(1)_{\rm PQ_{2}}$ symmetry transformations leads to $\theta_{1,2}\rightarrow \theta_{1,2}+\alpha_{1,2}$, with  $\alpha_{1,2}$ being the group parameters.
These general transformations are anomalous but, due the fact that $C_{2g}^\prime=9$, there is the following subset of discrete transformations leaving de vacuum invariant
\begin{equation}
Z_9\equiv e^{i2\pi\frac{k}{9}Q_2}\subset U(1)_{\rm PQ_2},
\label{Z9dw}
\end{equation}
with $k=0,\,...,8$, and $Q_2$  the $U(1)_{\rm PQ_2}$ PQ charge operator. Although $C_{1g}^\prime=2$ the group $Z_2\equiv e^{i\pi{k'}Q_1}$, with $k'=0,\,1$, acts trivially on the scalar and quark condensates so that it is not relevant here. Therefore, there are nine vacuum expectation values configurations  differing  from one to another
by action of Eq. \eqref{Z9dw}, being all degenerated with respect to the
vacuum state. Denoting collectively the vacuum expectation values of scalar fields and quark
condensates as the elements $\{\langle\varphi\rangle_k,\,\langle\overline{q_L} q_R\rangle_k\}$
in the set o degenerated vacua, the elements  are such that
\begin{eqnarray}
\langle \varphi \rangle_k & = & \langle \varphi \rangle_0e^{i2\pi\frac{k}{9}X_\varphi},\nonumber\\
\langle\overline{q_L} q_R\rangle_k & = &\langle\overline{q_L} q_R\rangle_0 \,e^{ i2\pi\frac{ k}{9}X_{q_R}}.
\label{vevsunific}
\end{eqnarray}
It happens that all elements in the set of vacua are connected. To see this we follow with the arguments in Ref. \cite{Barr:1986hs} observing first that since $\theta_i\equiv\theta_i+2\pi$ the space of possible vacua is a torus. To reach a degenerated point one has to go to a distance $\alpha_1=2\pi$ in the $\theta_1$ direction and $\alpha_2=2\pi/9$ in the $\theta_2$ direction. Also, there is a group $U(1)_\beta$  whose generator is defined in terms of the generators $Q_{1,2}$  of $U(1)_{\rm PQ_{1,2}}$ according to
\begin{eqnarray}
U(1)_\beta:& & \,\,\,\,  e^{i\beta Q_\beta}\,,\nonumber\\
& & Q_\beta=\beta_1 Q_1 + \beta_2 Q_2\,,
\label{ua}
\end{eqnarray}
where the integers are $(\beta_1,\,\beta_2)=(-4,\,1)$. These numbers are chosen in order that the $U(1)_\beta$ anomaly is equal to one,
\begin{equation}
N_\beta= \beta_1 C_{1g}^\prime +\beta_2 C_{2g}^\prime=1\,.
\label{uan}
\end{equation}
To perform a $2\pi$ rotation in $U(1)_\beta$ one needs a $-2\pi\times 4$ in the group $U(1)_{\rm PQ_1}$
and a $2\pi$ in the group $U(1)_{\rm PQ_2}$, but these two rotation brings back to the same point once
$\theta_i\equiv \theta_i+2\pi\beta_i$. Thus, all elements in the set of vacuum states are connected, and the degeneracy of the vacuum is just one by the reason that $N_\beta=1$, resulting that the domain wall number is $N_{\rm DW}=1$. The condition $N_\beta=1$
can always be reached when $C_{1g}^\prime$ and $C_{2g}^\prime$ are relative prime \cite{Barr:1986hs}, as is the case in the present model.

Another way to see that the elements
$\{\langle\varphi\rangle_k,\,\langle\overline{q_L} q_R\rangle_k\}$ are connected is to note that the symmetry
\begin{equation}
U(1)_a \equiv e^{i\gamma (9Q_1 - 2Q_2)},
\label{u1o}
\end{equation}
is free from QCD anomaly, and related to the Goldstone boson $a$. This symmetry can be used to connect all
elements transforming the scalar fields and ordinary quarks,
with $\gamma=\pi \frac{n}{9}$, with $n=0,\,...\,8$, so that there is just one vacuum ($e^{i\pi n Q_1}$ acts trivially on the vacuum).

The analysis above requires that the vacuum expectation values $\langle\sigma_1\rangle$ and
$\langle\sigma_2\rangle$ are the same. If $T_1$  is the temperature below which
$\sigma_1$ get non-vanishing, and  $T_2$  is the temperature below which
$\sigma_2$ get non-vanishing, it is shown in Ref. \cite{Barr:1986hs} that axion strings will form for
$T_1>T_2$, (or $T_1<T_2$). These "$\sigma_1$ axion strings" and "$\sigma_2$ axion strings" have
$n_1=C_{1g}^\prime/K_{\sigma_1}=1$ and $n_2=C_{2g}^\prime/X_{\sigma_2}=3$ axion domain walls, respectively.
Being $n_1$ and $n_2$ not equal to $\pm1$ is still a domain wall problem even if $N_\beta=1$
\cite{Barr:1986hs}. But a solution may still be found with the inclusion of extra quarks
with specific PQ charges in order to furnish  $n_2=1$ \cite{Georgi:1982ph}.

For the model in Sec. \ref{accion_laccion_model} only $U(1)_{\rm PQ_1}$ has the QCD anomaly. Therefore, the domain wall problem analysis goes as exposed in the beginning of this section (the same reasoning applies also to the model in Sec. \ref{yet_another_singlets_model}). The normalization leaving all the $U(1)_{\rm PQ_1}$ charges integers is such that for the model in Sec. \ref{accion_laccion_model} $K_{\sigma_1}=2$, resulting in the anomaly number $C_{1g}^\prime=2$. But in this model the discrete group
\begin{equation}
Z_2\equiv e^{i\pi k\, Q_1}\subset U(1)_{\rm PQ_1},
\label{Z2dwm2}
\end{equation}
with $k=0,\,1$, acts trivially on the vacuum defined by the condensates $\langle \sigma_1 \rangle$ and $\langle\overline{q_L} q_R\rangle$. Therefore, there is no domain wall problem in the model.

For the model in Sec. \ref{singlets_model} we also have that only $U(1)_{\rm PQ_1}$ has the QCD anomaly. But in this case the normalization $K_{\sigma_1}=2$ shows that $C_{1g}^\prime=4$. The vacuum is invariant under $Z_2\equiv Z_4/Z_2$. Therefore, the domain wall number is $N_{\rm DW}=2$. In order to avoid a domain wall in this model  we could still add a second quark singlet $Q_{2\,L,R}$ with the interaction $\sigma_1^*\overline{Q_{2L}} Q_{2R}$. This additional quark  would lead to $C_{1g}^\prime=2$, so that  $N_{\rm DW}=1$, eliminating the domain wall formation.

In our analysis here it was considered that the only explicit breaking of $U(1)_{\rm PQ_1}$ and $U(1)_{\rm PQ_2}$ is through QCD anomalies. Taking into account the gravitational interactions through Planck scale suppressed nonrenormalizable operators violating global symmetries, degenerated vacuum configurations for values within the periods of $a_{1,2}^\prime$ may not arise due the corrections for the axion potential like the second term in Eq. (\ref{Veff}). In fact, a solution for the domain wall problem in axion models is to allow for some explicitly breakdown of the PQ symmetry \cite{Sikivie:1982qv}. In our case the large discrete symmetries make a  small explicit breaking of $U(1)_{\rm PQ_1}$ and $U(1)_{\rm PQ_2}$. On the other hand, it is argued in  \cite{Hiramatsu:2012sc} that with only highly suppressed Planck scale suppressed operators long lived domain walls may still be present, posing problems for the standard cosmology in what concerns the Universe evolution.

%%%%%%%%%%%%%%%%%%%%%%%%%%%%%%%%%%%%%%%%%%%

%%%%%%%%%%%%%%%%%%%%%%%%%%%%%%%%%%%%%%%%%%%%%%%%%
\end{document}